\documentclass [11pt, twoside] {uwthesis}

\usepackage{tikz}
\usetikzlibrary{arrows}
\usepackage{amsmath,amssymb,graphicx}
 \usepackage[square, numbers, sort&compress]{natbib}
\usepackage[breaklinks]{hyperref}
\usepackage{appendix}

\begin{document}
%\preprint{INT-PUB-09-007, NT@UW-09-04}
%
% ==========   Preliminary pages 
%
\prelimpages
 
% ==========   Preliminary pages
%

%
% ----- title page
%
\Title{Relativistic Heavy Ion Collisions:  \\
Viscous Hydrodynamic Simulations and Final State Interactions}
\Author{Matthew W. Luzum}
\Year{2009}
\Program{Department of Physics}
 \titlepage  

%% --- sample stuff only -----
%% unusual footnote not found in a real thesis
%{\Degreetext{A dissertation
%%  \footnote[2]{an egocentric imitation, actually}
%  submitted in partial fulfillment of\\
%  the requirements for the degree of}
% \def\thefootnote{\fnsymbol{footnote}}
% \let\footnoterule\relax
% \titlepage
% }
%\setcounter{footnote}{0}
%% --- end-of-sample-stuff ---
 
%
% ----- signature page (put real names in these)
%

\Chair{Gerald A. Miller}{Professor}{Department of Physics}
\Signature{Gerald A. Miller}
\Signature{Paul Romatschke}
\Signature{Andreas Karch}
\signaturepage

% ----- quoteslip
%

%% --- sample stuff only -----
%\setcounter{page}{-1}
%\quoteslip{%
%Extensive copying of this demonstration thesis,
%including its input files and macro package,
%is allowable for scholarly purposes, consistent with ``fair use'' as
%prescribed in the U.S. Copyright Law.
%Requests for copying or reproduction of this thesis
%may be avoided by a simple connection to the author's web site at

%\begin{center}
%\texttt{http://staff.washington.edu/fox/tex/uwthesis.html}
%\end{center}

%\noindent
%where all the necessary files and documentation
%may be found.
%}
%% --- end-of-sample-stuff ---

% These are the real quote slips (choose one)
 
%  \thesisquoteslip

 \doctoralquoteslip

% \doctoralabstractquoteslip

%
% ----- abstract
%
\setcounter{page}{-1}
\abstract{%
In this dissertation I introduce relativistic heavy ion collisions and describe theoretical approaches to understanding them---in particular, viscous hydrodynamic simulations and investigations of final state interactions.  

The successful ideal hydrodynamic models of the collisions at the Relativistic Heavy Ion Collider (RHIC) were extended by performing \textsl{viscous} hydrodynamic simulations.  This was done by making use of the recently derived full conformally invariant second order relativistic viscous hydrodynamic equations.  % as well as two types of models for initial conditions (Glauber and Color-Glass-Condensate).  
Results for multiplicity, radial flow and elliptic flow in $\sqrt{s_{NN}}=200$ GeV Au+Au RHIC collisions are presented and the range of the ratio of shear viscosity over entropy density $\frac \eta s$ for which our hydrodynamic model is consistent with experimental data is quoted.

In addition, simulations were performed of the planned $\sqrt{s_{NN}}=5.5$ TeV Pb+Pb and $\sqrt{s}=14$ TeV p+p  collisions at the Large Hadron Collider (LHC).  The elliptic flow coefficient $v_2$ is predicted to be $10$\% larger for the Pb+Pb collisions compared to top energy RHIC collisions, and is predicted to be consistent with zero for proton collisions unless $\frac \eta s < 0.08$.
%
%Viscous hydrodynamic simulations were performed for Au-Au collisions at the Relativistic Heavy Ion Collider (RHIC) as well as the planned Pb-Pb collisions at the Larch Hadron Collider (LHC).  Results for the RHIC calculation are presented and com

Finally, final state interactions were investigated within the distorted wave emission function (DWEF) model.   Work is presented on an improved understanding of the DWEF model, and the potential effect of final state interactions in the form of a pion optical potential on the elliptic flow coefficient $v_2$ was calculated to be at the $\sim 20\%$ level.
%
%Results are presented for viscous hydrodynamic simulations of gold-gold collisions at the Relativistic Heavy Ion Collider as well as the planned lead-lead collisions at the Large Hadron Collider.  In addition, results of the calculation of elliptic flow in the distorted wave emission function (DWEF) model are presented along with work developing a better understanding of the model itself.
}
%
% ----- contents & etc.
%
\tableofcontents
\listoffigures
\listoftables

\acknowledgments{% \vskip2pc
  % {\narrower\noindent
The author wishes to express sincere appreciation to
all those who helped make graduate school bearable:
\begin{itemize}
\item
to my advisor Jerry Miller for his patient guidance% and his willingness to share his wisdom
\item
to my collaborator and mentor Paul Romatschke for (also patiently) putting up with all my stupid questions and his willingness to share his knowledge
\item
to my fellow physics graduate students for making grad student life much more interesting
\end{itemize} 
  % \par}
}

%
% ----- dedication
%
\dedication{\begin{center}to my parents\end{center}}

%
% end of the preliminary pages 
%
% ==========      Text pages
%
\textpages
% ========== Chapter 1 - Introduction to relativistic heavy ion collisions
\chapter{Prologue: Introduction, Motivation and Philosophy}
%\chapter{Introduction to Relativistic Heavy Ion Collisions}
The structure of this dissertation is as follows.  This chapter consists of a nontechnical and general discussion of the motivation behind relativistic heavy ion collisions.  Chapter \ref{experiment} briefly describes the experiments and introduces a few measured quantities that will be important for the theoretical work that is presented in the remaining chapters.  The framework of hydrodynamic theory in general is introduced in \autoref{hydro}.  Following these introductory chapters is the main body, which presents original work (collaboratively) done by the author.  In \autoref{RHIChydro}, hydrodynamic models of collisions at the Relativistic Heavy Ion Collider are introduced and results of these simulations are presented (corresponding to Refs.~\cite{Luzum:2008cw, Luzum:2008cwErr}).  Simulations of collisions at the Large Hadron Collider are given in \autoref{LHC} (corresponding to Ref.~\cite{Luzum:2009sb}).   Chapters \ref{DWEF} and \ref{DWEFv2} describe the DWEF model investigations of final state interactions (roughly corresponding to Refs.~\cite{Luzum:2008tc, Luzum:2008kt}).  Appendix \ref{v2appendix} offers additional details of the DWEF calculation of $v_2$ while appendix \ref{notation} contains a list of the conventions and notation used throughout this manuscript.
%
%All quantities will be reported using a system of units such that $c = \hbar = k_B = 1$ (``natural units").  I.e., all velocities are measured as fractions of the speed of light $c$, etc.  The convention for the space-time metric will be $g_{\mu\nu}$ = diag(1, -1, -1, -1).

A digital version of this manuscript, including high quality color figures, will be available online at 
\url{http://arxiv.org/a/luzum_m_1}.  At the time of this writing, source code and results from the viscous hydrodynamic simulations presented in chapters \ref{RHIChydro} and \ref{LHC} can be found on Paul Romatschke's webpage: \url{http://hep.itp.tuwien.ac.at/~paulrom/}.
\section{Introduction}
The goal of the work described in this dissertation is to better understand how the world works on the most fundamental levels.  By studying very small and/or simple physical systems, we can extract information about the fundamental laws that govern the world we live in---including (presumably) the behavior of the much more complex systems that we typically encounter in day-to-day life.  We do this because of simple human curiosity and a natural desire for knowledge, but also because this knowledge tends to be very useful.  When we have a detailed understanding of how the world works, we can often manipulate it for our benefit.

Of course this particular line of scientific inquiry is only one of many that are both useful and necessary.  More complex systems must be studied on their own as well.  For example, it is neither interesting nor practically possible to calculate the fundamental interactions of every molecule in a bridge when trying to determine whether it will support a load without collapsing (let alone all the atoms and electrons or quarks and gluons).  Even in the work described herein, hydrodynamic equations will be used extensively.  These equations describe a sort of coarse-grained behavior on a scale that is large compared to the fundamental microscopic dynamics to which hydrodynamic behavior is largely insensitive.  Indeed, many would argue that the study of larger scale and perhaps less fundamental behavior---e.g., chemistry, biology, materials science, medicine, etc.---is more important.  Nevertheless, I would argue that it is still very important to study these fundamental laws of nature---even in such exotic regimes as extremely high temperature nuclear matter, and even if it doesn't seem to have any obvious practical applications.  A hundred years ago, there was no reason to think that understanding the weird quantum mechanical behavior of tiny particles would be of any practical use.  On the contrary, almost none of the current technology that we all take for granted---and that enable much of the progress in other sciences---would be possible without the insights gained from these seemingly esoteric studies.
%
%By studying the fundamental laws that govern the world we live in, we are able to gain an understanding 
%More specifically, this is done here by 
%We study physics 
%By studying the behavior of matter on the smallest length scales, one hopes to uncover the fundamental laws that govern the world we live in.
%Why study physics?  
%How do we learn about the world around us?
\section{Strong Interactions, QCD Phase Diagram and the Quark-Gluon Plasma}
\subsection{What do we know about the world?}
The world as we know it is made up of matter and the forces that interact with and hold this matter together.  These interactions are usually classified into four known fundamental forces:  gravity, electricity and magnetism (electromagnetism), the weak nuclear force, and the strong nuclear force.  These fundamental forces are listed here in order of increasing strength, and therefore also generically of increasing importance as one considers behavior at smaller and smaller length scales.

For example, gravity is important for describing the movement of large collections of matter that have essentially no net electric charge, such as planets moving through the solar system.  If we want to study how atoms form into molecules and solids, on the other hand, gravity has essentially no effect because electromagnetic interactions are so much stronger and are much more important to the movement of electrically charged matter.  Going further down in scale, the structure of nuclei inside atoms is dominated by the strong and weak nuclear forces.

The goal here will be to study particular aspects of the strong nuclear force, often referred to by physicists simply as strong interactions.  Correspondingly, it will be necessary to look at very very small length scales.   This will---due to Heisenberg's uncertainty principle---involve studying the behavior of matter at very large energies that are obtained, perhaps unsurprisingly, by smashing things together in accelerators. 
\subsection{The strong force and quantum chromodynamics}
The Standard Model of elementary particle physics is composed of well-tested quantum field theories that describe all of the fundamental forces except gravity.  The strong interactions in particular are well described by a theory called Quantum Chromodynamics (QCD).  QCD describes the interaction of fundamental fields called quarks and gluons.  These quarks and gluons form the protons and neutrons that, along with electrons, form almost all of the matter we see around us.

In any particular interaction between fundamental particles in a quantum field theory such as QCD, there is a characteristic energy scale, and the strength of the interaction (quantified by a ``coupling" $g$) depends on this scale.  QCD has an unusual property called ``asymptotic freedom", which means that the strength of the interaction decreases as this energy scale increases, and vice versa.  An intrinsic energy scale for the strong interactions is $\Lambda_{QCD}$, where the coupling becomes order one.   At energies much larger than this, the coupling is small and one can usually use the familiar methods of perturbation theory to calculate various quantities in QCD.  Most of the precision tests of QCD have been done in this regime and it is relatively well understood.  When there are energy scales in a particular problem that are near or below $\Lambda_{QCD}$ (even if the energies are very large compared to atomic energy scales), things typically become much more difficult, and this will be a hindrance in the study of heavy ion collisions.
%
%The strong interactions are well described by a quantum field theory called Quantum Chromodynamics (QCD).  QCD is a non-abelian gauge theory that describes the interaction of quarks and gluons.  Quarks are described as massive spin-$\frac 1 2$ fields which transform in the fundamental representation of the SU(3) gauge group and quarks are massless spin-1 fields that transform according to the adjoint representation.
\subsection{Confinement, temperature, and phase transitions}
Related to asymptotic freedom (but on the other end of the energy spectrum) is the concept of color confinement, another property of the strong interactions.  Particles that participate in strong interactions have what's called a ``color" charge, analogous to electric charge for electromagnetic behavior (and completely unrelated to the color of visible light).  In loose terms, confinement means that it is impossible to isolate a color-charged particle such as a quark.  They are only found tightly bound together with other colored objects in overall color-neutral states.

As an example, think of a color-neutral pair of a quark and an antiquark.  Asymptotic freedom implies that if they are very close together they interact very weakly.  If one was to try to pull them apart, however, the attraction would become stronger and stronger such that it would in principle take an infinite amount of energy to completely separate them.  In reality there would eventually be so much energy between them that more quark-antiquark pairs would be created out of the vacuum, and you would just be left with multiple color-neutral states instead of the one you started with.

One can imagine, however, collecting together some strongly interacting matter and raising the temperature.  Asymptotic freedom implies that there exists some (extremely large) temperature at which the strong interactions would become so weak that individual quarks and gluons would no longer be confined.  
%Thus there is expected to exist a ``deconfined'' state of matter called the quark-gluon plasma (QGP).  
Thus, there is expected to exist a deconfinement phase transition, where the color-neutral hadrons would ``melt'' into a deconfined state of matter called the quark-gluon plasma (QGP).  

How might one go about studying such an exotic regime of physics?  Such huge temperatures are hard to come by in the current universe---the temperature required (a few trillion degrees) is perhaps 100000 times larger than that at the center of the sun.  The idea, then, is to actually create conditions in the laboratory sufficient to create a QGP.  By accelerating heavy ions (nuclei of large atoms) to extremely high energies and letting them collide, it was hoped that---if only for a tiny fraction of a second---the ``fireball'' created would be hot and dense enough to create this deconfined phase.  By carefully studying what comes out of such a collision, one could then in principle discern many properties of such a state of matter and of the strong interactions in general.
%
%Although there is much that can be learned about many facets of the strong interactions in the relativistic heavy ion collision experiments described herein, the main motivation is to investigate the properties of this novel phase of matter along with the properties of the expected phase transition itself.
%
% ========== Chapter 2 - Experiment
%
\chapter{Relativistic Heavy Ion Collisions}
%\section{Relativistic Heavy Ion Collisions}
\label{experiment} 
\section{AGS, SPS, RHIC and LHC}
%
%The first relativistic heavy ion collisions with enough energy to potentially create a quark-gluon plasma were located at the Super Proton Synchrotron (SPS), located at the European Organization for Nuclear Research, better known as CERN.   There, lead ions were collided at a top energy of 3.5 GeV per nucleon pair.
%
Once it was realized that QCD implies a deconfined state of matter, the idea of colliding heavy ions to study it was quickly adopted.  There already were relativistic heavy ion collisions being studied at the BEVALAC at Berkeley, but these were not at a sufficiently high energy to potentially create a QGP.  The first ultrarelativistic heavy ion collisions that were done to investigate the quark-gluon plasma were performed at existing particle colliders that were modified to accept heavy ion beams, most notably the Alternating Gradient Synchrotron (AGS) at Brookhaven and the Super Proton Synchrotron (SPS) at CERN \cite{Heinz:2008ds}.  These were both fixed-target experiments that, among other things, collided Au+Au at up to 11 GeV per nucleon beam energy (AGS) and Pb+Pb at up to 160 $A$ GeV (SPS).  These experiments revealed tantalizing evidence of a hot and dense state of matter that had not previously been seen \cite{Heinz:2000bk}.

The first facility specifically designed for colliding ultrarelativistic heavy ions was the Relativistic Heavy Ion Collider (RHIC) at Brookhaven National Laboratory.  There, collisions were performed by colliding beams with a \textit{center of mass} energy of up to 200 GeV per nucleon pair, leading to much more energy potentially being deposited in the collision region as well as the possibility of the resulting fireball remaining in the deconfined state for a longer period of time, compared to the earlier lower energy collisions.  All the experimental data analyzed in this dissertation are from runs at RHIC, and for simplicity the rest of chapter will focus on what has been done there.

The Large Hadron Collider (LHC) at CERN is the next generation collider.  Although most people know of it as a proton-proton collider, it was also designed to run heavy ion collisions part-time and the first runs should begin before long.  Ultimately it is planned to collide Pb+Pb beams at up to 5.5 TeV per nucleon pair.
\section{Some Relevant Observables}
\begin{figure}\begin{center}
 \includegraphics[width = 10cm]{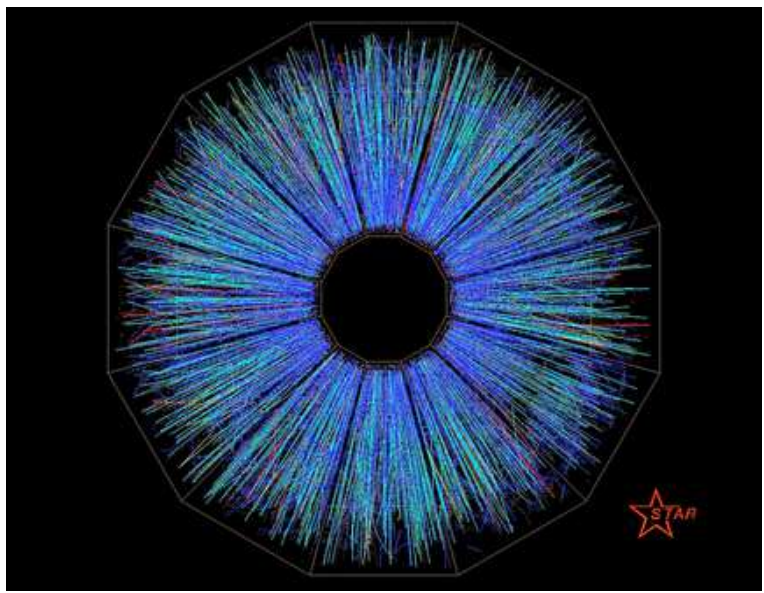}
\end{center}
\caption[Particle tracks from heavy ion collision event (STAR)]{\label{STAR}Thousands of particles are produced in a typical collision at RHIC.  Picture is of particle tracks in a $\sqrt{s}$=200 $A$ GeV Au+Au collision event from STAR collaboration \cite{STARpic}.}
\end{figure}
All the information that can be obtained from a given collision event comes from studying the thousands of produced particles that emerge from the collision region.  Any information about the evolution of the fireball system and its medium properties must be inferred by looking at these final products well after they have stopped interacting with each other, as they stream into one of the detectors surrounding the collision region  (see \autoref{STAR}).  Much of what a theorist would ideally like to measure, therefore, may be inaccessible to direct measurement.  However, a surprising amount about the collision can still be learned this way (see Refs. \cite{Arsene:2004fa, Back:2004je, Adams:2005dq, Adcox:2004mh} for an overview from each of the main detector collaborations at RHIC of the first four years of results).  The following describes the particular measured quantities that will be important for the work comprising this dissertation.
%
%The data that will be relevant here is the distribution of detected hadrons.  Photon and lepton measurements will not be discussed.
\subsection{Single particle spectra and elliptic flow}
\label{spectra}
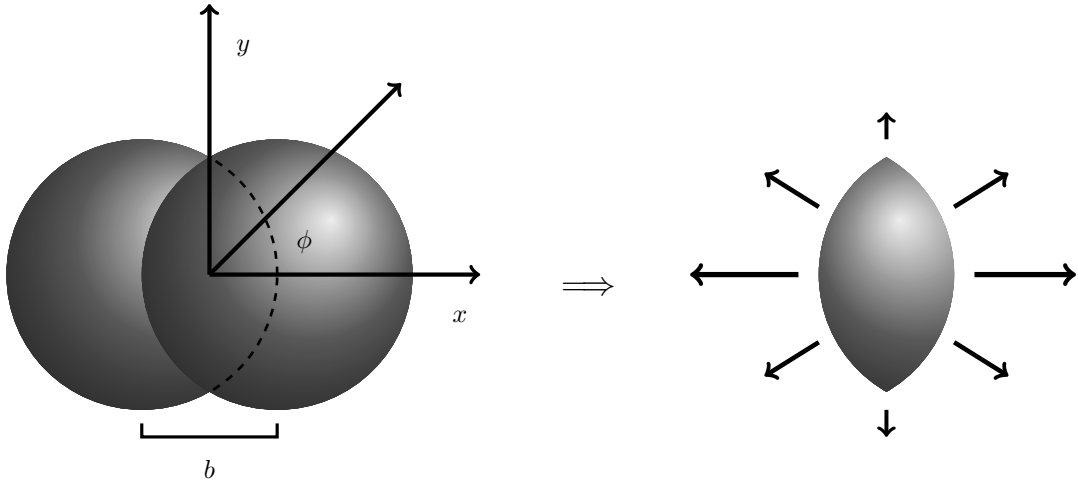
\begin{figure}
\begin{center}
\begin{tikzpicture}
 \pgftransformscale{0.9}
    {
%     \shade[bottom color=blue,top color=blue!25] (-1,0) circle (2cm);
%     \shade[bottom color=blue,top color=white] (1,0) circle (2cm);

%     \fill[shade, ball color = blue!70, shading angle = -90] (-1,0) circle (2cm);
%     \fill[shade, ball color=blue!70, shading angle = -90] (1,0) circle (2cm);
     \fill[shade, ball color = gray, shading angle = -90] (-1,0) circle (2cm);
     \fill[shade, ball color = gray, shading angle = -90] (1,0) circle (2cm);

%     \draw[dashed] (-1,0) circle (2cm);
%     \draw (1,0) circle (2cm);
   \begin{scope}
   \clip (1,0) circle (2cm);
   \draw[dashed, line width = 1 pt]  (-1,0) circle (2cm);
   \end{scope}
     \draw[very thick] (-1,-2.2) -- (-1,-2.4) -- (1,-2.4) -- (1,-2.2);
      \pgftext[base,x=0cm, y=-3.0cm] {$b$}
%   \begin{scope}
%    \clip (-1,0) circle (2cm);
%    \fill[color=DarkBlue!80] (1,0) circle (2cm);
%%    \fill[shade, bottom color=blue, top color=lightgray] (.5,0) circle (1cm);
%   \end{scope}
}   {
  \draw[->,ultra thick] (0,0) -- (4,0);
   \draw[->,ultra thick] (0,0) -- (0,4);
   \draw[->,ultra thick] (0,0) -- (45:4);
   \pgftext[base,x=1.4cm, y=.4cm] {$\phi$}
   \pgftext[base,x=3.7cm, y=-.7cm] {$x$}
   \pgftext[base,right, x=.6cm, y=3.3cm] {$y$}
%   \pgftext[base, left , x=-2cm, y = 3cm] {$\odot z$}
   \pgftext[base,x= 5.6cm, y=-.3cm] {\Large{$\Longrightarrow$}}
   
      \begin{scope}
    \clip (9,0) circle (2cm);
    \fill[shade, ball color=gray] (11,0) circle (2cm);
   \end{scope}
   \draw[->, line width = 1.5 pt] (10,2.) -- (10,2.4);
   \draw[->, line width = 1.5 pt] (10,-2) -- (10,-2.4);
   \draw[->, line width = 2 pt] (11.3,0) -- (12.8,0);
   \draw[->, line width = 2 pt] (8.7,0) -- (7.1,0);
   
   \draw[->, line width = 1.75 pt] (11,1) -- (11.8,1.5);
   \draw[->, line width = 1.75 pt] (9,1) -- (8.2,1.5);
   \draw[->, line width = 1.75 pt] (11,-1) -- (11.8,-1.5);
   \draw[->, line width = 1.75 pt] (9,-1) -- (8.2,-1.5);
   
%  \fill[shade, ball color = blue, shading angle = -90] (9,0) ellipse (1 and 1.8);
}
%\useasboundingbox (-3,5,4) rectangle (20,5);
  \end{tikzpicture}
\end{center}
  \caption[Diagram of coordinate definitions for elliptic flow]{\label{cartoon}Cartoon of a heavy ion collision.  The beam direction is the $z$ direction (out of the page) while the $x$ direction is in principle defined by the impact parameter $b$ [left side of figure].  Anisotropic pressure gradients in the material deposited in a non-central ($b$ $\neq$ 0) collision can lead to elliptic flow [right].}
\end{figure}

First, note some standard definitions:  The beam direction defines the $z$-axis, and the $x$ direction is defined such that the $x$-$z$ plane is the collision plane as seen in \autoref{cartoon}.  It is often useful to use polar coordinates in the transverse ($x$-$y$) plane, where the azimuthal angle $\phi$ is measured with respect to the $x$-axis.  One can also define an impact parameter $b$ connecting the lines-of-flight of the centers of mass of the colliding nuclei.  

%
%
%One should note that the colliding nuclei consist of a collection of well-localized nucleons and so are not as smoothly distributed as \autoref{cartoon} might suggest.  Therefore there can be some ambiguity in defining, e.g., the collision plane.  The theoretical calculations presented here, however, will model the collisions with smooth initial conditions and these definitions are then completely unambiguous.  The difficulty then comes in knowing which experimental results to compare the theoretical results to.  This will be discussed briefly in \autoref{RHIChydro}.

The detected particles are characterized by their momentum $p$ after they exit the collision region.  Instead of the longitudinal component of momentum $p_z$ or the polar angle with respect to the beam $\theta$, what is more commonly reported is the particle rapidity $Y \equiv {\rm arctanh}(p_z/E)$ or pseudorapidity $\eta \equiv -{\rm ln}[\tan(\frac \theta 2)]$.  (Note that rapidity and pseudorapidity are equivalent for a relativistically moving particle ($m\to 0$) since $Y = \frac 1 2 \ln \left[ \frac {E + p_z} {E - p_z} \right]$ and $\eta = \frac 1 2 \ln \left[ \frac {|\vec{p}| + p_z} {|\vec{p}| - p_z} \right]$.  Thus they are often used interchangeably, although one should always keep in mind the limits to the validity of this equivalence.)
%Only mid-rapidity ($Y$ = 0) data will be considered in the work presented here and $\eta$ will always refer to shear viscosity rather than pseudorapidity.

The probability of detecting a particle in a given event with a given rapidity and transverse momentum is given by the distribution $\frac {dN} {dY\, d^2p_T} = E \frac {dN} {d^3p}$ such that
\begin{equation}
N = \int dY\,d^2p_T \frac {dN} {dY\, d^2p_T},
\end{equation}
where $N$ is the total number of particles in the event and $p_T = \sqrt{p_x^2 + p_y^2}$ is the transverse momentum.  This distribution can refer to a particular species of identified particle (e.g., protons) or a composite measurement such as all charged hadrons combined.  From this, one can calculate the mean transverse momentum
\begin{equation}
\langle p_T \rangle = \left ( \int d^2p_T\, p_T\, \frac {dN} {dY\, d^2p_T} \right) / \left ( \int d^2p_T\, \frac {dN} {dY\, d^2p_T} \right).
\end{equation}

It is useful to break up the distribution into its Fourier components with respect to the azimuthal angle of the outgoing particle's momentum $\phi_p \equiv \tan^{-1}(\frac {p_y} {p_x})$:
\begin{equation}
\frac {dN} {dY\, d^2p_T} = v_0 \left[1 + \sum_{n = 1}^\infty 2 v_n\, \cos(n\, \phi_p) \right].
\end{equation} 
The sine terms vanish due to the reflection symmetry about the collision ($x$-$z$) plane, and for collisions of identical nuclei the reflection symmetry about the $y$-$z$ plane causes the odd cosine moments to vanish.  $v_0(p_T)$ is referred to as radial flow and the next lowest non-vanishing coefficient $v_2(p_T)$ is called the elliptic flow coefficient.   Explicitly we have

\begin{equation}
v_0 \equiv \int \frac {d\phi_p} {2\pi} \frac {dN} {dY\, d^2p_T},
\end{equation}
\begin{equation}
v_2 \equiv \langle \cos(2\phi_p) \rangle = \frac 1 {v_0} \int \frac {d\phi_p} {2\pi} \cos(2\phi_p) \frac {dN} {dY\, d^2p_T}.
\end{equation}

The momentum integrated elliptic flow coefficient is denoted
\begin{equation}
v_2^{int} \equiv \frac{\int d^2p_T\, v_2\, v_0}
{\int d p_T\, v_0 },
\end{equation}
and the minimum bias $v_2$ is defined by averaging over all collisions in a given run---i.e., integrating over impact parameter \cite{Kolb:2001qz}
\begin{equation}
v_2^{{\rm mb}} = \frac{\int db\, b\, v_2(b)\, v_0(b)}
{\int db\, b\, v_0 (b)}.
\end{equation}

One should note that the colliding nuclei consist of a collection of well-localized nucleons and so are not as smoothly distributed as \autoref{cartoon} might suggest.  Therefore there can be some ambiguity in defining, e.g., the collision plane.  (Also note that the odd moments of the particle distribution are not exactly zero).  The theoretical calculations presented here, however, will model the collisions with smooth initial conditions and these definitions are then completely unambiguous.  The difficulty then comes in knowing which experimental results to compare the theoretical results to.  (See, e.g.,  \autoref{RHIChydro}, comparing minimum bias $v_2$ results to different experimental extractions that attempt to remove ``non-flow'' effects).
%From this distribution one can derive several useful quantities:
%\begin{align}
%v_0 &\equiv \int \frac {d\phi_p} {2\pi} \frac {dN} {dY\, d^2p_T}\\
%\langle p_T \rangle &\equiv \frac {\int d^2p_T\, p_T\, v_0} {\int d^2p_T\, v_0}\\
%v_2 \equiv 
%\end{align}
%
%
\subsection{Two-particle correlations---Hanbury Brown/Twiss interferometry}
\label{HBT}
In the 1950's, Robert Hanbury Brown and Richard Q. Twiss began using a method of intensity  interferometry to measure the sizes  of various objects---most notably measuring the angular size of the star Sirius in 1956 \cite{HanburyBrown:1954wr, HanburyBrown:1956pf}.  It turns out that by correlating the intensity of light emitted incoherently from a source---even without measuring  any information about phase (a mandatory ingredient of typical amplitude interferometry)---one can directly measure information about its size, as well the time dependence of a source that varies in time.  This effect can be thought of as being caused by the symmetrization of the wavefunction for identical bosons, such as the photons emitted from a star, which causes enhancement of coincident measurement of pairs of these bosons with small momentum difference.  (Similarly, an ``anti-bunching'' effect is present for identical fermions because of the antisymmetric nature of their wavefunction).

Despite early skepticism, this technique was quickly employed to measure space-time properties of collision systems in the laboratory by analyzing two-particle correlations of identical particles emitted from such collisions \cite{Goldhaber:1960sf}, and has since been used extensively to study heavy ion collisions \cite{Lisa:2005dd}.  Such analyses of two-particle correlations are often generically referred to as Hanbury Brown/Twiss (HBT) interferometry, or simply femtoscopy.

Explicitly, the quantity constructed to analyze RHIC events is the ratio of the two-particle inclusive and single-particle inclusive spectra:
\begin{equation}
C(p_1,p_2) \equiv \frac {dN/(d^3p_1 d^3p_2)} {(dN/d^3p_1)(dN/d^3p_2)}.
\end{equation}
The numerator measures the probability that two particles of momentum $p_1$ and $p_2$ are detected in the same event, while the denominator is the product of the familiar spectra from \autoref{spectra}.  

Define the average momentum $K \equiv (p_1 + p_2)/2$ and the momentum difference $q \equiv (p_1 - p_2)$.  Then one can define the directions $L$ (longitudinal), $O$ (out), and $S$ (side) as the directions parallel to the beam, parallel to $K_T \equiv \sqrt{K_x^2 + K_y^2}$, and perpendicular to both the beam and $K_T$, respectively.

The correlation function can then be parameterized as a Gaussian with parameters that are fit to data:% \cite{Bertsch:1988db}:
 \begin{equation}
 C(p_1,p_2) = C(K, q) \approx 1 + \lambda \exp (- R_O^2 q_O^2 - R_S^2 q_S^2 - R_L^2 q_L^2),
 \end{equation}
 or, for small $q$
 \begin{equation}
 \label{smallq}
  C(K, q) \approx 1 + \lambda (1 - R_O^2 q_O^2 - R_S^2 q_S^2 - R_L^2 q_L^2).
 \end{equation}
 For a static Gaussian source, these HBT radii ($R_O$, $R_S$, $R_L$)  would be independent of $|{\vec K}|$, and would reveal the spatial extent of the source.  For a general dynamic source, the radii can be complicated functions of $K$, and their interpretation more complicated.
%It can also be useful to look at multi-particle correlations---i.e., what is the probability of $n$ identical particles with specified momenta to be detected in the same collision event.  In fact, it is in this way that one can actually gain much more direct information about the space-time structure of the ``fireball'' created in a heavy ion collision than can be gained by focusing only on single particle properties.

%
% ========== Chapter 3 - Hydrodynamics
\chapter{Hydrodynamics}
\label{hydro}
\section{Introduction}
Before delving in to the specifics of viscous hydrodynamic models of relativistic heavy ion collisions, it is useful to know a bit about the theory of hydrodynamics in general and in particular the development of relativistic viscous hydrodynamics (see Paul Romatschke's lecture notes \cite{Romatschke:2009im} for a nice, more detailed, treatment).

Hydrodynamics---also known as fluid dynamics---is the theory governing the motion of fluids.  As the name implies, it was initially developed to describe the dynamics of water, but can be applied to fluid behavior of a wide range of materials.  It can be thought of as an effective theory describing the long wavelength behavior of a system that has sufficient separation of scales such that this macroscopic motion is so slowly varying in space and time so as to be insensitive to the microscopic dynamics.  In the case of water, for example, if the macroscopically-averaged quantities such as pressure and temperature change very slowly in space compared to the average distance between molecules and very slowly in time compared to the scattering rate of the individual molecules, it will behave according to the equations of hydrodynamics.

Likewise, there should be hydrodynamic regimes for systems consisting of a collection of hadrons or even a quark gluon plasma.  One can imagine a hypothetical large system that has had enough time to everywhere come very close to thermal equilibrium, yet still has a temperature gradient that slowly varies across the system.  This system would behave hydrodynamically, even if it consists of a very hot and dense collection of strongly interacting matter.

It is a more difficult question, however, whether the medium created in a relativistic heavy ion collision interacts strongly enough to behave like a fluid for the short period of time before it flies apart.  Although it now appears from the success of hydrodynamical models that this is likely the case, it was not at all obvious that the hydrodynamic description should be correct and it was originally believed by many to be unlikely.
%
%
%
%
%\section{The Hydrodynamic Equations}
\section{Non-Relativistic Fluid Dynamics}
\label{nonrel}
The conventional (non-relativistic) formulation of the hydrodynamic equations describes the evolution of the fluid velocity ${\vec v}(t,{\vec x})$,
the pressure $p(t,{\vec x})$, and the mass density $\rho(t,{\vec x})$ of a fluid at each point in space and time via the equations \cite{Euler},\cite{Landau}\S2,
\begin{align}
\partial_t \rho + \rho\, {\vec \partial}\cdot {\vec v}+{\vec v}\cdot {\vec \partial} \rho &= 0\ .\label{Continuityeq}\\
\partial_t {\vec v}+\left({\vec v}\cdot {\vec \partial}\right) {\vec v}&= -\frac{1}{\rho} {\vec \partial} p\,,
 \label{Eulereq}
\end{align}
These equations are referred to as the continuity equation (\ref{Continuityeq}) and Euler equations (\ref{Eulereq}), respectively, and are simply the statements of conservation of mass and momentum for a continuous fluid without dissipation (an ``ideal" fluid).  To close the set of equations another relation is needed---usually given as an equation of state of the material $p=p(\rho)$. The Euler equations can be generalized to treat dissipative effects 
\begin{align}
\frac{\partial v^i}{\partial t}+v^k \frac{\partial v^i}{\partial x^k}&=-\frac{1}{\rho} \frac{\partial p}{\partial x^i}
-\frac{1}{\rho}\frac{\partial \Pi^{k i} }{\partial x^k}\,,
\label{NS}\\
\Pi^{ki}&=-\eta \left(\frac{\partial v^i}{\partial x^k}+\frac{\partial v^k}{\partial x^i}-\frac{2}{3}
\delta^{k i} \frac{\partial v^l}{\partial x^l}\right)-\zeta\, \delta^{i k} \frac{\partial v^l}{\partial x^l}\ .
\label{visctens}
\end{align}
These are called the Navier-Stokes equations  \cite{Navier},\cite{Landau}\S15.  The coefficients in the viscous stress tensor $\Pi^{ki}$ are termed the shear viscosity ($\eta$) and bulk viscosity ($\zeta$).  Their values, like the form of the equation of state, depend on the specific fluid in question and encode information about the microscopic dynamics of that material.
\section{Relativistic Ideal Hydrodynamics}
\label{rel}
%For a relativistic system one must generalize this formulation.  Note that this is the case even if the fluid velocity is always much less than the speed of light, providing there is sufficient microscopic motion such that the mass density $\rho(t,{\vec x})$ is no longer a useful degree of freedom.
%
Let us generalize this so that it can be applied to a relativistic system.  Any system can be characterized by its energy-momentum tensor $T^{\mu\nu}(x)$, which is a symmetric tensor that describes the distribution of energy and momentum in the system.  In a given reference frame the time-time component $T^{00}$ is the energy density, the time-space component $T^{0i} = T^{i0}$ is the $i$'th component of the momentum density, and the space-space component $T^{ik}$ is the flux of $i$'th momentum across the $x^k$ surface.

In relativistic notation, the statement of conservation of energy and momentum is simply
\begin{equation}
\label{cons}
\partial_\mu T^{\mu\nu} = 0.
\end{equation}
Any other conserved quantity (e.g., electric charge, baryon number, etc.) is characterized by a conserved current $j^\mu(x)$ that describes its charge density and current.  The conservation equations then include an additional equation, 
\begin{equation}
\partial_\mu j_n^\mu = 0,
\end{equation}
for each conserved quantity (labeled by $n$).  For simplicity it will be assumed that these additional conservation equations are unimportant to the motion of the fluid and so will be neglected in the following.

Define the local fluid rest frame at each point in space-time as the zero-momentum frame, $T^{0i}(x) = 0$.  The velocity of this local rest frame with respect to a fixed lab frame defines a fluid 4-velocity $u^\mu(x)$ such that in the rest frame $u_{rest}^\mu = (1, 0, 0, 0)$ (recall that for a 4-velocity, $u^2\equiv u^\mu u_\mu = 1$).  So then $u_\mu T^{\mu\nu} = \epsilon\, u^\mu$, where $\epsilon (x)$ is defined as the energy density in this fluid rest frame.

The equations of ideal (relativistic) hydrodynamics then emerge from the conservation equations \ref{cons} when the energy-momentum tensor has the property that it is isotropic (i.e., rotationally invariant) in the local rest frame:
$$
T^{\mu\nu}_{ideal, rest} = \left( \begin{array}{cccc}
          \epsilon & 0 & 0 & 0 \\
                      0 & p & 0 & 0 \\
                      0 & 0 & p & 0 \\
                      0 & 0 & 0 & p 
                     \end{array} \right).
$$
In an arbitrary fixed reference frame, in covariant notation, this is
\begin{equation}
T_{ideal}^{\mu\nu} =  (\epsilon+p)\ u^{\mu} u^{\nu} - p\ g^{\mu\nu} = \epsilon\ u^\mu u^\nu- p\ \Delta^{\mu \nu},
\end{equation}
where $p(x)$ is then the isotropic pressure in the rest frame and $g^{\mu\nu} ={\rm diag}(1, -1, -1, -1)$ is the metric tensor.   $\Delta^{\mu \nu} \equiv (g^{\mu \nu}-u^\mu u^\nu)$ is a projection operator on the space orthogonal to the fluid velocity.  It has the properties $\Delta^{\mu \nu}u_\mu = \Delta^{\mu \nu}u_\nu=0$
and $\Delta^{\mu \nu} \Delta_\nu^\alpha=\Delta^{\mu \alpha}$.  It is often useful to use this projector to express the hydrodynamic equations projected into directions parallel 
($u_\nu \partial_\mu T^{\mu \nu}$) and perpendicular 
($\Delta^\alpha_\nu \partial_\mu T^{\mu \nu}$) 
to the fluid velocity.  Explicitly they are:
\begin{gather}
u_\nu \partial_\mu T_{ideal}^{\mu \nu} = (\epsilon + p)\partial_\mu u^\mu+u^\mu \partial_\mu \epsilon = (\epsilon+p)\partial_\mu u^\mu + D\epsilon  = 0 \ ,\label{pe1}\\
\Delta^\alpha_\nu \partial_\mu T_{ideal}^{\mu \nu} = (\epsilon+p)\, u^\mu \partial_\mu u^\alpha- \Delta^{\mu \alpha} \partial_\mu p = (\epsilon+p)D u^\alpha-\nabla^\alpha p = 0\ , \label{pe2}
\end{gather}
where we have also introduced shorthand notation for the projection of derivatives parallel ($D\equiv u^\mu \partial_\mu$) and perpendicular ($\nabla^\alpha = \Delta^{\mu \alpha} \partial_\mu$) to the fluid velocity.

This system of equations is closed by specifying the equation of state $p = p(\epsilon)$.  If this is the equilibrium equation of state for the system in question, these equations describe a system that is everywhere in local thermal equilibrium.  Thus, the language used when talking about hydrodynamics is often that of thermal equilibrium, but note that all that is required for Equations \ref{pe1} and \ref{pe2} to be valid is isotropy in the fluid rest frame.

For fluid velocities much less than the speed of light [$u^\mu \simeq (1,{\vec v})$], and when the energy density is dominated by the mass density ($\epsilon \simeq \rho$), these relativistic ideal hydrodynamic equations reduce to the non-relativistic Euler and continuity equations from \autoref{nonrel} \cite{Romatschke:2009im}.

It should be noted that these (non-linear) ideal hydrodynamic equations contain instabilities that make them impossible to solve numerically---at least using the most na\"{\i}ve of algorithms.  To avoid these problems, numerical algorithms are used that contain what amounts to ``numerical viscosity'' that dampens the instabilities \cite{Romatschke:2009im}.  Adding viscous terms to the equations also fixes these instabilities, and so solving the viscous hydrodynamic equations will not require these algorithms.
\section{Relativistic Viscous Hydrodynamics}
\label{vischydro}
These equations can be generalized to include dissipative, or viscous, effects.  This is done by allowing for a more general energy-momentum tensor:
\begin{equation}
T^{\mu \nu}=T^{\mu \nu}_{ideal}+\Pi^{\mu \nu}.
\end{equation}
$\Pi^{\mu \nu}$ is the viscous stress tensor
that includes the contributions to $T^{\mu \nu}$ from 
dissipation.  The hydrodynamic equations then become \cite{Romatschke:2009im}
\begin{align}
D\epsilon +(\epsilon+p)\partial_\mu u^\mu -\Pi^{\mu \nu}\nabla_{(\mu} u_{\nu)} &= 0\,,\nonumber\\
(\epsilon+p)Du^\alpha-\nabla^\alpha p + \Delta^\alpha_\nu \partial_\mu \Pi^{\mu \nu} &= 0\ ,
\end{align}
where the $(\ldots)$ denote symmetrization, e.g.,
$$
\nabla_{(\mu} u_{\nu)}=\frac{1}{2}\left(\nabla_\mu u_\nu+\nabla_\nu u_\mu\right)\ .
$$
Of course, one must still specify the form of $\Pi^{\mu \nu}$.  Determining the correct form to use turns out to be less straightforward than in the non-relativistic case, and until recently there has been a number of versions in use.
\subsection{Relativistic Navier-Stokes equations}
If the macroscopically-averaged quantities from ideal hydrodynamics ($\epsilon$, $p$, $u^\mu$) vary slowly in space and time, it  can be useful to build a controlled gradient expansion of $T^{\mu\nu}$; i.e., an expansion in powers of derivatives of these quantities.  Using this perspective, $T_{ideal}^{\mu\nu} = T_0^{\mu\nu}$ is just the zeroth order term in such an expansion, and $\Pi^{\mu \nu}$ contains first and higher order derivative terms.

To first order in gradients, there are only two independent terms that are consistent with the symmetries of $\Pi^{\mu \nu}$ (it must be symmetric, $\Pi^{\mu \nu} = \Pi^{\nu \mu}$, so that $T^{\mu\nu}$ also remains symmetric, and it must be transverse to the fluid velocity, $u_\mu \Pi^{\mu \nu} = 0$, to retain the definition of the zero-momentum rest frame).  They are usually separated into a traceless term $\pi^{\mu\nu}$ and the remainder $\Pi$
\begin{equation}
\Pi^{\mu\nu} = T_1^{\mu \nu} = \pi^{\mu\nu} + \Delta^{\mu \nu} \Pi = \eta \nabla^{\langle\mu} u^{\nu\rangle} + \zeta\, \Delta^{\mu \nu} \nabla_\alpha u^\alpha.
\end{equation}
The angle brackets define a quantity that is traceless, symmetric, and transverse
$$
\nabla_{\langle\mu} u_{\nu\rangle}\equiv2 \nabla_{(\mu} u_{\nu)}-\frac{2}{3} \Delta_{\mu \nu} \nabla_\alpha u^\alpha\,,
$$
or, in general
\begin{equation}
A^{\langle\alpha}B^{\beta\rangle} \equiv P^{\alpha \beta}_{\mu \nu} A^\mu B^\nu\,,
\end{equation}
with $P^{\mu \nu}_{\alpha \beta}=\Delta^{\mu}_\alpha \Delta^\nu_\beta+\Delta^{\mu}_\beta \Delta^\nu_\alpha-\frac{2}{3} \Delta^{\mu \nu} \Delta_{\alpha \beta}$.
Upon plugging into the conservation equations, this results in what are termed the relativistic Navier-Stokes equations, since they reduce to Navier-Stokes in the non-relativistic limit.  $\eta$ is then identified as the shear viscosity and $\zeta$ the bulk viscosity.  Ideal hydrodynamics is recovered when these transport coefficients are set to zero (appropriate when gradients are so small compared to the coefficients that the terms can be neglected).  These terms are often derived in the literature by demanding that the second law of thermodynamics always be obeyed, $\partial_\mu s^\mu \ge 0$, where $s^\mu = s\, u^\mu $ is the entropy density current \cite{Romatschke:2009im}, rather than using the perspective of a gradient expansion to first order.
%
%Each additional conserved current and corresponding conservation equation would also add a transport coefficient (e.g., heat conductivity), but again we neglect this for simplicity.
%
%
%
%
 \subsection{Causality restored: second-order relativistic viscous hydrodynamics}
There are problems with the relativistic Navier-Stokes equations, however, which in general make them difficult to solve numerically.   This is caused by the presence of acausal signal propagation, and associated instabilities.

The problem can be illustrated by considering small perturbations of the energy density and fluid velocity, and tracking how these disturbances travel through the medium.  If one decomposes the perturbations into Fourier modes, one finds that the diffusion speed of a particular mode increases linearly with wavenumber.  A mode with arbitrarily large wavenumber will have an arbitrarily large speed (larger than the speed of light), and causality is violated \cite{Romatschke:2009im}.

On the surface this shouldn't necessarily be worrisome.  Large wavenumber (or short wavelength) modes are outside the realm of applicability of hydrodynamics---if there is significant short distance behavior, the gradient expansion will not converge and hydrodynamics is not an appropriate description of the system anyway.  In practice, however, this acausal behavior causes instabilities that make constructing a numerical solution with arbitrary initial conditions impossible \cite{Hiscock:1985zz}.

It turns out that, as with the (unrelated) instabilities of ideal hydrodynamics, these instabilities can be removed by adding higher order gradient terms to the equations.

At second order in gradients, one can construct 15 independent terms (in addition to the zeroth and first order terms already mentioned) \cite{Romatschke:2009kr}.   In the shear (traceless) sector we have
\begin{align}
\label{generalsecondorder}
\pi^{\mu \nu} =\, &\eta \sigma^{\mu \nu}-\eta\tau_\pi \left[^\langle D \sigma^{\mu \nu\rangle}+\frac 4 3 (\nabla\cdot u) \sigma^{\mu \nu}\right] 
 +\frac \kappa 2 \left[R^{\langle\mu \nu\rangle}+2 u_\alpha u_\beta R^{\alpha \langle\mu \nu\rangle\beta}\right]
\nonumber\\
& - \frac {\lambda_1} {2} \sigma^{\langle\mu}_{\quad \lambda}\sigma^{\nu\rangle \lambda}+
\frac {\lambda_2} {2} \sigma^{\langle\mu}_{\quad \lambda} \omega^{\nu\rangle \lambda}
- \frac {\lambda_3} {2} \omega^{\langle\mu}_{\quad \lambda}\omega^{\nu\rangle \lambda}\nonumber\\
& -\kappa^* u_\alpha u_\beta R^{\alpha \langle\mu \nu\rangle \beta} - \eta \tau_\pi^* \frac 4 3 (\nabla\cdot u) \sigma^{\mu \nu}
+ \frac {\lambda_4} {2} \nabla^{\langle\mu} \ln s \nabla^{\nu\rangle}\ln s\,,
\end{align}

%\begin{align}
%\label{generalsecondorder}
%\pi^{\mu \nu} =\, &\eta \sigma^{\mu \nu}-\eta\tau_\pi \left[^\langle D \sigma^{\mu \nu\rangle}+\frac{\nabla\cdot u}{3} \sigma^{\mu \nu}\right]
%+\kappa\left[R^{\langle\mu \nu\rangle}+2 u_\alpha u_\beta R^{\alpha \langle\mu \nu\rangle\beta}\right]
%\nonumber\\
%& -\lambda_1 \sigma^{\langle\mu}_{\quad \lambda}\sigma^{\nu\rangle \lambda}+\lambda_2 \sigma^{\langle\mu}_{\quad \lambda} \omega^{\nu\rangle \lambda}
%-\lambda_3 \omega^{\langle\mu}_{\quad \lambda}\omega^{\nu\rangle \lambda}\nonumber\\
%& -\kappa^* u_\alpha u_\beta R^{\alpha \langle\mu \nu\rangle \beta} - \eta \tau_\pi^* \frac{\nabla\cdot u}{3} \sigma^{\mu \nu}
%+ \frac {\lambda_4} {2} \nabla^{\langle\mu} \ln s \nabla^{\nu\rangle}\ln s\,,
%\end{align}
%
while in the bulk sector the most general form is
\begin{align}
\Pi =\, & \zeta\left(\nabla \cdot u\right)-\zeta \tau_\Pi D \left(\nabla\cdot u\right)
-\xi_1 \sigma^{\mu \nu} \sigma_{\mu \nu}-\xi_2 \left(\nabla \cdot u\right)^2
\nonumber\\
& -\xi_3 \omega^{\mu \nu} \omega_{\mu \nu}
+\xi_4 \nabla_\mu \ln s \nabla^\mu \ln s+\xi_5 R
-\xi_6 u^\alpha u^\beta R_{\alpha \beta}\,,
\end{align}
%
%with notational definitions
%\begin{align}
%\sigma^{\mu \nu} $= \nabla_{\langle\mu} u_{\nu\rangle} $ D \equiv u^\mu D_\mu
%\end{align}
where $s$ is the entropy density.  Since we are assuming vanishing conserved charge densities (i.e., zero chemical potential), this is given by $s = \frac {\epsilon + p} {T}$.  
Here we have also introduced the notation $\sigma^{\mu\nu} \equiv \nabla^{\langle\mu} u^{\nu\rangle} $,  and the fluid vorticity is defined as $\omega^{\mu \nu} \equiv -\nabla^{[\mu} u^{\nu]}$.  Square brackets indicate an antisymmetrized quantity
$$
A^{[\mu} B^{\nu]} \equiv \frac 1 2 \left( A^\mu B^\nu - A^\nu B^\nu \right) .
$$
In a general curved space-time the Riemann tensor is non-zero
$$
R^\lambda_{\ \mu \sigma \nu}\equiv\partial_\sigma \Gamma^\lambda_{\mu \nu}-\partial_\nu \Gamma^\lambda_{\mu \sigma}
+\Gamma^\kappa_{\mu \nu} \Gamma^\lambda_{\kappa \sigma}-\Gamma^\kappa_{\mu \sigma} \Gamma^\lambda_{\kappa \nu}\,,
$$
with 
$
\Gamma^\lambda_{\mu \nu}=\frac{1}{2} g^{\lambda \rho}\left(
\partial_\mu g_{\rho \nu}+\partial_\nu g_{\rho \mu} - \partial_{\rho} g_{\mu \nu}\right),
$
as well as the Ricci tensor $R_{\mu \nu}=R^\lambda_{\ \mu \lambda \nu}$ and Ricci scalar $R=R^\mu_\mu$.

Therefore, without any additional information, there are in principle 15 possible second-order transport coefficients multiplying these terms: $\tau_\pi$, $\tau_\pi^*$, $\kappa$, $\kappa^*$, $\lambda_1$, $\lambda_2$, $\lambda_3$, $\lambda_4$,  $\tau_\Pi$, $\xi_1$, $\xi_2$, $\xi_3$, $\xi_4$, $\xi_5$ and $\xi_6$, in addition to the first order transport coefficients $\eta$ and $\zeta$.   If one is lucky, these transport coefficients can be calculated from the underlying microscopic theory (e.g., QCD in the case of relativistic heavy ion collisions).  Often, however, this is not feasible (as is the case for QCD at the moment, although there is ongoing work).   If they cannot be computed from first principles, the transport coefficients can be treated as free parameters that are then constrained by experimental data.  In this case, it becomes important to understand which terms are necessary (or appropriate) to keep in a hydrodynamic simulation of a given problem.

One can use general arguments to reduce the number of free parameters: requiring the positivity of the divergence of the entropy current provides 2 extra constraints, leaving only 13 completely independent transport coefficients \cite{Romatschke:2009kr}.  Also, when space-time can be assumed flat, the terms involving the Riemann tensor and the Ricci tensor and scalar drop out, reducing the number of independent second order transport coefficients to 11, in addition to the 2 first order coefficients.

For most problems, however, this situation is still not satisfactory.  Having so many free parameters (as well as the fact that their temperature dependence is \textit{a priori} unknown) would significantly reduce the predictive power of any simulation.  On the other hand, while it is true that one can eliminate the problems of the relativistic Navier-Stokes equations by adding only a single second order term, the precise form of the term used can affect how sensitive the results are to the value of the corresponding transport coefficient (as well as affecting the interpretation of the resulting value in the context of the underlying theory), and so one must be convinced that keeping only that part of the second order expansion is justified.
\subsection{M\"uller-Israel-Stewart theory}
This issue of a causal set of relativistic viscous hydrodynamic equations was originally studied in depth by M\"uller \cite{Mueller1, Mueller}, and separately by Israel and Stewart \cite{IS0a, IS0b, Israel:1979wp}.  Thus, many of the second order extensions to viscous hydrodynamics are commonly referred to as M\"uller-Israel-Stewart theory, although the exact form of the equations that have been used varies.
An extensive history of the various forms used and derivations thereof is beyond the aim of this summary, although some generic comments can be made.

In all cases one has a term whose coefficient acts as a relaxation time which, when it is larger than a certain value, eliminates any possible acausality.  In the shear sector this relaxation time is usually called $\tau_\pi$, and it typically multiplies some combination that includes the term $^\langle D \sigma^{\mu \nu\rangle}$ (e.g. in \autoref{generalsecondorder}).

Besides causality, another important consideration is the second law of thermodynamics, 
\begin{equation}
\label{secondlaw}
 \partial_\mu s^\mu \ge 0\ .
\end{equation}
which should be obeyed in any physical system.  In the absence of dissipation, the entropy current is given by $s^\mu = s\, u^\mu$.  In ideal hydrodynamics, the inequality is saturated, and the entropy does not increase with time.  Assuming this form for the entropy current, one can show that the relativistic Navier-Stokes equations always obey \autoref{secondlaw} for any (non-negative) value of the transport coefficients \cite{Romatschke:2009im}.

When there is dissipation, however, the entropy current can also have gradient terms.  If one assumes that the entropy current has to be algebraic in the hydrodynamic degrees of freedom, and that the deviations from equilibrium are small enough that higher
order corrections can be neglected, it can be shown that the entropy current has to be of the form \cite{Mueller1,IS0a,Muronga:2003ta}
\begin{equation}
\label{entropycurrent}
s^\mu=s u^\mu-\frac{\beta_0}{2 T} u^\mu \Pi^2 -\frac{\beta_2}{2 T} u^\mu \pi_{\alpha \beta} \pi^{\alpha \beta}
+ {\cal O}(\Pi^3)\ .
\end{equation}
One can then plug this into \autoref{secondlaw} and derive a form for the second order hydrodynamic terms that is then guaranteed to always obey the second law of thermodynamics \cite{Romatschke:2009im}.  There is doubt, however, that the assumption that the entropy current has to be algebraic in the hydrodynamic degrees of freedom is necessarily true and therefore also that \autoref{entropycurrent} is the really the most general form \cite{Loganayagam:2008is, Romatschke:2009kr}.  Also, it may be an unnecessarily strong requirement to demand that the structure of the hydrodynamic equations must be such that \autoref{secondlaw} is satisfied in any regime and for any combination of values of the transport coefficients---it is only necessary that the second law of thermodynamics be obeyed in actually physically realized (or at least realizable) systems.

Another tack that is often taken is to look to the kinetic theory of gases for guidance.  Kinetic theory is a description of a system in terms of collisions of dilute particles (or quasi-particle states that are long lived compared to the scattering rate).  By considering small departures from equilibrium, one can derive another version of M\"uller-Israel-Stewart theory \cite{Romatschke:2009im}.  Unfortunately, although kinetic theory has a large range of applicability and can thus often provide much insight into various physical behavior, such a particle-based description is not valid for, e.g., the strongly coupled non-abelian plasma which may exist for part of the evolution of a relativistic heavy ion collision system.  Therefore, it is uncertain %(and perhaps even doubtful) 
that the resulting second-order viscous hydrodynamic equations are general enough to describe the evolution of such a system.
\subsection{Conformal relativistic viscous hydrodynamics}
If there is reason to believe that a system is approximately scale-invariant, the form of the hydrodynamic equations are greatly restricted.  To be precise, if one assumes a conformal symmetry in the underlying physics, the total number of possible independent first and second order transport coefficients is reduced to 6---and only 4 in the case of flat space.

The conformal group is the set of symmetry transformations that consists of scale transformations and special conformal transformations.  A theory that is conformally symmetric (a.k.a. ``conformal'') has the property that its action is invariant under a Weyl transformations of the metric,
\begin{equation}
g_{\mu \nu}\rightarrow \bar{g}_{\mu \nu}=e^{-2 w(x)} g_{\mu \nu}\, ,
\end{equation}
where $w(x)$ is an arbitrary function of $x$.

This implies that (to second order in derivatives) the trace of the energy-momentum tensor vanishes, and that $T^{\mu\nu}$ transforms under a Weyl rescaling as 
\begin{equation}
T^{\mu\nu}\rightarrow \bar{T}^{\mu \nu}=e^{6 w(x)} T^{\mu\nu}.
\end{equation}
Imposing these conditions gives as the most general form \cite{Baier:2007ix}
\begin{align}
\Pi^{\mu\nu} = \pi^{\mu \nu} =\, &\eta \sigma^{\mu \nu}-\eta\tau_\pi \left[^\langle D \sigma^{\mu \nu\rangle}+\frac 4 3 (\nabla\cdot u) \sigma^{\mu \nu}\right] \nonumber\\
& +\frac \kappa 2 \left[R^{\langle\mu \nu\rangle}+2 u_\alpha u_\beta R^{\alpha \langle\mu \nu\rangle\beta}\right]
\nonumber\\
& - \frac {\lambda_1} {2} \sigma^{\langle\mu}_{\quad \lambda}\sigma^{\nu\rangle \lambda}+
\frac {\lambda_2} {2} \sigma^{\langle\mu}_{\quad \lambda} \omega^{\nu\rangle \lambda}
- \frac {\lambda_3} {2} \omega^{\langle\mu}_{\quad \lambda}\omega^{\nu\rangle \lambda}\ .
\end{align}
Note in particular that there is no bulk viscosity.  To this order in derivatives, it is valid to replace $\eta \sigma^{\mu\nu} \to \pi^{\mu\nu}$, and so the form that is more convenient to actually solve numerically is
\begin{align}
\Pi^{\mu\nu} =\ & \eta \nabla^{\langle \mu} u^{\nu\rangle}
- \tau_\pi \left[ \Delta^\mu_\alpha \Delta^\nu_\beta D\Pi^{\alpha\beta} 
 + \frac 4{3} \Pi^{\mu\nu}
    (\nabla_\alpha u^\alpha) \right] \nonumber\\
  & 
  + \frac{\kappa}{2}\left[R^{\langle \mu\nu\rangle}+2 u_\alpha R^{\alpha\langle \mu\nu\rangle\beta} 
      u_\beta\right]\nonumber\\
  & -\frac{\lambda_1}{2\eta^2} {\Pi^{\langle \mu}}_\lambda \Pi^{\nu\rangle\lambda}
  +\frac{\lambda_2}{2\eta} {\Pi^{\langle \mu}}_\lambda \omega^{\nu\rangle\lambda}
  - \frac{\lambda_3}{2} {\omega^{\langle \mu}}_\lambda \omega^{\nu\rangle\lambda}\, .
\end{align}
In the expectation that the medium created in a relativistic heavy ion collision is approximately conformal, this is the form that will be used in the viscous hydrodynamic simulations of the following chapters.
%
%
% ========== Chapter 4 - Hydrodynamic models of heavy ion collisions
\newcommand{\beq}{\begin{equation}}
\newcommand{\eeq}{\end{equation}}
\newcommand{\bqa}{\begin{eqnarray}}
\newcommand{\eqa}{\end{eqnarray}}
\chapter{Modeling Heavy Ion Collisions Using Viscous Hydrodynamics}
\label{RHIChydro}
%\section{Introduction}
%The relativistic ideal hydrodynamic equations 
\section{Anatomy of a Heavy Ion Collision}
\begin{figure}[h]
\begin{center}\begin{tikzpicture}
\pgftransformscale{.85}
%\pgfsetarrows{stealth-stealth}
 \draw[line width = 5 pt] (-4,-4) -- (4,4);
 \draw[line width = 5 pt] (-4,4) -- (4,-4);
\fill[shade, ball color=yellow] (4,-4) ellipse (.1 and .6);
\fill[shade, ball color=yellow] (-4,-4) ellipse (.1 and .6);
% \draw[line width = 4 pt,  -angle 60] (0,-4) -- (0,4);
% \draw[line width = 4 pt,  -angle 60] (-4,0) -- (4,0);
% \fill[color=violet!50] (-1.9,2) .. controls (-.4,.5) and (-.2,.5) .. (0,.5) .. controls (.2,.5) and (.4,.5) .. (1.9,2) -- (2,2) -- (0,0) -- (-2,2);
% \fill[color=blue!50] (-1.9,2) .. controls (-.4,.5) and (-.2,.5) .. (0,.5) .. controls (.2,.5) and (.4,.5) .. (1.9,2) -- (1.8,2) .. controls (.8,1) and (.2,1) .. (0,1) .. controls (-.2,1) and (-.8,1) .. (-1.8,2);
%  \fill[color=green!50]  (1.8,2) .. controls (.8,1) and (.2,1) .. (0,1) .. controls (-.2,1) and (-.8,1) .. (-1.8,2) --  (-1.6,2) .. controls (-1.1,1.6) and (-.3,1.5) .. (0,1.5) .. controls (.3,1.5) and (1.1,1.6) .. (1.6,2);
% \fill[color=red!50] (-1.6,2) .. controls (-1.1,1.6) and (-.3,1.5) .. (0,1.5) .. controls (.3,1.5) and (1.1,1.6) .. (1.6,2);
 \shade[top color=blue!25, bottom color=blue!75] (-3.8,4) .. controls (-.8,1) and (-.4,1) .. (0,1) .. controls (.4,1) and (.8,1) .. (3.8,4) -- (3.6,4) .. controls (1.6,2) and (.4,2) .. (0,2) .. controls (-.4,2) and (-1.6,2) .. (-3.6,4);
  \shade[top color=violet!25, bottom color=violet!75] (-3.8,4) .. controls (-.8,1) and (-.4,1) .. (0,1) .. controls (.4,1) and (.8,1) .. (3.8,4) -- (4,4) -- (0,0) -- (-4,4);
  \shade[top color=blue!25, bottom color=blue!75]  (3.6,4) .. controls (1.6,2) and (.4,2) .. (0,2) .. controls (-.4,2) and (-1.6,2) .. (-3.6,4) --  (-3.2,4) .. controls (-2.2,3.2) and (-.6,3) .. (0,3) .. controls (.6,3) and (2.2,3.2) .. (3.2,4);
  \shade[top color=purple!25, bottom color=purple!75] (-3.2,4) .. controls (-2.2,3.2) and (-.6,3) .. (0,3) .. controls (.6,3) and (2.2,3.2) .. (3.2,4);
% \shade (-1.9,2) .. controls (-.4,.5) and (-.2,.5) .. (0,.5) ..  (0,1) .. controls (-.2,1) and (-.8,1) .. (-1.8,2);
% \draw[blue, thick] (-1.9,2) .. controls (-.4,.5) and (-.2,.5) .. (0,.5) .. controls (.2,.5) and (.4,.5) .. (1.9,2);
% \draw[green, thick] (-1.8,2) .. controls (-.8,1) and (-.2,1) .. (0,1) .. controls (.2,1) and (.8,1) .. (1.8,2);
% \draw[red, thick] (-1.6,2) .. controls (-1.1,1.6) and (-.3,1.5) .. (0,1.5) .. controls (.3,1.5) and (1.1,1.6) .. (1.6,2);
 \draw[line width = 3 pt,  -angle 60] (0,-4.3) -- (0,4.3);
 \draw[line width = 3 pt,  -angle 60] (-4.3,0) -- (4.3,0);
  \pgftext[base,x=3.6cm, y=-.8cm] {\large{$z$}}
  \pgftext[base,x=.4cm, y=3.8cm] {\large{$t$}}
%  \pgftext[base,left,x=1.7cm, y=.95cm] {\color{black}{\tiny{thermalization}}}
%  \pgftext[base,left,x=2.7cm, y=1.95cm] {\color{black}{\tiny{phase transition}}}
%  \pgftext[base,left,x=3.7cm, y=2.95cm] {\color{black}{\tiny{freeze-out}}}
%  
% \draw[line width = 1 pt, -angle 60] (1.5, 1) -- (.3,1);
% \draw[line width = 1 pt, -angle 60] (2.5,2) -- (.3,2);
% \draw[line width = 1 pt, -angle 60] (3.5,3) -- (.3,3);
 
 \pgftext[base,right,x=-1.7cm, y=.95cm] {\color{black}{{thermalization}}}
  \pgftext[base,right,x=-2.7cm, y=1.95cm] {\color{black}{{phase transition}}}
  \pgftext[base,right,x=-3.7cm, y=2.95cm] {\color{black}{{freeze-out}}}
  
 \draw[line width = 1 pt, -angle 60] (-1.5, 1) -- (-.3,1);
 \draw[line width = 1 pt, -angle 60] (-2.5,2) -- (-.3,2);
 \draw[line width = 1 pt, -angle 60] (-3.5,3) -- (-.3,3);
 
 \pgftext[base,left,x=1 cm, y=.45cm] {\color{violet}{{pre-equilibrium}}}
 \pgftext[base,left,x=2.5 cm, y=1.95cm] {\color{blue}{{hydrodynamics}}}
 \pgftext[base,left,x=4 cm, y=3.45cm] {\color{purple}{{free streaming}}}
 
\end{tikzpicture}\end{center}
\caption[Space-time cartoon diagram of a heavy ion collision]{Space-time cartoon diagram of a heavy ion collision, indicating the pre-equilibrium stage, hydrodynamic stage, and the system after it has frozen out.  Here it is assumed that there is hydrodynamic evolution on both sides of the phase transition as well as approximately boost invariant evolution.}
\end{figure}
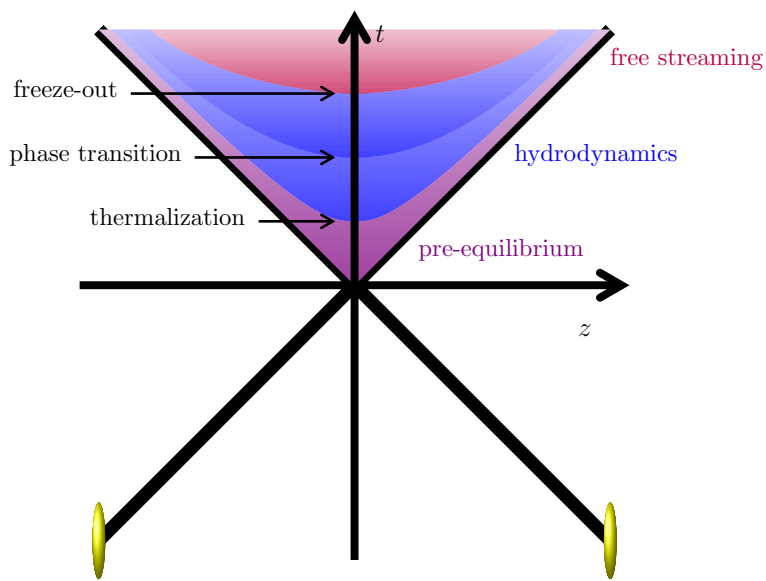
Even if a heavy ion collision system does behave hydrodynamically for a significant period of its evolution, that is not the whole story, of course.  Once the proper hydrodynamic equations are set, one must still specify the boundary conditions.  There is a finite period of time at the beginning of a collision before which the system can equilibrate (or at least isotropize---recall \autoref{rel}) and begin behaving hydrodynamically.  Likewise, as the system expands and cools, there will be some point at which the system no longer interacts strongly enough for hydrodynamics to be a valid description.  Eventually, the particles get so far apart that they completely cease interacting and these free particles are what ultimately get detected.  Therefore one must define the process by which the system ``freezes out''.  Once the initial conditions and freeze out algorithm have been specified, the simulations can be run numerically and experimental observables calculated.  Details of these choices will be given later in this chapter when describing the viscous hydrodynamic simulations done by the author in collaboration with Paul Romatschke.

Ideal hydrodynamic simulations were previously done, however, offering a remarkably good description of the experimental data for bulk properties 
(multiplicity, radial and elliptic flow) of low $p_T$ particles for heavy-ion
collisions at RHIC 
\cite{Teaney:2000cw,Huovinen:2001cy,Kolb:2001qz,Hirano:2002ds,Kolb:2002ve}.  
%
%\section{Initial Conditions}
%\subsection{Glauber}
%\subsection{Color-Glass-Condensate}
%\section{Freeze Out}
%\section{Results}
%\subsection{RHIC}
%\section{Introduction}

%The experimental program at the
%Relativistic Heavy-Ion Collider (RHIC) at Brookhaven has
%generated a wealth of data \cite{Adcox:2004mh,Back:2004je,Arsene:2004fa,Adams:2005dq} on QCD matter
%at the highest energy densities obtained in the laboratory.
%Remarkably, ideal hydrodynamics seems to offer a sensible
%description of the experimental data for bulk properties 
%(multiplicity, radial and elliptic flow) of low $p_T$ particles for heavy-ion
%collisions at RHIC 
%\cite{Teaney:2000cw,Huovinen:2001cy,Kolb:2001qz,Hirano:2002ds,Kolb:2002ve}.

Upon closer inspection, however, not all of this success can be attributed
to modeling the system as an ideal fluid.
For instance, the energy density
distribution which is used as an initial condition for the
hydrodynamic equations is customarily chosen such that the output from
the hydrodynamic model matches the experimental data for the 
multiplicity. Furthermore, the time where the hydrodynamic
model is initialized as well as the temperature (or energy
density) at which the hydrodynamic evolution is stopped
are typically chosen such that the model output matches the 
experimental data for the radial flow.
After these parameters have been fixed, only
the good description of experimental data
for the elliptic flow coefficient can be considered
a success for ideal hydrodynamics (in the sense that
it is parameter-free).

In order to make progress and learn more about the properties
of matter created at RHIC, the task is now to both test
and improve this ideal hydrodynamic model.
The obvious framework for this task is dissipative 
hydrodynamics, since it contains ideal
hydrodynamics as the special case when all dissipative 
transport coefficients (such as shear and bulk viscosity
and heat conductivity) are sent to zero.
If the value of the transport coefficients were
known (e.g. by some first principle calculation),
then one could use dissipative hydrodynamics to
constrain e.g. the initial energy density distribution,
which is chosen conveniently in the ideal hydrodynamic
models. Or otherwise, choosing again 
physically acceptable initial conditions, one
is able to constrain the allowed ranges of the
transport coefficients. Despite recent progress
in first principle calculations \cite{Policastro:2001yc,Arnold:2003zc,Nakamura:2004sy,Arnold:2006fz,Huot:2006ys,Gagnon:2006hi,Aarts:2007wj,Meyer:2007ic,Buchel:2007mf,CaronHuot:2007gq},
the values of the hydrodynamic transport coefficients for
QCD in the relevant energy range are poorly 
constrained to date, so the second option
is currently the only viable possibility.

For RHIC, the first step in this direction was carried
out by Teaney \cite{Teaney:2003kp}, who provided estimates
for the sign and size of corrections due to
shear viscosity. This famous calculation,
however, did not provide a 
description of experimental data for non-zero viscosity,
because it was not dynamic and the initial conditions
could not be altered. Only very recently, the first
hydrodynamic calculations with shear viscosity describing
particle spectra for central and non-central collisions at RHIC
have became available 
\cite{Romatschke:2007jx,Romatschke:2007mq,Chaudhuri:2007qp}. 

Several other groups have produced numerical
codes capable of performing similar matching to data
\cite{Muronga:2004sf,Chaudhuri:2005ea,Muronga:2005pk,Chaudhuri:2006jd,Mota:2007tz,Song:2007fn,
Dusling:2007gi,Song:2007ux}.

However, the precise formulation of the viscous hydrodynamic
equations themselves has long been debated (recall \autoref{vischydro}). 
%To appreciate the complication,
%one first has to understand a hydrodynamic formulation
%for RHIC physics necessarily has to be fully relativistic,
%and that the relativistic generalization of the Navier-Stokes
%equations are acausal since they contain modes that
%transport information at superluminal speeds.
%These are high wavenumber modes and therefore
%in principle outside the range of validity of
%hydrodynamics, but in practice one has to find a way to
%deal with them in viscous hydrodynamic simulations.
%A possible solution to this problem is known as 
%M\"uller-Israel-Stewart theory, where for each 
%transport coefficient a corresponding relaxation time
%is introduced which controls the speed of signal
%propagation for the high wavenumber modes \cite{IS0a,IS0b,Israel:1979wp,Mueller}.
%For low momentum modes (up to first order in gradients),
%the M\"uller-Israel-Stewart theory is identical to
%the Navier-Stokes equations, but differs
%for higher order gradients.
%Unfortunately, this implied that 
%the resulting equations retained
%a certain degree of arbitrariness, as
%it was not clear which additional 
%terms of second or higher order in gradients
%either within the M\"uller-Israel-Stewart or
%other frameworks 
%(see e.g.\cite{Muronga:2003ta,Heinz:2005bw,Baier:2006um,Koide:2006ef}) 
%were allowed.
For the case of non-vanishing shear viscosity only,
it was shown recently \cite{Baier:2007ix} that the most general
form implies five independent terms of second order
in gradients. This form is general enough to describe
the hydrodynamic properties of (conformal) plasmas
both for weakly coupled systems
 describable by the Boltzmann
equation as well as infinitely strongly coupled
plasmas, which are accessible via Maldacena's conjecture
\cite{Maldacena:1997re}.

The aim of this chapter is to now apply this 
new set of equations for relativistic
shear viscous hydrodynamics to the problem of 
heavy-ion collisions at RHIC. 
In \autoref{sec:one}, we review the setup of conformal
relativistic viscous hydrodynamics and our numerics
for the simulation of heavy-ion collisions. In \autoref{sec:two},
details about the two main models of initial conditions
for hydrodynamics are given. Section \ref{sec:three} contains
our results for the multiplicity, radial flow and elliptic flow
in Au+Au collisions at top RHIC energies, as well as a note
on the notion of ``early thermalization''. We conclude in
\autoref{sec:four}.

\section{Setup}
\label{sec:one}

We use the most general form of the second order viscous hydrodynamic equations,
which as a reminder are given by (see \autoref{vischydro})
\begin{align}
\Pi^{\mu\nu} =\ & \eta \nabla^{\langle \mu} u^{\nu\rangle}
- \tau_\pi \left[ \Delta^\mu_\alpha \Delta^\nu_\beta D\Pi^{\alpha\beta} 
 + \frac 4{3} \Pi^{\mu\nu}
    (\nabla_\alpha u^\alpha) \right] \nonumber\\
  &
  + \frac{\kappa}{2}\left[R^{\langle \mu\nu\rangle}+2 u_\alpha R^{\alpha\langle \mu\nu\rangle\beta} 
      u_\beta\right]\nonumber\\
  & -\frac{\lambda_1}{2\eta^2} {\Pi^{\langle \mu}}_\lambda \Pi^{\nu\rangle\lambda}
  +\frac{\lambda_2}{2\eta} {\Pi^{\langle\mu}}_\lambda \omega^{\nu\rangle\lambda}
  - \frac{\lambda_3}{2} {\omega^{\langle\mu}}_\lambda \omega^{\nu\rangle\lambda}\, .
\label{maineq}
\end{align}
%where $\omega_{\mu \nu}=-\nabla_{[\mu} u_{\nu]}$ is the 
%fluid vorticity and $R^{\alpha \mu \nu \beta},R^{\mu \nu}$
%are the Riemann and Ricci tensors, respectively.
The coefficients $\tau_\pi,\kappa,\lambda_1,\lambda_2,\lambda_3$
are the five new coefficients controlling the size of
the allowed terms of second order in gradients.
Having an application to the problem of heavy-ion collisions
in mind, the above set of equations can be simplified:
for all practical purposes spacetime can be considered flat, such
that both the Riemann and Ricci tensors vanish identically.
Thus, only the four coefficients 
$\tau_\pi,\lambda_1,\lambda_2,\lambda_3$ enter the problem.

\subsection{A note on bulk viscosity and conformality}

Besides shear viscosity, QCD also has non-vanishing
bulk viscosity $\zeta$ which can be related to the 
QCD trace anomaly \cite{Kharzeev:2007wb}
\beq
\zeta\sim T^{\mu}_\mu=\epsilon-3 p.
\eeq
QCD lattice simulations seem to indicate that the ratio
bulk viscosity over entropy density $s$, $\zeta/s$,
is small compared to $\eta/s$ except for a small region
around the QCD deconfinement transition temperature, 
where it is sharply peaked \cite{Sakai:2007cm,Meyer:2007dy,Karsch:2007jc}.
If we are interested in describing effects
from shear viscosity only, we are led to consider $\zeta=0$, or
conformal fluids. This has been the main guiding principle
in Ref.~\cite{Baier:2007ix} and as a consequence \autoref{maineq}
obeys conformal invariance, unlike most other second-order 
theories\footnote{Note that Muronga derived a version of 
\autoref{maineq} in Ref.~\cite{Muronga:2003ta} that 
turns out to obey conformal symmetry.}.

\subsection{First steps: 0+1 dimensions}
\label{section01}

In order to get a crude estimate of the 
effect of viscous corrections, let us consider the 
arguably simplest model of a heavy-ion collision: 
a system expanding in a boost-invariant fashion along
the longitudinal direction and having uniform energy
density in the transverse plane.
Introducing the Milne variables proper time 
$\tau=\sqrt{t^2-z^2}$ and space-time rapidity $\xi={\rm arctanh}(z/t)$,
boost invariance simply translates to requiring all
hydrodynamic variables ($\epsilon,u^\mu,\Pi^{\mu \nu}$)
to be independent of rapidity, and tensor components
$u^\xi,\Pi^{\mu \xi}$ to vanish. Assuming uniformity
in the transverse plane furthermore requires independence
from the transverse coordinates ${\bf x}_T=(x,y)$.
Even though this means that all the velocity components
except $u^\tau$ are zero,
the system is nevertheless non-trivial in the 
sense that the %it is expanding since the 
sum over velocity gradients
does not vanish, $\nabla_\mu u^\mu=\frac{1}{\tau}$,
sometimes referred to as ``Bjorken flow''.

In a way one has modeled an expanding system in static
space-time by a system at rest
in an expanding space-time. This has been achieved by
transforming to the Milne coordinates $\tau,\xi$,
where the metric is 
$g_{\mu \nu}={\rm diag}(g_{\tau \tau},g_{xx},g_{yy},g_{\xi \xi})=
(1,-1,-1,-\tau^2)$.
Note that even though the spacetime in these coordinates
is expanding, it is nevertheless flat (e.g. has vanishing
Riemann tensor).

In this 0+1 dimensional toy model, the viscous hydrodynamic
equations become exceptionally simple \cite{Baier:2007ix},
\bqa
\partial_\tau \epsilon&=&-\frac{\epsilon+p}{\tau}+\frac{\Pi^\xi_\xi}{\tau}
\nonumber\\
\partial_\tau \Pi^\xi_\xi &=& -\frac{\Pi^\xi_\xi}{\tau_\pi}
+\frac{4 \eta}{3 \tau_\pi \tau}-\frac{4}{3 \tau} \Pi^\xi_\xi
-\frac{\lambda_1}{2\tau_\pi\eta^2} \left(\Pi^\xi_\xi\right)^2.
\label{0+1dsystem}
\eqa
The Navier-Stokes equations are recovered formally in the limit
where all second-order coefficients vanish 
(e.g. $\tau_\pi,\lambda_1\rightarrow 0$); then, one 
simply has 
\beq
\Pi^\xi_\xi=\frac{4 \eta}{3 \tau}.
\label{FOvalue}
\eeq
The equations (\ref{0+1dsystem}) can be solved numerically along the
lines of \cite{Muronga:2001zk,Baier:2006um}.
At very early times, where $\Pi^\xi_\xi>(\epsilon+p)$, 
the Navier-Stokes equations indicate an increase in energy density
and a negative effective longitudinal pressure $p-\Pi^\xi_\xi$.
Since gradients $\nabla_\mu u^\mu=1/\tau$ are strongest at early times,
this suggests that one is applying the Navier-Stokes equations
outside their regime of validity. Theories including second order
gradients may be better behaved at early times, but eventually
also have to break down when gradients become too strong. Here we want
to study the effects of the second order coefficients
on the value of the shear tensor at late times, where
a hydrodynamic approach should be valid.

To this end, let us study the deviation of the shear
tensor from its first order value, $\delta \Pi=\Pi^{\xi}_\xi-\frac{4 \eta}{3 \tau}$. At late times, \autoref{0+1dsystem} implies 
$\epsilon\sim \tau^{-4/3}$, so $\eta\sim \tau^{-1}$. Thus, if
$\delta \Pi$ is small compared to the first order value, from
\autoref{0+1dsystem} we find
\beq
\delta \Pi=\frac{4 \eta}{3 \tau}\left(\frac{2\tau_\pi}{3\tau}
-\frac{2\lambda_1}{3 \tau \eta}\right).
\eeq
For a strongly coupled ${\cal N}=4$ plasma 
\cite{Policastro:2001yc,Baier:2007ix,Bhattacharyya:2008jc,Natsuume:2007ty}, 
one has\footnote{For completeness, we also mention the results
$\kappa=\frac{\eta}{\pi T},\lambda_2=-\frac{\eta \ln 2}{\pi T},\lambda_3=0$
from \cite{Baier:2007ix,Bhattacharyya:2008jc}.}
\beq
\frac{\eta}{s}=\frac{1}{4 \pi},\qquad
\tau_\pi=\frac{2-\ln 2}{2 \pi T},\qquad 
\lambda_1=\frac{\eta}{2 \pi T},
\eeq
and thus $\Pi^\xi_\xi$ is larger than its first order value
by a factor of $1+\frac{1-\ln 2}{3 \pi T \tau}$.
For RHIC, $T \tau\gtrsim 1$ is a reasonable estimate, so one
finds that the second order corrections to $\Pi^\xi_\xi$ 
increase its value by a few percent over the first order result.

As an example on the importance of obeying conformal invariance, 
imagine dropping the term involving $\nabla_\alpha u^\alpha$ 
in the first line of \autoref{maineq}. Redoing the above 
calculation one finds
\beq
\delta \Pi_{NC}=\frac{4 \eta}{3 \tau}\left(\frac{2\tau_\pi}{\tau}
-\frac{2\lambda_1}{3 \tau \eta}\right),
\eeq
which indicates a nearly ten-fold increase of the size of $\delta \Pi$
for the non-conformal theory. For a weakly coupled plasma well described
by the Boltzmann equation \cite{Baier:2007ix}, where
one has $\tau_\pi=\frac{6 \eta}{s T}$, ($\lambda_1$ is unknown but
generally set to zero in M\"uller-Israel-Stewart theory ),
the effect may be less pronounced, but still one qualitatively
expects second-order effects to be anomalously large if conformal
invariance is broken in an ``ad-hoc'' manner.

%For a weakly-coupled plasma well described by the Boltzmann equation, 
%$\lambda_1=0$ and
%$\tau_\pi=6 \frac{\eta}{s T}$, implying also $\delta \Pi>0$.
%More precisely, even using conjectured minimal bound value 
%\cite{Kovtun:2004de} of
%$\eta/s=\frac{1}{4 \pi}$, this amounts to a $\Pi^\xi_\xi$ which
%is larger than the first order value by as much as $\sim 30$ 
%percent. 
%
%
%Before concluding this section, we would like
%to point out that even though the value of $\Pi^\xi_\xi$
%seems to be enlarged by second-order corrections
%at late times, the trend may be inversed at early times.
%Specifically, ignoring the correct starting value
%of the shear tensor $\Pi^\xi_\xi$, a ``minimalistic
%assumption'' is to set $\Pi^\xi_\xi=0$ initially.
%At late times, $\Pi^\xi_\xi$ has to approach its first-order
%value, however, which is positive, and decreasing with time.
%Therefore, $\Pi^\xi_\xi$ has to have a maximum at some finite
%time, where $\partial_\tau \Pi^\xi_\xi=0$.
%At this instance in time, Eq.~(\ref{0+1dsystem})
%then implies 
%\beq
%\delta \Pi<0
%\eeq
%since $\eta,\tau_\pi>0$ and all known $\lambda_1\ge 0$.
%Put differently, the second order coefficients tend to 
%reduce the value of the shear tensor below its first order
%(Navier-Stokes) value, bringing the system closer
%to an ideal fluid behavior. Using the values from above,
%one finds that this decrease is again as large as $40$ percent.

Clearly, the above estimates are not meant to be quantitative.
Indeed, even the sign of the correction may change when allowing
more complicated (e.g., three-dimensional) dynamics.
However, the lesson to be learned from this exercise is that
second-order gradients can and indeed do modify the 
shear tensor from its first order (Navier-Stokes) value.
This is physically acceptable, as long as the second-order
corrections are small compared to the first order ones (otherwise
the system is probably too far from equilibrium for even
a hydrodynamic description correct to second order in gradients 
to be valid). A practical means for testing this is 
calculating physical observables for different values of
the second-order coefficients and making sure that the
results do not strongly depend on the choice for these specific values.

\subsection{Including radial flow: lessons from 1+1 dimensions}

Some more insight on the effect of viscous corrections may be
gained by improving the model of the previous subsection
to allow for radially symmetric dynamics in the transverse plane
(but still assuming boost invariance).
This is most easily implemented by changing to polar coordinates
$(x,y)\rightarrow(r, \phi)$ with $r=\sqrt{x^2+y^2}$ and
$\phi={\rm arctan}(y/x)$. In this case, the only non-vanishing
velocity components are $u^\tau$ and $u^r$, and hence the
vorticity $\omega^{\mu \nu}$ vanishes identically.
Although non-trivial, the radially symmetric flow case is still 
a major simplification over the general form 
\autoref{maineq}, since again the terms involving
$\kappa,\lambda_2,\lambda_3$ drop out.

Such a formulation allows both for important code
tests \cite{Baier:2006gy} as well as realistic simulations
of central heavy-ion collisions \cite{Romatschke:2007jx}
(note that truncated versions of \autoref{maineq}
were used in these works).
The advantage of this formulation is that since the equations
are comparatively simple, it is rather straightforward to
implement them numerically and they are 
not very time consuming to solve since
only one dimensional (radial) dynamics is involved. 
The shortcoming of simulations with radially symmetric
flow profiles (``radial flow'') is that by construction
they cannot be matched to experimental data on the
impact-parameter dependence of multiplicity. Thus,
the considerable freedom in the initial/final conditions
inherent to all hydrodynamic approaches cannot be
eliminated in this case.

For this reason, we will choose not to discuss
the case of radial flow here in more detail, but
rather will comment on it later as a special
case of the more general situation.

\subsection{Elliptic flow: 2+1 dimensional dynamics}

Retaining the assumption of boost invariance, but
allowing for general dynamics in the transverse plane,
it is useful to keep Cartesian coordinates in the
transverse plane, and thus $u^\tau,u^x,u^y$
are the non-vanishing fluid velocities. The main reason
is that e.g. in polar coordinates
the equations
for the three independent components of $\Pi^{\mu \nu}$
would involve some extra non-vanishing Christoffel symbols
(other than $\Gamma^\tau_{\xi \xi}=\tau$,$\Gamma^\xi_{\tau \xi}=1/\tau$).

Fortunately, the case of two dimensions is special insofar
as the only nontrivial component of the vorticity tensor,
namely $\omega^{xy}$, fulfills the equation \cite{Romatschke:2007mq}
\beq
D \omega^{xy}+ \omega^{xy} \left[ \nabla_\mu u^\mu + \frac{D p}{\epsilon+p}
-\frac{D u^\tau}{u^\tau}\right]=\mathcal{O}(\Pi^3),
\label{vorteq}
\eeq
which can be derived by forming the combination 
$\nabla^x D u^y-\nabla^y D u^x$. The expression $\mathcal{O}(\Pi^3)$
denotes that the r.h.s. of \autoref{vorteq} is of third order in
gradients, and thus should be suppressed in the domain
of applicability of hydrodynamics.
For heavy-ion collisions, typically $\nabla_\mu u^\mu\ge\frac{1}{\tau}$,
so that for an equation of state with a speed of sound squared 
$c_s^2\equiv \frac{dp(\epsilon)}{d \epsilon}\sim \frac{1}{3}$,
\autoref{vorteq} translates to $\frac{D \omega^{xy}}{\omega^{xy}}<0$
unless $D \ln u^\tau\geq (1-c_s^2)\nabla_\mu u^\mu$.
In particular, this implies that in general 
$\omega^{xy}=0$ is a stable fix-point of the above equation and
hence we expect $\omega^{xy}$ to remain small throughout
the entire viscous hydrodynamic evolution if it is small initially.

Generically, one uses $u^{x,y}=0$ as an initial condition 
for hydrodynamics \cite{Huovinen:2006jp}, which implies $\omega^{xy}=0$
initially. Therefore, to very good approximation we can neglect the
terms involving vorticity in \autoref{maineq}, such that
again only the second-order coefficients 
$\tau_\pi,\lambda_1$ have to be specified.

The equations to be solved for 2+1 dimensional relativistic viscous hydrodynamics are
then (in components)
\bqa
(\epsilon+p) D u^i
%\left(u^\tau \partial_\tau u^i
%+u^j \partial_j u^i\right)
&=&
c_s^2 \left(g^{ij}\partial_j \epsilon-u^i u^\alpha \partial_\alpha \epsilon
\right)-\Delta^i_\alpha D_\beta \Pi^{\alpha \beta}\nonumber\\
%\left(u^\tau \partial_\tau \epsilon+u^i \partial_i \epsilon\right)
D \epsilon
&=&-(\epsilon+p) \nabla_\mu u^\mu+\frac{1}{2} \Pi^{\mu \nu}
\nabla_{\langle \mu} u_{\nu \rangle}\nonumber\\
D_\beta \Pi^{\alpha \beta}&=&
\Pi^{i \alpha}\partial_\tau \frac{u^i}{u^\tau}+
\frac{u^i}{u^\tau}\partial_\tau \Pi^{i \alpha}
%\partial_\tau \Pi^{\alpha \tau}
+\partial_i \Pi^{\alpha i}
+\Gamma^{\alpha}_{\beta \delta} \Pi^{\beta \delta}
+\Gamma^\beta_{\beta \delta} \Pi^{\alpha \delta}\nonumber\\
\partial_\tau \Pi^{i \alpha}&=&
-\frac{4}{3 u^\tau}\Pi^{i \alpha} \nabla_\beta u^\beta
+\frac{\eta}{\tau_\pi u^\tau}\nabla^{\langle i} u^{\alpha \rangle}
-\frac{1}{\tau_\pi u^\tau}\Pi^{i \alpha}
\nonumber\\
&&-\frac{u^i \Pi^{\alpha}_\kappa + u^\alpha \Pi^{i}_\kappa}{u^\tau}
%\left(u^\tau \partial_\tau u^\kappa+u^j \partial_j u^\kappa\right)
Du^\kappa-\frac{u^j}{u^\tau} \partial_j \Pi^{i \alpha}
-\frac{\lambda_1}{2 \eta^2 \tau_\pi u^\tau}
\Pi^{\langle i}_\lambda \Pi^{\alpha \rangle \lambda}\nonumber\\
\nabla_\mu u^\mu &=&\partial_\tau u^\tau+\partial_i u^i
+\frac{u^\tau}{\tau}\nonumber\\
\nabla_{\langle x} u_{x \rangle}&=&
2 \Delta^{\tau x}\partial_\tau u^x
+2 \Delta^{i x}\partial_i u^x-\frac{2}{3}\Delta^{xx} \nabla_\mu u^\mu
\nonumber\\
\nabla_{\langle x} u_{y \rangle}&=&
\Delta^{\tau x}\partial_\tau u^y
+\Delta^{\tau y}\partial_\tau u^x
+\Delta^{i x}\partial_i u^y
+\Delta^{i y}\partial_i u^x-\frac{2}{3}\Delta^{xy} \nabla_\mu u^\mu\nonumber\\
\nabla_{\langle \xi} u_{\xi \rangle}&=&
2 \tau^4 \Delta^{\xi \xi}\Gamma^\xi_{\tau \xi} u^\tau
-\frac{2}{3}\tau^4\Delta^{\xi\xi} \nabla_\mu u^\mu.
\label{manyeq}
\eqa
Here and in the following 
Latin indices collectively denote the transverse coordinates $x,y$ and
the relation $u_\mu \Pi^{\mu \nu}=0$ has been used to derive 
the above equations (similarly, 
$u^\mu \nabla_{\langle \mu} u_{\nu \rangle}=0$ can be used to obtain 
the other non-trivial components needed).
Note that this particular form of \autoref{manyeq} 
has not been simplified further since it 
roughly corresponds to the equations 
implemented for the numerics of \cite{Romatschke:2007mq},
and is meant to facilitate understanding of the code \cite{codedown}.
A simple algorithm to solve \autoref{manyeq} has been
outlined in \cite{Baier:2006gy} and will be reviewed in the next 
subsection for completeness.

\subsection{A numerical algorithm to solve relativistic viscous hydrodynamics}

The first step of the algorithm consists of choosing the 
independent degrees of freedom. For boost-invariant 2+1 dimensional dynamics,
a sensible choice for this set 
is e.g. $\epsilon$, $u^x$, $u^y$, $\Pi^{xx}$, $\Pi^{xy}$, $\Pi^{yy}$.
The pressure is then obtained via the equation of state $p(\epsilon)$,
and the only other non-vanishing velocity as $u^\tau=\sqrt{1+u^2_x+u^2_y}$.
Similarly, the other nonzero components of $\Pi^{\mu \nu}$ are calculated
using the equations $\Pi^{\mu}_\mu=0$, $u_\mu \Pi^{\mu \nu}=0$.

Given the value of the set of independent components at some
time $\tau=\tau_0$, the aim is then to construct an algorithm
from \autoref{manyeq} such that the new values of the set can be
calculated as time progresses. Note that in \autoref{manyeq}, time derivatives
of the independent component set enter only linearly.
Therefore, \autoref{manyeq} may be written as a matrix equation
for the derivatives of the independent component set,
\beq
\left(
\begin{array}{cccc}
a_{00} & a_{01} & \ldots  &a_{05}\\
a_{10} & a_{11} & \ldots  &a_{15}\\
\multicolumn{4}{c}\dotfill\\
a_{50} & a_{51} & \ldots  &a_{55}
\end{array}
\right)
\cdot
\left(\begin{array}{c}
\partial_\tau \epsilon\\
\partial_\tau u^x\\
\ldots\\
\partial_\tau \Pi^{yy}
\end{array}\right)
=\left(\begin{array}{c}
b_0\\
b_1\\
\ldots\\
b_6
\end{array}\right).
\eeq
Denoting the above matrix and vector as ${\bf A}$ and ${\bf b}$, respectively,
a straightforward way to obtain the time derivatives is via 
numerical matrix inversion,
\beq
\left(\begin{array}{c}
\partial_\tau \epsilon\\
\partial_\tau u^x\\
\ldots\\
\partial_\tau \Pi^{yy}
\end{array}\right) = {\bf A}^{-1} \cdot {\bf b}.
\label{simpleeq}
\eeq
Choosing a naive discretization of derivatives
\beq
\partial_\tau f(\tau)=\frac{f(\tau+\delta \tau)-f(\tau)}{\delta \tau},
\qquad
\partial_x f(x)=\frac{f(x+a)-f(x-a)}{2a},
\eeq
which is first order accurate in the temporal grid spacing $\delta \tau$ 
and second order accurate in the spatial grid spacing $a$,
one can then directly calculate the new values of the independent
component set from \autoref{simpleeq}. 

Note that for ideal hydrodynamics, the algorithm \autoref{simpleeq} 
would fail for this naive discretization \cite{NR}.
The reason is that ideal hydrodynamics is inherently unstable
to high wavenumber fluctuations (which can be thought of as the basis
for turbulence). For ideal hydrodynamics, one thus has to
use a discretization which amounts to the introduction of
numerical viscosity to dampen these fluctuations.
Luckily, viscous hydrodynamics does not suffer from this
problem because it has real, physical viscosity inbuilt.
It is because of this reason that the naive discretization
can be used in the algorithm \autoref{simpleeq}
without encountering the same problems as in ideal hydrodynamics,
as long as a finite value for the viscosity $\eta$ is used\footnote{
In practice we have used $\frac{\eta}{s}>10^{-4}$. Typically,
between $\frac{\eta}{s}=10^{-2}$ and $\frac{\eta}{s}=10^{-4}$ there 
are no significant changes in the hydrodynamic results and we 
refer to $\frac{\eta}{s}=10^{-4}$ as ``ideal hydrodynamics''.}.
While applicable to sufficiently smooth initial conditions,
the above algorithm is too simple to treat strong gradients
such as the propagation of shocks, and should be replaced
by a more involved scheme in such cases.

Since matrix inversions are computationally intensive,
one can speed up the numerics by expressing 
$\partial_\tau \Pi^{\mu \nu}$ in terms of 
$\partial_\tau u^i,\partial_\tau \epsilon$. Inserting these
in the equations for $D u^\mu$ and $D \epsilon$,
one only needs to invert a $3\times3$ matrix to obtain
the new values of the energy density and fluid velocities.
This approach has been used in 
\cite{Romatschke:2007jx,Romatschke:2007mq,Baier:2006gy}.

\subsection{Initial conditions and equation of state}

As outlined at the beginning of the chapter, any hydrodynamic
description of RHIC physics relies on given initial 
energy density distributions. Two main classes of models
for boost-invariant setups exist: the Glauber models
and the Color-Glass-Condensate (CGC) models.

As will be shown in the following, both model classes can give a reasonable
description of the experimentally found multiplicity distribution,
but they differ by their initial spatial eccentricity.
A detailed discussion of the initial conditions will be given
in subsequent sections.

Besides an initial condition for the energy density,
one also needs to specify an initial condition for
the independent components of the fluid velocities and the
shear tensor. For the fluid velocities we will follow
the standard assumption that these vanish initially
\cite{Huovinen:2006jp}. Finally, when using the set of equations
(\ref{manyeq}), one also has to provide initial values for
the independent components of $\Pi^{\mu \nu}$. Extreme choices
are $\Pi^{\mu \nu}=0$ and a shear tensor so large that
a diagonal component of the energy-momentum tensor vanishes
in the local rest frame (e.g. $\Pi^\xi_\xi=p$, or zero longitudinal
effective pressure),
with the physical result expected somewhere in between (see e.g.
the discussion in \cite{Dumitru:2007qr}).

Once the initial conditions for the independent hydrodynamic
variables have been specified, one needs the equation of
state to solve the hydrodynamic equations (\ref{manyeq}).
Aiming for a description of deconfined nuclear matter at zero chemical potential,
a semi-realistic equation of state has to incorporate
evidence from lattice QCD calculations \cite{Aoki:2006we} that the transition
from hadronic to deconfined quark matter is probably 
an analytic crossover, not a first or second order phase transition
as often used in ideal hydrodynamic simulations.
On the other hand, continuum extrapolations for the value of 
the energy density and pressure
for physical quark masses are still not accessible with high precision
using current lattice methods. For this reason, we will employ
the equation of state by Laine and Schr\"oder \cite{Laine:2006cp},
which is derived from a hadron resonance gas at low temperatures,
a high-order weak-coupling perturbative QCD calculation at high temperatures,
and an analytic crossover regime interpolating between the high
and low temperature regime, respectively.
\begin{figure}
\center
\includegraphics[width=.5\linewidth]{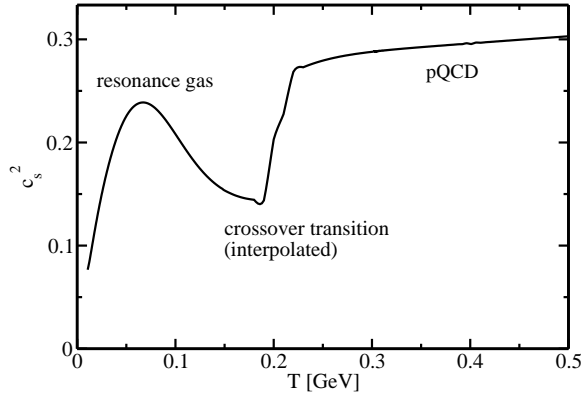}
\caption[Speed of sound squared used in hydrodynamic simulations]{The speed of sound squared from Ref.~\cite{Laine:2006cp},
used in the hydrodynamic simulations. See text for details.}
\label{figcs2}
\end{figure}
For hydrodynamics, an important quantity is the speed of sound
squared extracted from the equation of state, 
$c_s^2\equiv \frac{d p(\epsilon)}{d \epsilon}$. For completeness,
we reproduce a plot of this quantity in \autoref{figcs2}.

\subsection{Freeze-out}
\label{sec:fo}

At some stage in the evolution of the matter produced in 
a heavy-ion collision, the system will become too dilute
for a hydrodynamic description to be applicable.
This ``freeze-out'' process is most probably happening 
gradually, but difficult to model realistically.
A widely-used approximation is therefore to assume
instantaneous freeze-out whenever a certain fluid cell
cools below a certain predefined temperature or energy
density (see \cite{Hung:1997du,Dusling:2007gi} for different approaches). 
The standard prescription for this freeze-out
process is the Cooper-Frye formula \cite{Cooper:1974mv}, which allows
conversion of the hydrodynamic variables (energy density, 
fluid velocity,...) into particle distributions.

Specifically, in the case of isothermal freeze-out at a
temperature $T_f$, the conversion from hydrodynamic
to particle degrees of freedom will have to take place
on a three-dimensional freeze-out hypersurface $\Sigma$,
which can be characterized by its normal four-vector, and parametrized by three space-time 
variables \cite{Ruuskanen:1986py,Rischke:1996em}. The spectrum
for a single particle on mass shell with four momentum 
$p^\mu=(E,{\bf p})$ and degeneracy $d$ is then given by
\beq
E\frac{d^3 N}{d^3 {\bf p}}\equiv 
\frac{d}{(2 \pi)^3} \int p_\mu d\Sigma^\mu f\left(x^\mu,p^\mu\right),
\label{CF}
\eeq
where $d\Sigma^\mu$ is the normal vector on the hypersurface 
$\Sigma$ and $f$ is the off-equilibrium distribution function.

Originally, the Cooper-Frye prescription was derived for
systems in thermal equilibrium, where $f$ is
built out of a Bose or Fermi distribution function 
$f_0(x)=\left(\exp[(x)\pm 1]^{-1}\right)$,
depending on the statistics of the particle under consideration.
In order to generalize it to systems out of equilibrium,
one customarily relies on the ansatz used in the derivation
of viscous hydrodynamics from kinetic theory 
\cite{deGroot},
\beq
f\left(x^\mu,p^\mu\right)=f_0\left(\frac{p_\mu u^\mu}{T}\right)
+f_0\left(\frac{p_\mu u^\mu}{T}\right) \left[1\mp 
f_0\left(\frac{p_\mu u^\mu}{T}\right)\right] \frac{p_\mu p_\nu \Pi^{\mu \nu}}
{2 T^2 (\epsilon+p)}.
\label{fullfansatz}
\eeq
For simplicity, in the following we approximate 
$f_0(x)\sim \exp(-x)$, so similarly
\beq
f\left(x^\mu,p^\mu\right)=\exp\left(-p_\mu u^\mu/T\right)
\left[1+
\frac{p_\mu p_\nu \Pi^{\mu \nu}}
{2 T^2 (\epsilon+p)}\right].
\eeq
The effect of this approximation will be commented on 
in the following sections.

In practice, for boost-invariant 2+1 dimensional hydrodynamics,
the freeze-out hypersurface 
$\Sigma^\mu=\left(\Sigma^t,\Sigma^x,\Sigma^y,\Sigma^z\right)=(t,x,y,z)$
can be parametrized either
by $\tau,\xi$ and the polar angle $\phi$, or by $x,y,\xi$:
\beq
\begin{array}{c}
t=\tau \cosh \xi\\
x=x(\tau, \phi)\\
y=y(\tau, \phi)\\
z=\tau \sinh \xi
\end{array}\quad,\qquad \qquad
\begin{array}{c}
t=\tau(x,y) \cosh \xi\\
x=x\\
y=y\\
z=\tau(x,y) \sinh \xi
\end{array}.
\eeq
The normal vector on $\Sigma^\mu$ is calculated by
\begin{align}
d\Sigma_\mu(\tau,\phi,\xi) &= \varepsilon_{\mu \alpha \beta \gamma}
\frac{\partial \Sigma^\alpha}{\partial \tau}
\frac{\partial \Sigma^\beta}{\partial \phi}
\frac{\partial \Sigma^\gamma}{\partial \xi}
d\tau d\phi d\xi\nonumber\\
d\Sigma^\mu(\tau,\phi,\xi)&= -\tau \left(\cosh \xi 
\left(\frac{\partial x}{\partial \tau}\frac{\partial y}{\partial \phi}-
\frac{\partial y}{\partial \tau}\frac{\partial x}{\partial \phi}\right),
\frac{\partial y}{\partial \phi},-\frac{\partial x}{\partial \phi},
\sinh \xi 
\left(\frac{\partial x}{\partial \tau}\frac{\partial y}{\partial \phi}-
\frac{\partial y}{\partial \tau}\frac{\partial x}{\partial \phi}\right)
\right) d\tau d\phi d\xi \nonumber
\end{align}
%\textbf{SIGN?} 
and similarly for the other 
parametrization \cite{Kolb:2003dz}.

\begin{figure}
\center
\includegraphics[width=.5\linewidth]{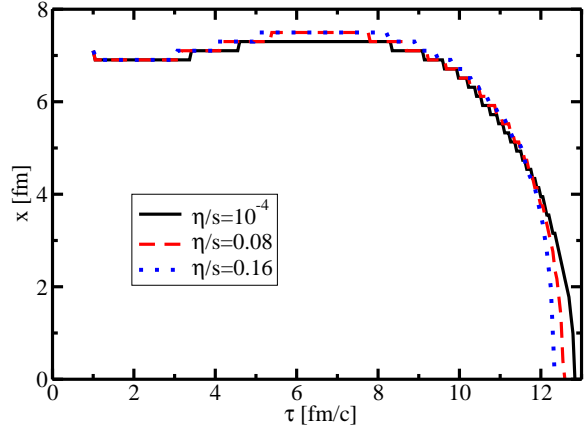}
\caption[Slice of freeze-out hypersurface]{Space-time cut through the three-dimensional hypersurface
for a central collision within the Glauber model. Simulation parameters
used were $a=1$ GeV$^{-1}$, $\tau_0=1$ fm/c, $T_i=0.36$ GeV, 
$T_f=0.15$ GeV, $\tau_\pi=6 \frac{\eta}{s}$ and $\lambda_1=0$ (see
next sections for definitions). As can be seen from the figure,
inclusion of viscosity only slightly changes the form of the surface.}
\label{freeze}
\end{figure}

%\textbf{Plot of freeze-out hypersurface}
For a realistic equation of state, at early times
the freeze-out hypersurface will contain the same transverse coordinate values 
$(x,y)$ for different times $\tau$ (see \autoref{freeze}). Therefore, the 
parametrization in terms of $(x,y,\xi)$ cannot be used for early times. 
On the other hand,
the parametrization in terms of $(\tau,\phi,\xi)$ 
contains derivatives of $(x,y)$
with respect to $\tau$, which become very large at late times (see
\autoref{freeze}).
Numerically, it is therefore not advisable to use this parametrization
at late times.
As a consequence, we use the one parametrization at
early times but switch to the other parametrization at late times,
such that the integral in \autoref{CF} is always defined and
numerically well-behaved\footnote{It may be possible that other
parametrizations may turn out to be more convenient. For instance,
it is conceivable that performing a triangulation of the three-dimensional
hypersurface and replacing the integral in (\ref{CF}) by a sum
over triangles could turn out to be numerically superior to our method.}.

In order to evaluate the integral (\ref{CF}), it is useful
to express $p^\mu$ also in Milne coordinates, 
\beq
p^\mu=(p^\tau,p^x,p^y,p^\xi)=(m_T \cosh (Y-\xi),p^x,p^y,  
\frac{m_T}{\tau}\sinh (Y-\xi)),
\eeq
where $m_T=\sqrt{m^2+p_x^2+p_y^2}=\sqrt{E^2-p_z^2}$. Here and
in the following $Y={\rm arctanh}(p^z/E)$ is the rapidity, 
and $m$ is the rest mass of the
particle under consideration.
Then the $\xi$ integration can be carried out analytically using
\beq
\frac{1}{2}
\int_{-\infty}^\infty d\xi \cosh^n(Y-\xi) \exp[-z \cosh(Y-\xi)]=
(-1)^n \partial_z^n K_0(z) \equiv K(n,z),
\eeq
where $K_0(z)$ is a modified Bessel function.
One finds
\begin{align}
E \frac{d^3 N}{d^3{\bf p}}&=
\frac{2 d}{(2 \pi)^3} \int d\tau d\phi \exp{[(p^x u^x+p^y u^y)/T]}\times
\nonumber\\
& \left[m_T 
\left(\frac{\partial x}{\partial \tau}\frac{\partial y}{\partial \phi}-
\frac{\partial y}{\partial \tau}\frac{\partial x}{\partial \phi}\right)
\left(T_1 K(1,m_T u^\tau/T)+T_2  K(2,m_T u^\tau/T)+T_3 K(3,m_T u^\tau/T)
\right)\right.\nonumber\\
& \left.-\left(p^x \frac{\partial y}{\partial \phi}-
p^y \frac{\partial x}{\partial \phi}\right)
\left(T_1 K(0,m_T u^\tau/T)+T_2  K(1,m_T u^\tau/T)+T_3 K(2,m_T u^\tau/T)
\right)\right]\ ,\nonumber\\
T_1&= 1+\frac{m_T^2 \Pi^\xi_\xi+p_x^2 \Pi^{xx}+p_y^2 \Pi^{yy}+2 p_x p_y
\Pi^{xy}}{2 T^2 (\epsilon+p)}\ ,\nonumber\\
T_2&= -2 m_T  \frac{p^x \Pi^{x \tau}+p^y \Pi^{y \tau}}{2 T^2 (\epsilon+p)}\ ,
\nonumber\\
T_3&= m_T^2\frac{\Pi^{\tau \tau}-\Pi^\xi_\xi}{2 T^2 (\epsilon+p)}\ ,
\end{align}
for the $(\tau,\phi,\xi)$ parametrization,
and a similar result for the other parametrization of the hypersurface.
The remaining integrals for the particle spectrum have 
to be carried out numerically unless one is considering
the case of a central collision 
\cite{Baier:2006gy,Romatschke:2007jx} where the 
integral has an additional symmetry in $\phi$.

For the simulation of a heavy-ion collision,
one then also needs to take into account the 
feed-down process of particle resonances that
decay into lighter, stable particles 
\cite{Sollfrank:1990qz,Sollfrank:1991xm}. 
Therefore, we calculate the spectra for particle
resonances with masses up to $\sim 2$ GeV and
then use available routines from
the AZHYDRO package \cite{OSCAR} to determine the spectra
of stable particles including these feed-down
contributions. Ultimately, one would be interested in
describing the last stage of the evolution by coupling the hydrodynamics 
to a hadronic cascade code \cite{Bass:2000ib,Teaney:2001av,Hirano:2005xf,Nonaka:2006yn}.
We leave this for future work.

The particle spectra $E\frac{d N_{\rm corr}}{d^3 {\bf p}}$ including
feed-down contributions
can then be used to calculate experimental observables at central
rapidity $Y=0$ ,
such as radial and elliptic flow coefficients, $v_0,v_2$, respectively, and the mean transverse momentum $\langle p_T \rangle$, as defined in \autoref{spectra}.
%These are defined as
%\begin{align}
%v_0(p_T,b)=& \int \frac{d \phi_p}{2 \pi} E\frac{d N_{\rm corr}}{d^3 {\bf p}}\ ,
%&%\qquad
%E\frac{d N_{\rm corr}}{d^3 {\bf p}}=&v_0(p_T,b) \left[1+2 v_2 (p_T,b) \cos (2 \phi_p)
%+\ldots\right],
%\end{align}
%where $\phi_p=\arctan (p^y/p^x)$ and $p_T=\sqrt{p_x^2+p_y^2}$. Furthermore, 
%the total multiplicity per unit rapidity $\frac{dN}{dy}$ and 
%the mean transverse momentum $<p_T>$ 
%are then given by
%\beq
%\frac{d N}{dy}\equiv 2\pi \int dp_T\ p_T\ v_0(p_T,b)\ ,
%\qquad
%<p_T>\equiv \frac{\int d p_T\  p_T^2 \ v_0(p_T,b)}
%{\int d p_T\ p_T\ v_0 (p_T,b)}.
%\eeq
%The $p_T$ integrated elliptic flow coefficient is defined as
%\beq
%v_2^{\rm int}(b)=\frac{\int d p_T\ p_T v_2(p_T,b) v_0(p_T,b)}
%{\int d p_T\ p_T\ v_0 (p_T,b)}
%\eeq
%and the minimum bias elliptic flow coefficient as \cite{Kolb:2001qz}
%\beq
%v_2^{{\rm mb}}(p_T)=\frac{\int db\ b\ v_2(p_T,b)\ v_0(p_T,b)}
%{\int db\ b\ v_0 (p_T,b)}.
%\eeq

\subsection{Code tests}

It is imperative to subject the numerical implementation
of the relativistic viscous hydrodynamic model to several tests.
The minimal requirement is that the code is stable for a range
of simulated volumes and grid spacings $a$, such that 
an extrapolation to the continuum may be attempted (keeping the 
simulated volume fixed but sending $a\rightarrow 0$).
Our code fulfills this property.

Furthermore, one has to test whether 
this continuum extrapolation corresponds to the correct physical
result in simple test cases.
One such test case is provided by the 0+1 dimensional
model discussed in \autoref{section01}. Using initial
conditions of uniform energy density in the 2+1 dimensional
numerical code, the temperature evolution should match
that of \autoref{0+1dsystem}, for which it is straightforward 
to write an independent numerical solver. 
Our 2+1 dimensional code passes this test, for small and large
$\eta/s$ and different values for $\tau_\pi,\lambda_1$.

The above test is non-trivial in the sense that it
allows to check the implementation of
nonlinearities in the hydrodynamic model.
However, it does not probe the dynamics of the model,
since, e.g., all velocities are vanishing.
Therefore, another test that one can (and should!) conduct is to
study the dynamics of the model against that
of linearized hydrodynamics (this test was first
outlined in Ref.~\cite{Baier:2006gy}; see \cite{Bhalerao:2007ek} for similar
considerations). More specifically,
let us consider a viscous background ``solution'' with 
$u^i=0$ but non-vanishing $\epsilon(\tau),\Pi^\xi_\xi(\tau)$
obeying \autoref{0+1dsystem}. To first order in small fluctuations 
$\delta \epsilon,\delta u^\mu, \delta \Pi^{\mu \nu}$
around this background the set of equations (\ref{manyeq}) become
\begin{align}
\left[c_s^2 \partial_\tau \epsilon+\frac{1}{2}\partial_\tau \Pi^\xi_\xi
+\frac{3}{2 \tau} \Pi^\xi_\xi
+(\epsilon+p+\frac{1}{2}\Pi^\xi_\xi)\partial_\tau \right] \delta u^x
+c_s^2 \partial_x \delta \epsilon + \partial_i \delta \Pi^{x i}
&= 0\nonumber\\
\left[c_s^2 \partial_\tau \epsilon+\frac{1}{2}\partial_\tau \Pi^\xi_\xi
+\frac{3}{2 \tau} \Pi^\xi_\xi
+(\epsilon+p+\frac{1}{2}\Pi^\xi_\xi)\partial_\tau \right] \delta u^y
+c_s^2 \partial_y \delta \epsilon + \partial_i \delta \Pi^{y i}
&= 0\nonumber\\
\left[\partial_\tau +\frac{1+c_s^2}{\tau} \right]\delta\epsilon+\left[(\epsilon+p)+\frac{1}{2}\Pi^\xi_\xi\right] \partial_i \delta u^i
-\frac{1}{\tau}\delta \Pi^\xi_\xi
&= 0\nonumber\\
\left[\frac{4}{3\tau}+\frac{1}{\tau_\pi}+\partial_\tau\right]
\delta \Pi^\xi_\xi
-\left[\frac{4 \eta}{3 \tau \tau_\pi}+ \frac{1}{4\tau_\pi} \Pi^\xi_\xi\right] \frac{\delta \epsilon}{\epsilon}
+\left[\frac{2 \eta}{3 \tau_\pi} +\frac{4}{3} \Pi^\xi_\xi\right]
\partial_i \delta u^i&= 0\nonumber\\
\left[\frac{4}{3\tau}+\frac{1}{\tau_\pi}+\partial_\tau\right]
\delta \Pi^{xx}
-\left[\frac{2 \eta}{3 \tau_\pi \tau}+ \frac{1}{4\tau_\pi} \Pi^{xx}\right]\frac{\delta \epsilon}{\epsilon}
+\frac{2 \eta}{\tau_\pi}\partial_x \delta u^x
+\left[-\frac{2 \eta }{3 \tau_\pi}+\frac{4}{3} \Pi^{xx}\right]
\partial_i \delta u^i&= 0\nonumber\\
\left[\frac{4}{3\tau}+\frac{1}{\tau_\pi}+\partial_\tau\right]
\delta \Pi^{xy}
+\frac{\eta}{\tau_\pi}\left(\partial_x \delta u^y+\partial_y
\delta u^x\right)&= 0,\qquad \
\label{linearizedeq}
\end{align}
where we have put $\lambda_1=0$ and assumed a constant $c_s^2$ for simplicity. 
Noting that $\delta \Pi^{yy}=\delta \Pi^\xi_\xi-\delta \Pi^{xx}$
from $\delta \Pi^\mu_\mu=0$,  \autoref{linearizedeq} 
are a closed set of linear, but coupled differential equations
for the fluctuations 
$\delta \epsilon,\delta u^x,\delta u^y,\delta \Pi^\xi_\xi, \delta \Pi^{xx},\delta \Pi^{xy}$. Doing a Fourier transform,
\beq
\delta \epsilon(\tau,x,y)=\int \frac{d^2 {\bf k}}{(2 \pi)^2}
e^{i x k^x+i y k^y} \delta \epsilon(\tau,k^x, k^y)
\eeq
(and likewise for the other fluctuations), \autoref{linearizedeq}
constitute coupled ordinary differential equations for each
mode doublet ${\bf k}=(k^x,k^y)$, which again are straightforward
to solve with standard numerical methods \cite{codedown} (and analytically for 
ideal hydrodynamics).

A useful test observable is the correlation function
\beq
f(\tau,{\bf x_1},{\bf x_2})=\frac{\left<\delta \epsilon(\tau,{\bf x_1}) 
\delta \epsilon(\tau,{\bf x_2})\right>}{\epsilon(\tau)^2},
\eeq
where $\left< \right>$ denotes an ensemble average
over initial conditions $\left.\delta \epsilon\right|_{\tau=\tau_0}$.
In particular, let us study initial conditions where $\delta \epsilon$
is given by Gaussian random noise with standard deviation $\Delta$, 
\beq
f(\tau_0,{\bf x_1},{\bf x_2})=\Delta^2 \delta^2({\bf x_1-x_2})
\eeq
and all other fluctuations vanish initially. These initial
conditions are readily implemented both for the full 2+1 dimensional
hydrodynamic code as well as for the linearized system 
\autoref{linearizedeq}. As the system evolves to finite time $\tau$,
both approaches have to give the same correlation function $f$ as 
long as the linearized treatment is applicable, and hence
\autoref{linearizedeq} can be used to test the dynamics of 
the full numerical code.

\begin{figure}
\center
\includegraphics[width=.5\linewidth]{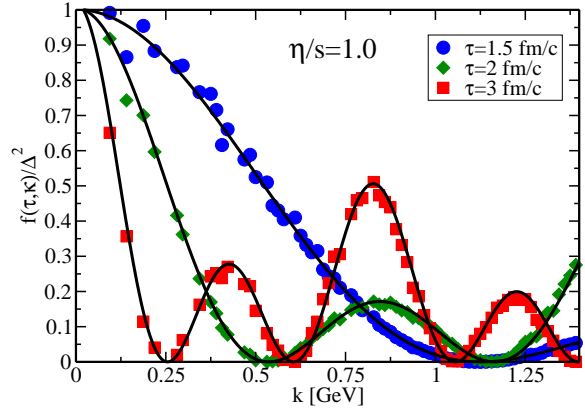}
\caption[Correlation function $f(\tau,{\bf k})$ 
as a function of momentum $k=|{\bf k}|$ in full hydrodynamic simulation compared to linearized hydrodynamics]{The correlation function $f(\tau,{\bf k})$ 
as a function of momentum $k=|{\bf k}|$
for a lattice with $a=1$ GeV$^{-1}$, $64^2$ sites and averaged over
30 initial configurations (symbols),
compared to the result from the linearized hydrodynamic
equations (lines).}
\label{figflucs}
\end{figure}

In practice, note that for the above construction $f$ can only
depend on the difference of coordinates,
\beq
\frac{\left<\delta \epsilon(\tau,{\bf x_1}) 
\delta \epsilon(\tau,{\bf x_2})\right>}{\epsilon(\tau)^2}
=f(\tau,{\bf x_1}-{\bf x_2})=\int \frac{d^2{\bf k}}{(2\pi)^2} e^{i {\bf k}\cdot
({\bf x_1-x_2})} f(\tau,{\bf k})
\eeq
and therefore in Fourier-space
\beq
 f(\tau,{\bf k}) \delta^2({\bf k^\prime}) =
\frac{\left<\delta \epsilon(\tau,{\bf k}) 
\delta \epsilon(\tau,{\bf k^\prime-k})\right>}{(2 \pi)^2 \epsilon(\tau)^2}.
\label{corrdef}
\eeq
In the full 2+1 dimensional numerical code which is discretized
on a space-time lattice, $\delta^2({\bf k^\prime})$
is regular for any finite $a$, and one can maximize the signal
for $f(\tau,{\bf k})$ by calculating the r.h.s. 
of \autoref{corrdef} for ${\bf k^\prime}=0$. 
Similarly, one solution $\delta \epsilon (\tau,{\bf k})$ per ${\bf k}$ mode
is sufficient calculate $f(\tau,{\bf k})$
for the linearized system \autoref{linearizedeq}.

The above initial conditions imply $f(\tau=\tau_0,{\bf k})=\Delta^2$,
but for finite times characteristic peaks develop as a function
of $|{\bf k}|$, whose position, height
and width are sensitive to the values of $c_s^2,\tau_\pi,\eta/s$ and
of course the correct implementation of the hydrodynamic equations.
The comparison between full numerics and linearized treatment
shown in \autoref{figflucs} suggests that our code also passes
this test\footnote{Note that a small numerical
error occurred in the linearized hydrodynamic solver 
and the corresponding figure in Ref.~\cite{Romatschke:2007mq}.
This error has been corrected in \autoref{figflucs}.}.

Finally, for the case of ideal hydrodynamics, 
analytic solutions to the hydrodynamic equations
are known \cite{Baym:1984sr,Chojnacki:2006tv,Nagy:2007xn}.
Specifically, the code for central 
collisions \cite{Baier:2006gy} has been found to agree with the results from
Ref.~\cite{Baym:1984sr} for ideal hydrodynamics. Since our code
agrees with Ref.~\cite{Baier:2006gy} in the case of central collisions
and when dropping the appropriate terms in the equations (\ref{maineq}),
this provides yet another test on our numerics.

To summarize, after conducting the above tests we are 
reasonably confident that our numerical
2+1 dimensional code solves the relativistic viscous hydrodynamic
equations (\ref{manyeq}) correctly.
This completes the setup of a viscous hydrodynamic description
of relativistic heavy-ion collisions. In the following sections,
we will review comparisons of viscous hydrodynamic simulations
to experimental data, for both Glauber and CGC initial conditions.

\section{Initial Conditions: Glauber Model vs. CGC}
\label{sec:two}

\subsection{The Glauber model}

In the Glauber model \cite{Kolb:2001qz}, the starting point is
the Woods-Saxon density distribution for nuclei,
\beq
\rho_A({\bf x})=\frac{\rho_0}{1+\exp{[(|{\bf x}|-R_0)/\chi]}},
\eeq
where for a gold nucleus with weight $A=197$ we use $R_0=6.4$ fm and
$\chi=0.54$ fm. The parameter $\rho_0$ is chosen such
that $\int d^3 {\bf x} \rho_A({\bf x})=A$.
One can then define the nuclear thickness function 
\beq
T_A(x^i)=\int_{-\infty}^\infty dz \rho_A({\bf x}) 
\eeq
and subsequently the number density of
nucleons participating in the collision ($n_{\rm Part}$)
and the number density of binary collisions ($n_{\rm Coll}$).
For a collision of two nuclei with weight A 
at an impact parameter $b$, one has
\begin{align}
n_{\rm Part}(x,y,b)&= T_A\left(x+\frac{b}{2},y\right)
\left[1-\left(1-\frac{\sigma T_A\left(x-\frac{b}{2},y\right)}{A}\right)^A
\right]\nonumber\\
&
+T_A\left(x-\frac{b}{2},y\right)
\left[1-\left(1-\frac{\sigma T_A\left(x+\frac{b}{2},y\right)}{A}\right)^A
\right],\nonumber\\
n_{\rm Coll}(x,y,b)&= \sigma T_A\left(x+\frac{b}{2},y\right) T_A\left(x-\frac{b}{2},y\right),
\end{align}
where $\sigma$ is the nucleon-nucleon cross section. We assume
\hbox{$\sigma\simeq 40$ mb} for Au+Au collisions at $\sqrt{s}=200$
GeV per nucleon pair.

While the total number of participating nucleons
$N_{\rm Part}(b)=\int dx dy n_{\rm Part}(x,y,b)$
will be used to characterize the centrality class of the 
collision, as an initial condition for the energy density
we will only use the parametrization
\beq
\epsilon(\tau=\tau_0,x,y,b)={\rm const}\times n_{\rm Coll}(x,y,b),
\label{edglauber}
\eeq
since it gives a sensible description of the multiplicity
distribution of experimental data, as will be discussed later on.
In the following, ``Glauber-model initial condition'' is
used synonymous to \autoref{edglauber}.

The constant in \autoref{edglauber} is chosen such that the 
central energy density for zero impact parameter,
$\epsilon(\tau=\tau_0,0,0,0)$ corresponds to a predefined
temperature $T_i$ via the equation of state. 
This temperature will be treated as a free parameter and
is eventually fixed by matching to experimental data on
the multiplicity.

\subsection{The CGC model}

The other model commonly used to obtain initial conditions for 
hydrodynamics is the so-called Color-Glass-Condensate approach, 
based on ideas of gluon saturation at high energies.  In particular, 
we use a modified version of the KLN (Kharzeev-Levin-Nardi) 
$k_T$-factorization approach \cite{Kharzeev:2002ei}, 
due to Drescher {\it et al.} \cite{Drescher:2006pi}. We follow exactly 
the procedure described in \cite{Dumitru:2007qr} and in fact we use the same 
numerical code, provided to us by the authors and only slightly modified 
to output initial conditions suitable for input into our viscous 
hydrodynamics program.%, and referred to here as fKLN.

In this model, the number density of gluons produced in a 
collision of two nuclei with atomic weight $A$ is given by
\beq
  \frac{dN_g}{d^2 {\bf x}_{T}dY} = {\cal N}
   \int \frac{d^2{\bf p}_T}{p^2_T}
  \int^{p_T} d^2 {\bf k}_T \;\alpha_s(k_T) \;
  \phi_A(x_1, ({\bf p}_T+{\bf k}_T)^2/4;{\bf x}_T)\;
              \phi_A(x_2, ({\bf p}_T - {\bf k}_T)^2/4;{\bf x}_T)
 \label{eq:ktfac}
\eeq
where ${\bf p}_T$ and $Y$ are the transverse momentum and 
rapidity of the produced gluons, respectively.  
$x_{1,2} = p_T\times\exp(\pm Y)/\sqrt{s}$ is the momentum fraction 
of the colliding gluon ladders with $\sqrt{s}$ the center of mass 
collision energy and $\alpha_s(k_T)$ is the strong coupling constant 
at momentum scale $k_T \equiv \left| {\bf k}_T \right|$.

The value of the normalization constant $\cal N$ is unimportant here, 
since as for Glauber initial conditions, we treat the overall normalization 
of the initial energy density distribution as a free parameter.
The unintegrated gluon distribution functions are taken as
\beq  \label{uGDF}
\phi (x,k^2_{T}; {\bf x}_{T}) =
\frac{1}{\alpha_s (Q^2_s)} \frac{Q^2_s}{\textrm{max}(Q^2_s,k^2_{T})}
\,P({\bf x}_{T})(1-x)^4~,
\eeq
$P({\bf x}_{T})$ is the probability of finding at least one nucleon at transverse position ${\bf x}_{T}$, taken from the definition for $n_{\rm Part}$
\beq
P({\bf x}_{T}) = 1-\left(1-\frac{\sigma T_A}{A}\right)^A,
\eeq
where $T_A$ and $\sigma$ are as defined in the previous section.

The saturation scale at a given
momentum fraction $x$ and transverse coordinate ${\bf x}_{T}$ is given by
\beq
  Q^2_{s}(x,{\bf x}_T) =
  2\,{\rm GeV}^2\left(\frac{T_A({\bf x}_T)/P({\bf x}_T)}{1.53/{\rm fm}^2}\right)
  \left(\frac{0.01}{x}\right)^\lambda~.
  \label{eq:qs}
\eeq
%$T_A$ is the nuclear thickness function as defined in the previous section.  
The growth speed is taken to be $\lambda = 0.288$.

%For simplicity, we take $\alpha_s$ to be constant, which has little effect on the resulting distribution, as compared to allowing the coupling to run according to LO pQCD [].

The initial conditions for hydrodynamic evolution require that we specify 
the energy density in the transverse plane at some initial proper time 
$\tau_0$ at which the medium has thermalized.  \autoref{eq:ktfac}, on the 
other hand, is in principle valid at a time $\tau_s = 1/Q_s$ at which the medium is likely not 
yet in thermal equilibrium.  To obtain the desired initial conditions, we 
again follow \cite{Dumitru:2007qr} and assume that the number of gluons is 
effectively conserved during the evolution from $\tau_s$ to $\tau_0$ and so 
the number density profile is the same at both times, scaled by the 
one-dimensional Bjorken expansion $n(\tau_0) = \frac {\tau_s}{\tau_0} n(\tau_s)$.  
The energy density can then be obtained from the number density through 
thermodynamic relations---it is proportional to the number density to the 4/3 power.  
Again, we take the overall normalization as a free parameter, so the initial 
energy density is finally given as
\beq
 \epsilon(\tau=\tau_0,{\bf x}_T,b)={\rm const}\times \left[ \frac{dN_g}{d^2 {\bf x}_{T}dY}({\bf x}_T,b)\right] ^{4/3}
\label{edCGC}
\eeq
where the number density is given by \autoref{eq:ktfac} evaluated at central rapidity $Y=0$.

As a final comment, it should be pointed out that the original version of the
CGC, the McLerran-Venugopalan model \cite{McLerran:1993ni,McLerran:1993ka}, 
differs from the KLN ansatz we used here, as will be discussed in the next-section.

\subsection{Spatial and momentum anisotropy}
\label{sec:aniso}

\begin{figure}
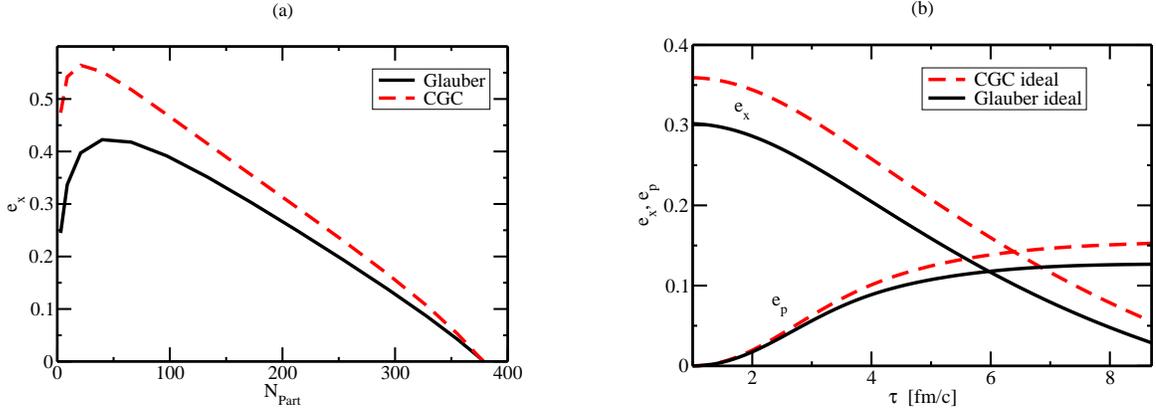

\includegraphics[width=.45\linewidth]{newdev/eccentricity2.eps}
\hfill
\includegraphics[width=.45\linewidth]{newdev/SpatialMomentumEccentricity.eps}
\caption[Initial spatial anisotropy for Glauber
and CGC models and time evolution of spatial and momentum anisotropy]{Left: The initial spatial anisotropy for the Glauber
and CGC model. Right: The time
evolution of the spatial and momentum anisotropy for a collision
with $b=7$ fm in ideal hydrodynamics.}
\label{fig:eccen}
\end{figure}

One of the key parameters discussed in the following is the
eccentricity (or spatial anisotropy) of the collision geometry.
Following \cite{Kolb:2001qz}, we define it as 
\beq
e_x\equiv 
\frac{\langle y^2-x^2\rangle_{\epsilon}}{\langle y^2+x^2\rangle_\epsilon},
\eeq
where $\langle \rangle_\epsilon$ denotes an averaging procedure over space 
with the energy density $\epsilon$ as a weighting factor. 
Shown in \autoref{fig:eccen}a, a plot of $e_x$ for different
centralities highlights the quantitative difference between the 
initial energy density from the Glauber and CGC model, 
\autoref{edglauber} and \autoref{edCGC}, respectively.
As can be seen from this figure, the CGC model generally
gives a higher spatial anisotropy than the Glauber model.
Note that the results for the CGC model shown here are
extreme in the sense that the McLerran-Venugopalan model
gives spatial eccentricities which essentially match
the ones from the Glauber model \cite{Lappi:2006xc}.
This allows us to use the difference between the CGC and 
Glauber models as an indication of the systematic theoretical 
error stemming from the ignorance of the correct physical
initial condition.

Hydrodynamics converts pressure gradients into fluid velocities,
and hence one expects the spatial anisotropy to decrease at the expense
of a momentum anisotropy (which is related to the magnitude 
of the elliptic flow). We follow \cite{Kolb:1999it}
in defining a momentum anisotropy according to
\beq
e_p\equiv\frac{\langle T^{xx}-T^{yy}\rangle}{\langle T^{xx}+T^{yy}\rangle},
\label{momaniso}
\eeq
where we stress that here $\langle\rangle$ denotes spatial averaging
with weight factor unity.  \autoref{fig:eccen}b shows the time
evolution in ideal hydrodynamics ($\eta/s\ll 1$)
of both the spatial and momentum anisotropies 
for a heavy-ion collision at $b=7$ fm modeled through 
Glauber and CGC initial conditions.
As one can see, for the same impact parameter,
the higher initial spatial anisotropy for the CGC
model eventually leads to a higher momentum anisotropy than
the Glauber model. Using a quasiparticle interpretation where
the energy momentum tensor is given by 
\beq
T^{\mu \nu}\propto \int \frac{d^3 {\bf p}}{(2 \pi)^3} \frac{p^\mu p^\nu}{E}
f\left(x^\mu, p^\mu\right),
\eeq
the momentum anisotropy $e_p$ can be approximately related to 
the integrated elliptic flow $v_2^{\rm int}(b)$,
with a proportionality factor of $\sim2$ \cite{Kolb:1999it,Ollitrault:1992bk}.
We find this proportionality to be maintained even for 
non-vanishing shear viscosity, as can be seen in \autoref{fig:v2int}.

\section{Results}
\label{sec:three}

\subsection{Which parameters matter?}

\begin{figure}
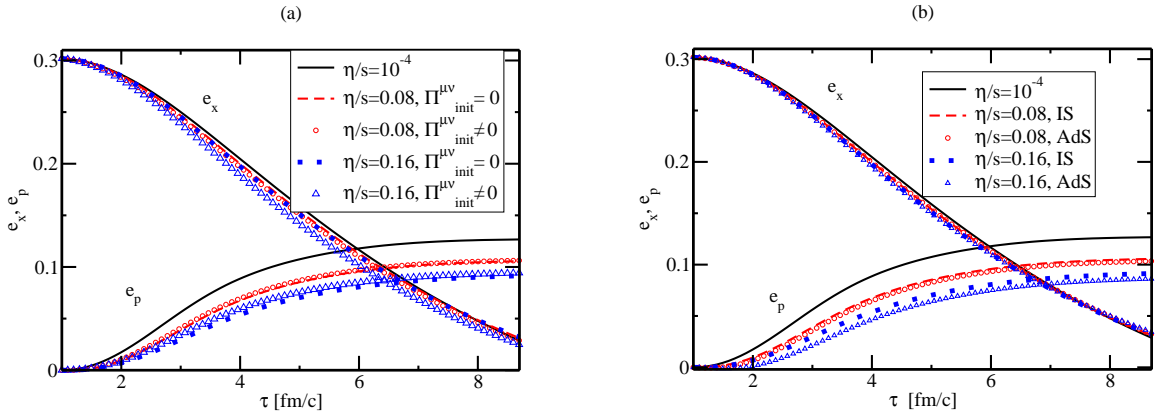

\includegraphics[width=.45\linewidth]{newdev/NZ.eps}
\hfill
\includegraphics[width=.45\linewidth]{newdev/sensi.eps}
\caption[Spatial and momentum anisotropy as a function of proper time for various values of the shear tensor initialization and second-order transport coefficients]{Spatial and momentum anisotropy for the Glauber model at $b=7$ fm
with $T_i=0.353$ GeV, $\tau_0=1$ fm/c and various values for the viscosity
(grid spacing $a=2\ {\rm GeV}^{-1}$). 
(a): The dependence on the 
initialization value of the shear tensor: shown are results
for vanishing initial value ($\Pi^{\mu\nu}_{\rm init}=0$) and 
Navier-Stokes initial value ($\Pi^{\mu\nu}_{\rm init}\neq 0$),
given in \autoref{FOvalue}.
(b): The dependence on the choice of value for $\tau_\pi,\lambda_1$:
shown are results for $\tau_\pi=\frac{6}{T}\frac{\eta}{s}$, 
$\lambda_1=0$ (labelled ``IS'') and 
$\tau_\pi=\frac{2(2-\ln2)}{T}\frac{\eta}{s}$, 
$\lambda_1=\frac{\eta}{2 \pi T}$ (labelled ``AdS''). 
For $\tau_\pi=\frac{2(2-\ln2)}{T}\frac{\eta}{s}$, 
the results for $\lambda_1=0$ (not shown) would be indistinguishable
by bare eye from those for $\lambda_1=\frac{\eta}{2 \pi T}$.}
\label{fignotmatter}
\end{figure}

In the following, we will attempt to obtain limits on
the mean value (throughout the hydrodynamic evolution) 
of the ratio $\eta/s$ from experimental data.
While e.g. temperature variations of $\eta/s$ are to
be expected in the real physical systems, probing for
such variations would invariable force us to introduce more
unknown parameters. We prefer to leave this program
for future studies once robust results for the mean value of $\eta/s$
exist.
Having fixed the equation of state and the freeze-out procedure
as explained in the previous sections, the remaining
choices that have to be made in the hydrodynamic model are the
\begin{itemize}
\item
Initial energy density profile: Glauber or CGC
\item
Initial value of shear tensor: vanishing or Navier-Stokes value
\item
Hydrodynamic starting time $\tau_0$
\item
Second-order coefficients: relaxation time $\tau_\pi$ and $\lambda_1$
\item
Ansatz for non-equilibrium particle distribution \autoref{fullfansatz}
\end{itemize}
where it is to be understood that we fix the initial energy density normalization ($T_i$) and the freeze-out temperature $T_f$ such that 
the model provides a reasonable description of the experimental data on 
multiplicity and $\langle p_T\rangle$.
Historically, a strong emphasis has been put on requiring
a small value of $\tau_0$ for ideal hydrodynamics \cite{Kolb:2000sd,Heinz:2004pj}. 
For this reason, we will discuss the dependence on $\tau_0$ separately
in \autoref{notherm}.
A good indicator for which parameters
matter is the momentum anisotropy since it is very sensitive
to the value of $\eta/s$. From \autoref{fig:eccen} one therefore
immediately concludes that the choice of Glauber or CGC initial
conditions is important since it has a large effect on $e_p$.
Fortunately, most of the other choices turn out not to have a 
strong influence on the resulting $v_2$ coefficient and hence
the extracted $\eta/s$.  In the following we test for this sensitivity
by studying $e_p$ for a ``generic'' heavy-ion collision of two gold nuclei,
modeled by Glauber initial conditions at an initial starting
temperature of $T_i=0.353$, an impact parameter of $b=7$ fm,
and various choices of the above parameters. 

\autoref{fignotmatter} shows the time evolution of $e_x,e_p$
for various values of $\eta/s$. From these plots, it can be
seen that $e_p$ (and hence $v_2$) clearly is sensitive to the value of
$\eta/s$, suggesting that it can be a useful observable
to determine the viscosity of the fluid from experiment.
However, in order to be a useful probe of the fluid viscosity,
the dependence of the final value of $e_p$ on other parameters
should be much weaker than the dependence on $\eta/s$.
In \autoref{fignotmatter}a we show $e_p$, calculated
for $\Pi^{\mu \nu}(\tau_0)=0$ and $\Pi^{\mu \nu}(\tau_0)$ equal
to the Navier-Stokes value, \autoref{FOvalue}.
As can be seen from this figure, the resulting
anisotropies are essentially independent of this choice, 
corroborating the finding in Ref.~\cite{Song:2007fn,Song:2007ux}.
Similarly, in \autoref{fignotmatter}b we show $e_p$
calculated in simulations where the values of the 
second-order transport coefficients were 
either those of a weakly-coupled M\"uller-Israel-Stewart theory ($\tau_\pi=6 \frac{\eta}{sT}$,
$\lambda_1=0$) or those inspired by a strongly coupled
${\cal N}=4$ SYM plasma ($\tau_\pi=2(2-\ln 2) \frac{\eta}{s T}$,
$\lambda_1=\frac{\eta}{2 \pi T}$). 
Again, the dependence
of $e_p$ on the choice of the values of $\tau_\pi,\lambda_1$
can be seen to be very weak for the values of $\eta/s$ shown here. 
This result is in stark contrast to the findings of 
Ref.~\cite{Song:2007fn}, where a large sensitivity on the value
of $\tau_\pi$ was found. However, recall that 
Ref.~\cite{Song:2007fn} used evolution equations that
differ from \autoref{maineq} and in particular
do not respect conformal invariance. As argued in
\autoref{section01}, it is therefore expected
to encounter anomalously large sensitivity to
the value of the second order transport coefficients.

\begin{figure}
\begin{center}
\includegraphics[width=.5\linewidth]{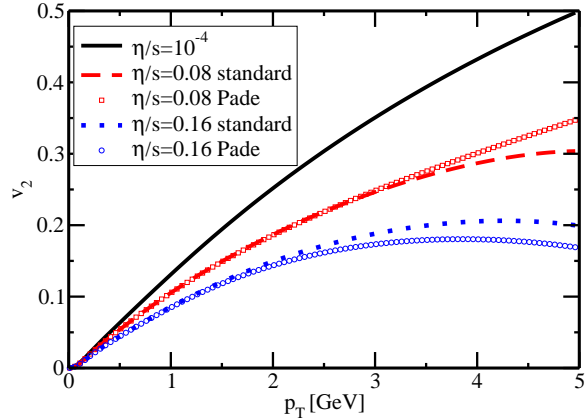}
\end{center}
\caption[Charged hadron elliptic flow using Pad\'e resummed particle spectra compared to our standard method]{Charged hadron elliptic flow for the Glauber model at $b=7$ fm with $T_i=0.353$ GeV,
$\tau_0=1$ fm/c and various viscosities.}
\label{fig:Pade}
\end{figure}

To study the dependence of results on the ansatz of
the non-equilibrium particle distribution function (\ref{fullfansatz}),
one would want to quantify the effect of neglecting terms
of higher order in momenta in \autoref{fullfansatz}. To estimate this,
let us rewrite $E d^3 N/d^3{\bf p}= E d^3 N^{(0)}/d^3{\bf p}+E d^3 N^{(1)}/d^3{\bf p}$,
where $N^{(0)}$ contains only the equilibrium part where 
$f(x^\mu,p^\mu)=f_0\left(\frac{p_\mu u^\mu}{T}\right)$, and perform a 
Pad\'e-type resummation,
\beq
E \frac{d^3 N^{\rm Pade}}{d^3{\bf p}}\equiv E \frac{d^3 N^{(0)}}{d^3{\bf p}}
%\frac{1}{1-\frac{d N^{(1)}}{d N^{(0)}}}.
\frac{1}{1-\frac{d^3 N^{(1)}}{d^3{\bf p}} \frac{d^3{\bf p}}{d^3 N^{(0)}}}.
\label{Padere}
\eeq
Since \autoref{Padere} contains powers of momenta to all orders when re-expanded,
the difference between the ansatz (\ref{fullfansatz}) and the Pad\'e resummed
particle spectra can give a handle on the systematic error of the truncation
used in \autoref{fullfansatz}. Shown in \autoref{fig:Pade}, this
difference suggests that this systematic error is small 
for momenta $p_T \lesssim 2.5$ GeV. Therefore, we do not
expect our results to have a large systematic uncertainty coming from 
the particular ansatz (\ref{fullfansatz}) for these momenta.

To summarize, for values of $\eta/s\lesssim 0.2$, the 
results for the momentum anisotropy are essentially 
insensitive to the choices for the second-order transport
coefficients $\tau_\pi,\lambda_1$ and the initialization
of the shear tensor $\Pi^{\mu\nu}(\tau=\tau_0)$.
Conversely, $e_p$ is sensitive to the value of viscosity
and the choice of initial energy density profile (initial eccentricity).
Since the physical initial condition is currently unknown, 
this dependence will turn out to be the dominant
systematic uncertainty in determining $\eta/s$ from 
experimental data.

\subsection{Multiplicity and radial flow}

As outlined in the introduction, we want to 
match the hydrodynamic model to experimental data
for the multiplicity, thereby fixing the constant in
Equations \ref{edglauber} and \ref{edCGC}. This translates to fixing an initial
central temperature $T_i$ for $b=0$, which we will quote 
in the following. 

For a constant speed of sound, the evolution for ideal
hydrodynamics is isentropic, while for viscous
hydrodynamics additional entropy is produced.
Since the multiplicity is a measure of the entropy of
the system, one expects an increase of multiplicity
for viscous compared to ideal hydrodynamic evolution.
This increase in final multiplicity has been measured as a 
function of $\eta/s$ for the semi-realistic speed of 
sound \autoref{figcs2} in central heavy-ion collisions 
in Ref.~\cite{Romatschke:2007jx},
and found to be approximately\footnote{
The quoted fraction is for a hydrodynamic starting time of $\tau_0=1$ fm/c.
Reducing $\tau_0$ leads to considerably larger entropy production.} 
a factor of $0.75 \eta/s$.
(See Ref.~\cite{Lublinsky:2007mm,Dumitru:2007qr} for 
related calculations in simplified models.)
Reducing $T_i$ accordingly therefore ensures that
for viscous hydrodynamics, 
the multiplicity in central collisions will stay close
to that of ideal hydrodynamics.

Hydrodynamics gradually 
converts pressure gradients into flow velocities,
which in turn relate to the mean particle momenta.
Starting at a predefined time $\tau_0$ and requiring the hydrodynamic model spectra to
match the experimental data on particle $\langle p_T\rangle$ 
then fixes the freeze-out temperature $T_f$.

\begin{figure}
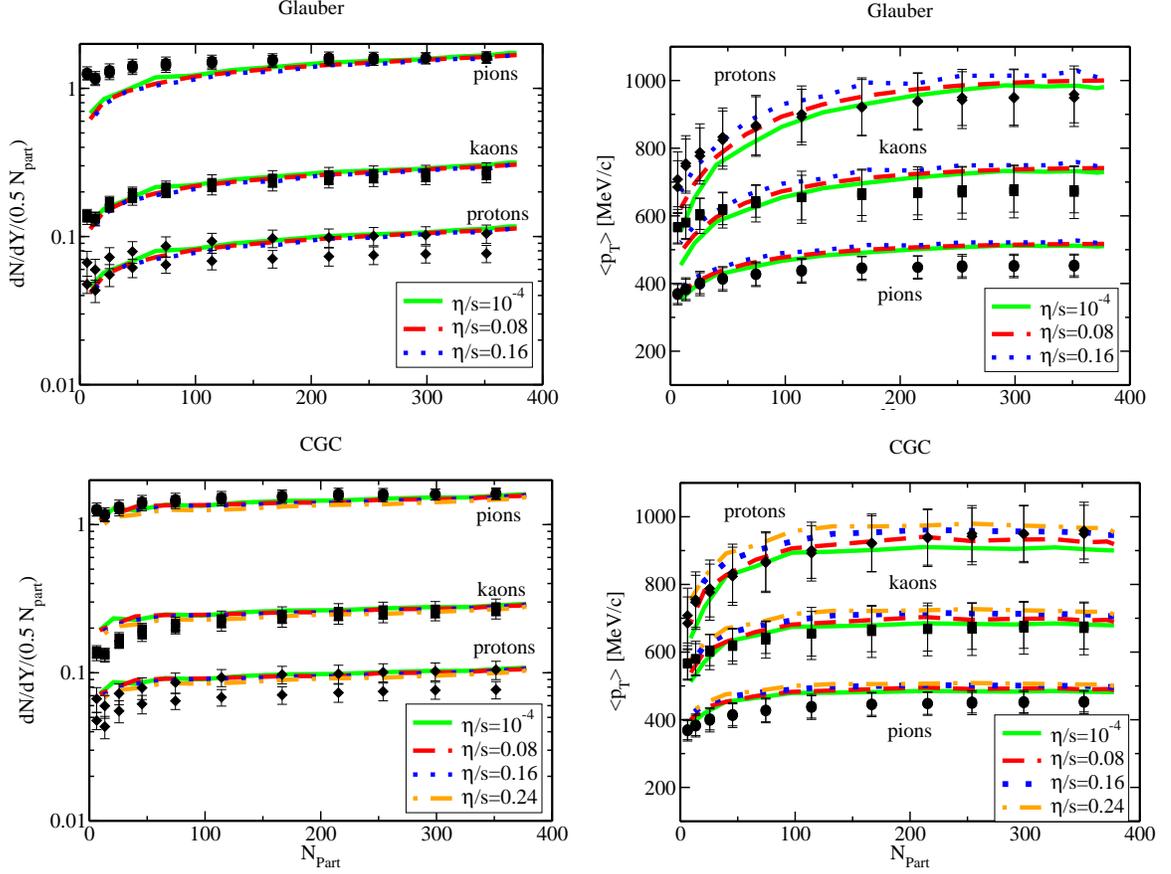

\includegraphics[width=.48\linewidth]{newdev/multGlauber.eps}
\hfill
\includegraphics[width=.48\linewidth]{newdev/meanptGlauber.eps}
\hfill\\
\vspace*{1cm}
\includegraphics[width=.48\linewidth]{newdev/multCGC.eps}
\hfill
\includegraphics[width=.48\linewidth]{newdev/meanptCGC.eps}

\caption[Centrality dependence of $dN/dY$ and $\langle p_T \rangle$ 
for identified particles compared to viscous hydrodynamic calculations]{Centrality dependence of total
multiplicity $dN/dY$ and $\langle p_T \rangle$ 
for $\pi^+$, $\pi^-$, $K^+$, $K^-$, $p$ and $\bar{p}$ 
from PHENIX \cite{Adler:2003cb} 
for Au+Au collisions at $\sqrt{s}=200$ GeV, compared to 
the viscous hydrodynamic model and various $\eta/s$, 
for Glauber initial conditions 
and CGC initial conditions. The model parameters used here are 
$\tau_0=1$ fm/c, $\tau_\pi=6 \eta/s$,
$\lambda_1=0$, $T_f=140$ MeV %, grid spacing $a=1$ GeV$^{-1}$
and adjusted $T_i$ (see \autoref{tab:par}). }
\label{mult1}
\end{figure}

For both Glauber-type and CGC-type model initial conditions, 
%For ideal hydrodynamics with $T_i=360$ MeV at $\tau_0=1$ fm/c 
%and $T_f=150$ MeV, 
the experimental impact parameter 
dependence of the multiplicity and $\langle p_T\rangle$
is reasonably well parametrized for both ideal hydrodynamics
as well as viscous hydrodynamics provided $T_i$ is adjusted
accordingly (see \autoref{mult1}). %For convenience,
The values for $T_i$ used in the simulations
are compiled in \autoref{tab:par}.
%are given in the following table:
We recall that no chemical potential is included in our
equation of state, prohibiting a distinction between particles
and anti-particles, and chemical and kinetic freeze-out of
particles occurs at the same temperature.
Furthermore, approximating the equilibrium particle-distributions
for bosons by a Boltzmann distribution (\ref{fullfansatz}) 
leads to small, but consistent underestimation of the multiplicity
of light particles, such as pions.
For these reasons, it does not make sense to attempt a precision fit 
to experimental data, especially for pions and protons. 
Rather, we have aimed for a sensible description of the 
overall centrality dependence of multiplicity and $\langle p_T\rangle$ of kaons.

Note that in particular for the CGC model one could achieve a better fit to
the data on mean $\langle p_T\rangle$ by increasing the freeze-out temperature
by $\sim 10$ MeV. This would also lead to a decrease in elliptic flow
for this model. However, to facilitate comparison between the CGC and 
Glauber initial conditions, we have kept $T_f$ the same for both models.

\begin{table}
\caption[Parameters used for the viscous hydrodynamics simulations of RHIC collisions]{Summary of parameters used for the viscous hydrodynamics
simulations}%, including grid spacing $a$.
\begin{center}
\begin{tabular}{|c|c|c|c|c|c|}
\hline
Initial condition & 
$\eta/s$ & 
$T_i$ [GeV] & 
$T_f$ [GeV] &
$\tau_0$ [fm/c] &  
a [GeV$^{-1}$]\\
\hline
Glauber& $10^{-4}$ & 0.340 & 0.14&1&2\\
Glauber& $0.08$ & 0.333 & 0.14&1&2\\
Glauber& $0.16$ & 0.327 & 0.14&1&2\\
CGC& $10^{-4}$ & 0.310 & 0.14&1&2\\
CGC& $0.08$ & 0.304 & 0.14&1&2\\
CGC& $0.16$ & 0.299 & 0.14&1&2\\
CGC& $0.24$ & 0.293 & 0.14&1&2\\
%
%
%Initial condition & 
%$\eta/s$ & 
%$T_i$ [GeV] & 
%$T_f$ [GeV] &
%$\tau_0$ [fm/c] &  
%a [GeV$^{-1}$]\\
%\hline
%Glauber& $10^{-4}$ & 0.36 & 0.15&1&1\\
%Glauber& $0.08$ & 0.353 & 0.15&1&1\\
%Glauber& $0.16$ & 0.346 & 0.15&1&2\\
%CGC& $10^{-4}$ & 0.34 & 0.15&1&2\\
%CGC& $0.08$ & 0.335 & 0.15&1&2\\
%CGC& $0.16$ & 0.33 & 0.15&1&2\\
%CGC& $0.24$ & 0.325 & 0.15&1&2\\
\hline
\end{tabular}
\end{center}
%\caption[Parameters used for the viscous hydrodynamics simulations of RHIC collisions]{Summary of parameters used for the viscous hydrodynamics
%simulations%, including grid spacing $a$.
%}
\label{tab:par}
\end{table}

\subsection{Elliptic flow}

\begin{figure}
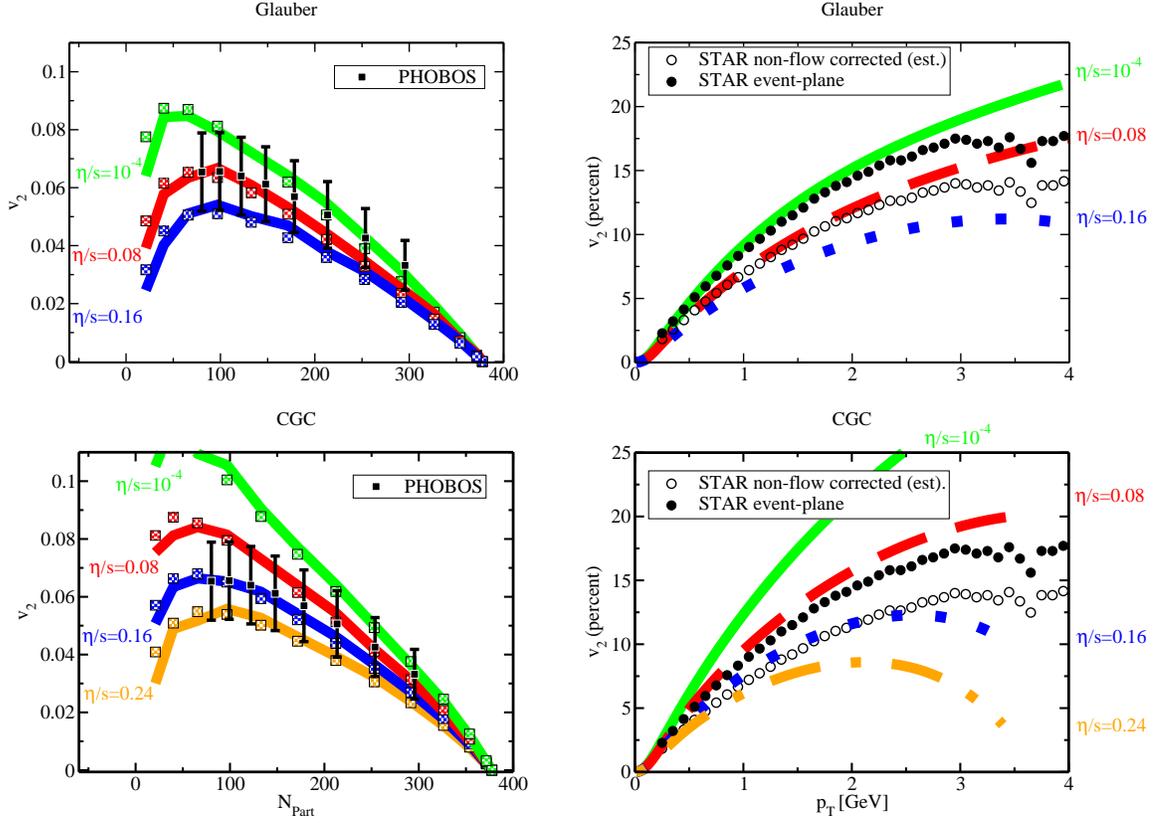

\includegraphics[width=.445\linewidth]{newdev/v2intGlauber.eps}
\hfill
\includegraphics[width=.49\linewidth]{newdev/v2mbGlauber.eps}
\hfill\\
\vspace*{1cm}
\includegraphics[width=.445\linewidth]{newdev/v2intCGC.eps}
\hfill
\includegraphics[width=.49\linewidth]{newdev/v2mbCGC.eps}
\hfill

\caption[Comparison of hydrodynamic models to
experimental data on charged hadron integrated 
and minimum bias elliptic flow]{Comparison of hydrodynamic models to
experimental data on charged hadron integrated (left)
and minimum bias (right) elliptic flow by PHOBOS \cite{Alver:2007qw} 
and STAR \cite{Abelev:2008ed}, respectively.
STAR event plane data has been reduced by 20 percent 
to estimate the removal of non-flow contributions \cite{Abelev:2008ed,Poskanzer}.
The line thickness for the hydrodynamic model curves is 
an estimate of the accumulated numerical error (due to, e.g., finite
grid spacing). The integrated $v_2$ coefficient 
from the hydrodynamic models
(full lines) is well reproduced by $\frac{1}{2} e_p$ (dots); indeed,
the difference between the full lines and dots gives an
estimate of the systematic uncertainty of the freeze-out prescription.}
\label{fig:v2int}
\end{figure}

Having fixed the parameters $\tau_0,T_i,T_f$ for a given $\eta/s$
to provide a reasonable description of the experimental data,
a sensible comparison between the model and experimental results
for the elliptic flow coefficient can be attempted.
For charged hadrons, the integrated and minimum-bias $v_2$ coefficients 
are shown in \autoref{fig:v2int} 
for Glauber and CGC initial conditions.
As noted in \autoref{sec:aniso}, charged hadron
$v_2^{\rm int}$ turns out to be very well reproduced by
the momentum eccentricity $\frac{1}{2}\ e_p$, evaluated when
the last fluid cell has cooled below $T_f$. This agreement
is independent from impact parameter or viscosity
and hence may serve as a more direct method on obtaining 
an estimate for $v_2^{\rm int}$
if one cannot (or does not want to) make use of the 
Cooper-Frye freeze-out procedure described in
\autoref{sec:fo}.

The comparison of the hydrodynamic model 
to experimental data with $90$\% confidence level
systematic error bars from PHOBOS \cite{Alver:2007qw} 
for the integrated elliptic flow in \autoref{fig:v2int}
suggests a maximum value of $\eta/s\sim0.16$ for Glauber-type
and $\eta/s\sim0.24$
for CGC-type initial conditions.
Whereas for Glauber initial conditions, ideal hydrodynamics 
($\eta/s\sim0$) gives results consistent with PHOBOS data,
for CGC initial conditions zero viscosity does not give a good
fit to the data, which is consistent with previous findings 
\cite{Hirano:2005xf}.

For minimum-bias $v_2$, to date only experimental data 
using the event-plane method are available, %. \textbf{cite someone},
where the statistical, but not the systematic error 
of that measurement is directly accessible.
The dominant source of systematic error is associated
with the presence of so-called non-flow effects \cite{Ollitrault:1995dy}.
Recent results from STAR suggest that removal of 
these non-flow effects imply a reduction of the event-plane minimum
bias $v_2$ by $20$ percent \cite{Abelev:2008ed,Poskanzer}.
For charged hadrons, a comparison of both the event-plane and the 
estimated non-flow corrected experimental data from STAR
with the hydrodynamic model is shown in \autoref{fig:v2int}.

For Glauber-type initial conditions, the data on
minimum-bias $v_2$ for charged hadrons is consistent with 
the hydrodynamic model for viscosities in the range
$\eta/s\in[0,0.1]$, while for the CGC case the respective
range is $\eta/s\in[0.08,0.2]$. It is interesting
to note that for Glauber-type initial conditions,
experimental data for both the integrated as well as
the minimum-bias elliptic flow coefficient (corrected for non-flow effects)
seem to be reproduced best\footnote{
In Ref.~\cite{Romatschke:2007mq} a lower value of $\eta/s$
for the Glauber model was reported. 
The results for viscous hydrodynamics shown in \autoref{fig:v2int} 
are identical
to Ref.~\cite{Romatschke:2007mq}, but the new STAR data with non-flow corrections
became available only after \cite{Romatschke:2007mq} had been
published.} by a hydrodynamic model with
$\eta/s=0.08\simeq\frac{1}{4\pi}$. This number has
first appeared in the gauge/string duality context 
\cite{Policastro:2001yc} and has
been conjectured to be the universal lower bound
on $\eta/s$ for any quantum field theory at finite temperature
and zero chemical potential \cite{Kovtun:2004de}. 
For CGC-type initial conditions, 
the charged hadron
$v_2$ data seems to favor a hydrodynamic model with
$\eta/s\sim 0.16$, well above this bound.

\subsection{Early vs. late thermalization}
\label{notherm}

\begin{figure}
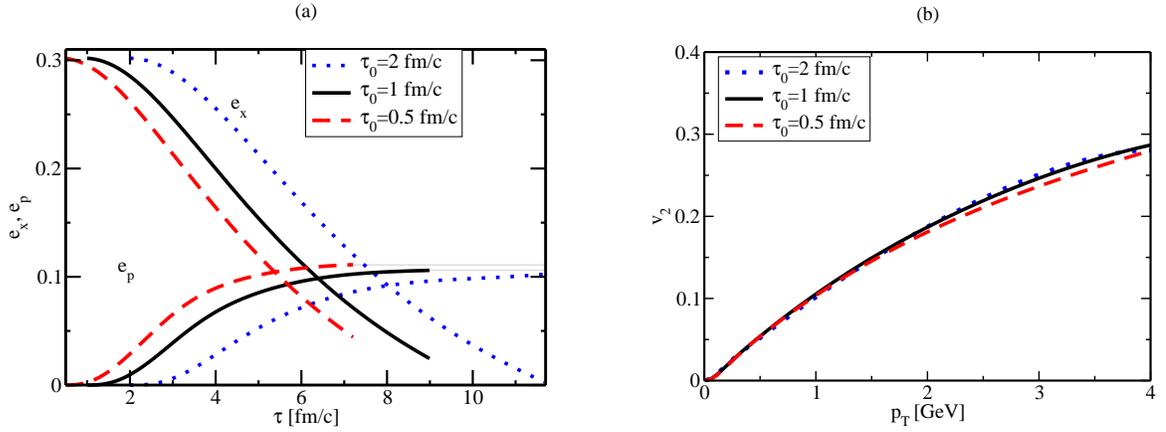

\includegraphics[width=.47\linewidth]{newdev/tdep2.eps}
\hfill
\includegraphics[width=.44\linewidth]{newdev/tdep3.eps}
\caption[Momentum anisotropy and elliptic flow for charged hadrons]{Momentum anisotropy (a) and elliptic flow for charged hadrons (b)
for $b=7$ fm, $\eta/s=0.08$ and different hydrodynamic initialization times $\tau_0$.
Horizontal light gray lines in (a) are visual aids to compare the final value of 
$e_p$. As can be seen from these plots, neither the final $e_p$ nor 
the charged hadron $v_2$ depend sensitively on the value of $\tau_0$
if the same energy distribution is used as initial condition at the respective
initialization times. Simulation parameters were 
$T_i=0.29\ {\rm GeV},\, T_f=0.14\ {\rm GeV}$ for $\tau_0=2$ fm/c,
$T_i=0.36\ {\rm GeV},\, T_f=0.15\ {\rm GeV}$ for $\tau_0=1$ fm/c,
and $T_i=0.43\ {\rm GeV},\, T_f=0.16\ {\rm GeV}$ for $\tau_0=0.5$ fm/c.
}
\label{fig:tdep}
\end{figure}

Currently, there seems to be a common misunderstanding in the
heavy-ion community that hydrodynamic models 
can universally only reproduce experimental data if they are initialized at 
early times $\tau_0<1$ fm/c. This notion has been labeled 
``early thermalization'' and continues to create a lot of confusion.
In this section, we argue that the matching 
of hydrodynamics to data itself does not require $\tau_0<1$ fm/c.
It is the additional assumptions about pre-equilibrium dynamics 
that lead to this conclusion for the Glauber initial conditions.

Performing hydrodynamic simulations in the way we have described earlier,
the energy density distribution is specified by either the Glauber or CGC
model at an initial time $\tau_0$. In \autoref{fig:tdep} we show the result for the 
elliptic flow coefficient (or the momentum anisotropy) for three different
values of $\tau_0$, namely $0.5,1$ and $2$ fm/c, where also $T_i$ and $T_f$
have been changed in order to obtain roughly the same multiplicity and
mean $p_T$ for each $\tau_0$. As can be seen from this figure,
the resulting final elliptic flow coefficient is essentially independent
of the choice of $\tau_0$. In particular, this implies that experimental
data for bulk quantities 
can be reproduced by hydrodynamic models also for large initialization times,
so no early thermalization assumption is needed.

\begin{figure}
\begin{center}
\includegraphics[width=.5\linewidth]{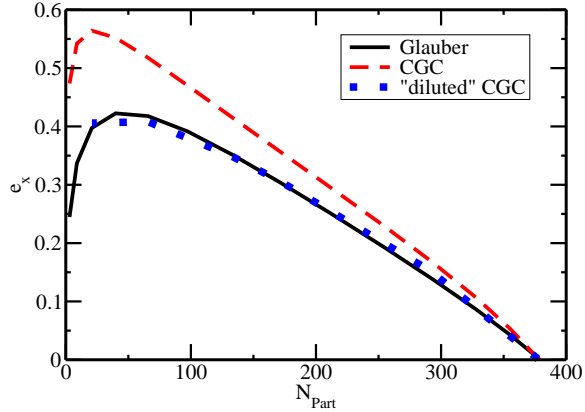}
\end{center}
\caption[Spatial eccentricity for the Glauber and CGC model compared
to evolving the CGC model according to \autoref{dilution} 
for $\tau=1.5$ fm/c]{Spatial eccentricity for the Glauber and CGC model compared
to evolving the CGC model according to \autoref{dilution} 
for $\tau=1.5$ fm/c. This implies that starting with Glauber-type initial
conditions at $\tau_0>1$ fm/c may not be unreasonable.}
\label{fig:dilu}
\end{figure}

However, it is true that the above procedure assumes that the energy
density distribution remains unchanged up to the starting time
of hydrodynamics, which arguably becomes increasingly inaccurate for larger $\tau_0$.
It has therefore been suggested \cite{Kolb:2000sd}
to mimic the pre-hydro time evolution of the
energy density distribution by assuming free-streaming of partons.
Assuming free-streaming gives the maximal contrast to assuming hydrodynamic evolution,
since the latter corresponds to very strong interactions while the former corresponds
to no parton interactions at all. Indeed, one can calculate the 
effect of the free-streaming evolution on the spatial anisotropy, 
finding \cite{Kolb:2000sd}
\beq
e_x(\tau)=\frac{e_x(0)}{1+\frac{\tau^2}{3 \langle R^2\rangle}},\quad
\langle R^2\rangle =\frac{\int d^2 {\bf x} \epsilon(\tau=0)}{\int d^2 {\bf x} \frac{(x^2+y^2)}{2}\epsilon(\tau=0)}.
\label{dilution}
\eeq
This implies that the spatial anisotropy decreases with time, whereas one
can show that free-streaming does not lead to a build-up of $e_p$. In other words,
the eccentricity gets diluted without producing elliptic flow, such that 
once hydrodynamic evolution starts, it will not lead to as much $v_2$
as it would have without the dilution effect\footnote{
It seems that if one forces the energy-momentum tensor at the end of
free-streaming period to match to that of \emph{ideal} hydrodynamics
(instantaneous thermalization), the resulting fluid velocities
are anisotropic, i.e. correspond to a non-vanishing elliptic flow coefficient
\cite{Pasi,Broniowski:2008vp}. It is possible that this effect stems
from neglecting velocity gradients (viscous hydrodynamic corrections) 
in the matching process. We ignore the complications of the detailed
matching from free-streaming to hydrodynamics in the following.}. 
It is tempting to
conclude from this that by comparing to experimental data on elliptic flow 
one could place an upper bound on the maximally allowed dilution time,
and interpret this as the thermalization time of the system.
One should be aware, however, that this bound will depend
on the assumption made about the pre-hydro evolution. Furthermore, 
one should take into account the fact that the initial
state of the system remains unknown. For instance, the system could
start with an energy density distribution similar to the CGC model,
which has a fairly large eccentricity. \autoref{fig:dilu} shows that when allowing
the eccentricity to get diluted according to \autoref{dilution},
it takes a time of $\tau\sim 1.5$ fm/c until the eccentricity
has shrunk to that of the Glauber model. This implies that 
even when assuming no particle interactions (no elliptic flow build-up)
for the first stage of the system evolution, one can get
eccentricities which are Glauber-like after waiting for 
a significant fraction of the system life time. Allowing at least
some particle interactions (which is probably more realistic),
one expects some build-up of elliptic flow already in the dilution (or pre-equilibrium)
phase, and therefore dilution (or ``thermalization'') times of $\tau\sim 2$ fm/c seem not 
to be incompatible with the observed final elliptic flow even for non-vanishing
viscosity.

\section{Summary and Conclusions}
\label{sec:four}

We applied conformal relativistic viscous hydrodynamics
to simulate Au+Au collisions at RHIC at energies of $\sqrt{s}=200$ GeV per
nucleon pair. Besides one first-order transport
coefficient (the shear viscosity) in general there are five
second-order transport coefficients in this theory, for which
one would have to supply values. We provided arguments that 
physical observables in the parameter range accessible to hydrodynamics
(low momenta, central to semi-central collisions) do not seem to be 
strongly dependent on specific (reasonable) choices for any of these
second-order coefficients. On the other hand, we do find a pronounced
dependence of the elliptic flow coefficient on the ratio of shear viscosity over
entropy density, which suggests that by combining 
viscous hydrodynamics and experimental data a measurement of the
quark-gluon plasma viscosity may not be futile.
However, we have shown that our ignorance about the precise distribution
of energy density at the earliest stages of a heavy-ion collision
introduces a large systematic uncertainty in the final
elliptic flow of the hydrodynamic model. Adding to this is
the considerable experimental uncertainty pertaining to the removing
of non-flow contributions to the elliptic flow.
For these reasons, we are unable to make precise statements
about the value of the shear viscosity of the quark-gluon plasma
and in particular cannot place a firm lower bound on $\eta/s$.
Indeed, our hydrodynamic models seem to be able to 
consistently describe experimental data for multiplicity,
radial flow and elliptic flow of bulk charged hadrons for a wide
range of viscosity over entropy ratios,
\beq
\frac{\eta}{s}=0.1\pm0.1({\rm theory})\pm0.08({\rm experiment}),
\label{eq:result}
\eeq
where we estimated the systematic uncertainties for both
theory and experiment from the results shown in \autoref{fig:v2int}. 
We stress that \autoref{eq:result} does not account for
physics not included in our model, such as finite chemical potential,
bulk viscosity, heat flow, hadron cascades, three-dimensional
fluid dynamic effects and possibly many more. Consistent inclusion
of all these may result in changes of the central value and
theory uncertainty in \autoref{eq:result}. Nevertheless,
none of the mentioned refinements 
is currently expected to dramatically increase the elliptic
flow coefficient (though some increase may be expected
when e.g. implementing partial chemical equilibrium \cite{Hirano:2005wx}). 
Therefore, we seem to be able to 
exclude viscosities of $\eta/s\gtrsim 0.5$ 
with high confidence, which indicates that the quark-gluon
plasma displays less friction than any other known laboratory 
fluid \cite{Kovtun:2004de,Schafer:2007pr}. 
Other groups have come to similar conclusions 
\cite{Gavin:2006xd,Adare:2006nq,Drescher:2007cd}.

To better quantify the shear viscosity of the quark-gluon plasma
at RHIC calls for more work, both in theory and experiment.
On the theory side, a promising route seems to be the study
of fluctuations and comparing to existing experimental data 
\cite{Adams:2005aw,Gavin:2006xd,Sorensen:2006nw,Alver:2007qw,Vogel:2007yq,
Drescher:2007ax,Trainor:2007ny,Hama:2007dq}. For instance,
it might be interesting to investigate the critical value of $\eta/s$
for the onset of turbulence in heavy-ion collisions and explore possible consequences
of fully developed turbulence \cite{Romatschke:2007eb}.
However, maybe most importantly, 
a more thorough understanding of the earliest stages of a heavy-ion 
collision, in particular thermalization, 
could fix the initial conditions for hydrodynamics and hence
dramatically reduce the theoretical uncertainty in final observables.

Leaving these ideas for future work, we stress that with the advent
of conformal relativistic viscous hydrodynamics at least
the uncertainties of the hydrodynamic evolution itself
now seem to be under control. We hope that this serves
as another step towards a better understanding of the dynamics
of relativistic heavy-ion collisions.
%\section{LHC}
%\input{LHC}
%
% ========== Chapter 4 - Viscous Hydrodynamic Predictions for LHC
\chapter{Viscous Hydrodynamic Predictions for the LHC}
\label{LHC}
\section{Introduction}
%Much work has been done recently using viscous hydrodynamics to study 
%the properties of gold-gold collisions at the Relativistic Heavy Ion Collider 
%(RHIC)~ \cite{Luzum:2008cw,Dusling:2007gi,Song:2008si,Huovinen:2008te}.  A 
%measurement of particular interest is the elliptic flow coefficient $v_2$, 
%the second moment in the azimuthal angle of the distribution of emitted 
%particles (cf.~\cite{Kolb:1999it}), which allows to extract information about
%material constants (such as viscosity) of the high density nuclear matter created at RHIC.
Using the knowledge gained from viscous hydrodynamic simulations for RHIC, it should be possible to
predict experimental results at the Large Hadron Collider (LHC),
which will collide lead ions at a maximum center of 
mass energy of $\sqrt{s}=5.5$ TeV per nucleon pair compared to $\sqrt{s}=200$ GeV gold ions at RHIC.
%Depending on whether the experimental data on e.g. $v_2$ from LHC will be close 
%or far away from the hydrodynamic model prediction, it should be possible to
%either confirm that real progress has been made in understanding nuclear 
%matter at extreme energy densities, or conclude that the successful hydrodynamic
%description of experimental data from RHIC might have been a coincidence.
%
If experimental data on, e.g., $v_2$ from LHC is close to 
the hydrodynamic model prediction, this would confirm that 
real progress has been made in understanding nuclear matter at extreme energy densities;
if far away, it may indicate that the successful hydrodynamic
description of experimental data from RHIC was a coincidence.

Regardless of the outcome, the advent of the RHIC experiments clearly has
lead to major progress in the theory and application of hydrodynamics to heavy-ion collisions.
A few years ago the form of the hydrodynamic equations in the presence 
of shear viscosity $\eta$ was still unresolved, with different groups 
keeping some terms while neglecting others 
\cite{Muronga:2003ta,Heinz:2005bw,Baier:2006um,Koide:2006ef}.
For the case of approximately conformal theories, where the viscosity coefficient 
for bulk---but not shear---becomes negligible, all possible terms to 
second order in gradients were derived in Ref.~\cite{Baier:2007ix} (see \autoref{hydro}), and their relative
importance investigated in Ref.~\cite{Luzum:2008cw} (\autoref{RHIChydro}). Three of the groups
performing viscous hydrodynamic simulations now agree on these terms 
\cite{Luzum:2008cw,Song:2008si,Huovinen:2008te}, 
while another group \cite{Dusling:2007gi} uses a different formalism that 
gives matching results. While this development still leaves out the consistent
treatment of bulk viscosity, the quantitative suppression of elliptic
flow by shear viscosity is therefore essentially understood. 
As shown in \autoref{RHIChydro}, from comparison of viscous hydrodynamic simulations to experimental 
data \cite{Alver:2007qw,Abelev:2008ed}, one can
infer an upper limit of the ratio of shear viscosity over entropy density,
$\eta/s<0.5$, for the matter produced in Au+Au collisions at 
$\sqrt{s}=200$ GeV \cite{Luzum:2008cw}, which is in agreement to extractions
by other methods \cite{Gavin:2006xd,Adare:2006nq,Drescher:2007cd}.
%sevral extractions from data 
%based on other methods \cite{Gavin:2006xd,Adare:2006nq,Drescher:2007cd}.
A sizeable uncertainty for this limit comes from the 
fact that the initial conditions for the hydrodynamic evolution are 
poorly known, with the two main models, the Glauber and Color-Glass-Condensate (CGC)
models, giving different results for the elliptic flow coefficient
\cite{Luzum:2008cw}. This difference can be understood to originate from
the different initial spatial eccentricity $e_x$ in the Glauber/CGC models, which we recall is defined as
%The eccentricity is defined as
%\begin{equation}
%e_x\equiv 
%%\frac{\int dx dy\ (y^2-x^2)\ \epsilon(x,y)}
%%{\int dx dy\ (y^2+x^2)\ \epsilon(x,y)}
%\frac{\left<y^2-x^2\right>}
%{\left<y^2+x^2\right>}\, ,
%\label{exdef}
%\end{equation}
%where the symbols $\left<\right>$ denote
%averaging over the initial energy density in the transverse plane, $\epsilon(x,y)$.

%
%Recall from \autoref{RHIChydro} that an important aspect of the hydrodynamic initial conditions is the initial spacial eccentricity
%
\begin{equation}
\label{exdef}
e_x\equiv 
%\frac{\int dx dy\ (y^2-x^2)\ \epsilon(x,y)}
%{\int dx dy\ (y^2+x^2)\ \epsilon(x,y)}
\frac{\left\langle y^2-x^2\right\rangle_\epsilon}
{\left\langle y^2+x^2\right\rangle_\epsilon}\, ,
\end{equation}
where the symbols $\left<\right>_\epsilon$ denote
averaging over the initial energy density in the transverse plane, $\epsilon(x,y)$.

Indeed, it had been suggested  \cite{Alt:2003ab} that the elliptic flow coefficient $v_2$
at the end of the hydrodynamic evolution would be strictly proportional to the initial 
spatial eccentricity, $v_2/e_x\propto {\rm const.}$, if the fluid was evolving
without any viscous stresses for an infinitely long time. 
This is to be contrasted with experimental data indicating
% that seems to indicate
a proportionality factor of total multiplicity over
overlap area %$\frac{v_2}{e_x}\propto\frac{1}{S_{\rm overlap}}\frac{dN}{dY}$ 
$v_2/e_x\propto dN/dY/S_{\rm overlap}$
\cite{Alt:2003ab}.
%when using the Glauber model for $e_x$.
Total multiplicity $\frac{dN}{dY}$ here refers to the total number of 
observed particles $N$ per unit rapidity $Y$, 
while the overlap area is calculated as
\begin{equation}
S_{\rm overlap}=\pi \sqrt{\left<x^2 \right>\left<y^2\right>}\, .
\label{Sodef}
\end{equation}

%Ideal fluid dynamics is not an adequate description of the latest stage following
%a heavy-ion collision, the hadronic gas, because of the sizeable viscosity 
%coefficient in this stage \cite{Prakash:1993bt}.
Ideal fluid dynamics does not adequately describe the latest stage of 
a heavy-ion collision (the hadron gas), because of the large viscosity 
coefficient in this stage \cite{Prakash:1993bt}.
Therefore, the hydrodynamic stage lasts only for a finite time
(e.g., until all fluid cells have cooled below the decoupling temperature),
resulting in a dependence of $v_2/e_x$ on $dN/dY$. Also, viscous effects affect
the proportionality between $v_2$ and $e_x$,
leading to a behavior that is qualitatively similar to that observed in the data
\cite{Song:2008si}.

One of the objectives of this work is to extend the energy range for 
fluid dynamic results of $v_2/e_x$ from Au+Au collisions at
top RHIC to Pb+Pb collisions at top LHC energies, as well
as to study the dependence on shear viscosity. If in the future
either $e_x$ or the mean $\eta/s$ becomes known, these results can thus be used
to constrain the respective other quantity from experimental data. 
On the other hand, the values of shear viscosity 
for which the Glauber/CGC models match to experimental data at top RHIC energies 
have been extracted in Ref.~\cite{Luzum:2008cw, Luzum:2008cwErr} for Au+Au 
collisions.
Since $\eta/s$ averaged over the system evolution 
is not expected to be dramatically different for Pb+Pb collisions at the LHC,
another objective of this work is to obtain a prediction for the elliptic flow
coefficient for the LHC based on the best-fit values to RHIC.

Finally, the feasibility of detecting elliptic flow in 
p+p collisions at $\sqrt{s}=14$ TeV at the LHC is being discussed~\cite{Lpriv}.
As a reference for other approaches and experiment,
it interesting to study the possible size and viscosity dependence of $v_2$ 
under the hypothetical assumption that the bulk evolution following p+p collisions could be 
captured by fluid dynamics.

\section{Setup}
To make predictions for nuclear collisions at LHC energies, we use our hydrodynamic
model that successfully described experimental data at RHIC \cite{Luzum:2008cw,Luzum:2008cwErr}
and make modifications 
to the input parameters appropriate for the higher collision energies at the LHC.

As a reminder, the hydrodynamic model \cite{Luzum:2008cw} is based on the 
conservation of the energy momentum tensor
\cite{Baier:2007ix}
\begin{align}
T^{\mu \nu} =\ &\epsilon u^\mu u^\nu- p \Delta^{\mu \nu}+\Pi^{\mu \nu}\, ,\nonumber\\
\Pi^{\mu\nu}  =\ & \eta \nabla^{\langle \mu} u^{\nu\rangle}
- \tau_\pi \left[ \Delta^\mu_\alpha \Delta^\nu_\beta D\Pi^{\alpha\beta} 
 + \frac 4{3} \Pi^{\mu\nu}
    (\nabla_\alpha u^\alpha) \right] \nonumber\\
%  &&\quad 
%  + \frac{\kappa}{2}\left[R^{<\mu\nu>}+2 u_\alpha R^{\alpha<\mu\nu>\beta} 
%      u_\beta\right]\nonumber\\
  & -\frac{\lambda_1}{2\eta^2} {\Pi^{\langle\mu}}_\lambda \Pi^{\nu\rangle\lambda}
  +\frac{\lambda_2}{2\eta} {\Pi^{\langle\mu}}_\lambda \omega^{\nu\rangle\lambda}
  - \frac{\lambda_3}{2} {\omega^{\langle\mu}}_\lambda \omega^{\nu\rangle\lambda}\, ,
\nonumber
\end{align}
where $\epsilon,p$ and $u^\mu$ are the energy density, pressure,
and fluid 4-velocity, respectively. $D\equiv u^\mu D_\mu$ and 
$\nabla_\alpha\equiv \Delta_\alpha^\mu D_\mu$
are time-like and space-like projections of the covariant derivative $D_\mu$,
where $\Delta^{\mu \nu}=g^{\mu \nu}-u^\mu u^\nu$
and we remind the compact notations\\ 
$A_{\langle \mu} B_{\nu\rangle} \equiv \left(\Delta^\alpha_\mu \Delta^\beta_\nu + 
\Delta^\alpha_\nu \Delta^\beta_\mu-\frac{2}{3} \Delta^{\alpha \beta} 
\Delta_{\mu \nu}\right) A_\alpha B_\beta$ and 
%$\omega_{\mu \nu}\equiv - \nabla_{[\mu}u_{\nu]}$.
$\omega_{\mu \nu}\equiv \frac 1 2 \left( \nabla_\nu u_\mu - \nabla_\mu u_\nu \right)$.
For relativistic nuclear
collisions it is convenient to follow Bjorken \cite{Bjorken:1982qr} 
and use Milne coordinates proper time 
$\tau=\sqrt{t^2-z^2}$ and spacetime rapidity $\xi={\rm atanh}\frac{z}{t}$,
in which the metric becomes $g_{\mu \nu}={\rm diag}(1,-1,-1,-\tau^2)$,
and assume that close to %central rapidities, 
$\xi=0$, the hydrodynamic
degrees of freedom are approximately boost-invariant ($\xi\simeq Y$).

The hydrodynamic equations $D_\mu T^{\mu \nu}=0$ 
then constitute an initial value problem in proper time 
and transverse space, and are solved numerically 
%using the method outlined in \cite{Luzum:2008cw}. 
(see Ref.~\cite{Luzum:2008cw}). The input parameters for 
hydrodynamic evolution are the equation of state $p=p(\epsilon)$ 
%which fixes the speed of sound squared $c_s^2=\frac{d p}{d\epsilon}$
and the first (second) order hydrodynamic transport coefficients $\eta$
($\tau_\pi,\lambda_1,\lambda_2,\lambda_3$).
The values for $\lambda_{1,2,3}$ have been found 
to hardly affect the boost-invariant hydrodynamic evolution for Au+Au collisions
at RHIC \cite{Luzum:2008cw}, so here they are generally set to zero.

\begin{table}
\caption[Parameters used for the viscous hydrodynamics
simulations of RHIC heavy ion collisions and LHC heavy ion and proton-proton collisions]{\label{tab:parLHC} Central collision parameters used for the viscous hydrodynamics
simulations ($T_f=0.14$ GeV for all).
}
 \begin{center}
\begin{tabular}{|cccccc|}
\hline
Beam &
Initial cond. & 
$\frac{dN_{\rm ch}}{dY}$ & 
$T_i$ [GeV] & 
$\sqrt{s}$ [GeV] &
$\tau_0$ [fm/c] \\
\hline
Gold& Glauber& 800 & 0.34 & 200&1\\
Gold& CGC& 800 & 0.31 & 200&1\\
Lead& Glauber& 1800& 0.42& 5500&1\\
Lead& CGC& 1800& 0.39& 5500&1\\
Protons& Glauber& 6& 0.400& 14000& 0.5\\
Protons& Glauber& 6& 0.305& 14000& 1\\
Protons& Glauber& 6& 0.270& 14000& 2\\
\hline
\end{tabular}
 \end{center}
%\caption[Parameters used for the viscous hydrodynamics
%simulations of RHIC heavy ion collisions and LHC heavy ion and proton-proton collisions]{Central collision parameters used for the viscous hydrodynamics
%simulations ($T_f=0.14$ GeV for all).
%}
%\label{tab:parLHC}
\end{table}

The mechanisms leading to thermalization (the onset of hydrodynamic behavior) 
are not well understood in nuclear collisions.  Therefore, %it is also not understood
it is not known 
how the thermalization time $\tau_0$ at which hydrodynamic evolution 
is started depends on the collision energy.  Barring further insight, 
%the model for hydrodynamic evolution at LHC is started at the same
we start hydrodynamic evolution for the LHC at the same
time as for the RHIC simulations ($\tau_0=1$ fm/c). At this time,
%
%At the initial time $\tau_0$, where 
%the hydrodynamic evolution is started, 
the initial conditions for the transverse
energy density $\epsilon(x,y)$ are given by the Glauber or CGC model, respectively,
the fluid velocities are assumed to vanish, and the shear tensor $\Pi^{\mu \nu}$
is set to zero (other values for $\Pi^{\mu \nu}$ do not seem to
affect the final results \cite{Song:2007ux,Luzum:2008cw}). 
For brevity, we refer to %Ref.~\cite{Luzum:2008cw}
\autoref{RHIChydro} for the details of
the Glauber and CGC models, but  note
that we use the Woods-Saxon parameters of radius $R_0=6.4\, (6.6)$ fm and skin 
depth $\chi=0.54\,(0.55)$ fm  for gold (lead), 
and assume a nucleon-nucleon cross section of $\sigma= 40\,(60)$ mb 
for $\sqrt{s}=200\,(5500)$ GeV collisions.

The overall normalization
of the initial energy density (parametrized by the initial temperature at the center $T_i$) 
was adjusted to match the experimentally observed
multiplicity at RHIC; by analogy, for LHC the normalization
is adjusted to match the predicted multiplicity 
\cite{Kharzeev:2004if,Armesto:2008fj,Abreu:2007kv,Busza:2007ke}. 
Since we lack detailed knowledge about its temperature dependence, 
the ratio of shear viscosity to entropy density $\eta/s$ is set to be constant
during the hydrodynamic evolution
(equal to the average over the spacetime evolution of the system).
The relaxation time coefficient $\tau_\pi$ is expected \cite{Baier:2007ix,York:2008rr}
to lie in the range
$\frac{\tau_\pi}{\eta} (\epsilon+p)\simeq2.6-6$.
The equation of state (EoS) can in principle be provided by lattice QCD.
While at present there are points of disagreement between lattice groups about, e.g.,  
the precise location of the QCD phase transition, 
there is consensus that it is an analytic crossover 
\cite{Aoki:2009sc, Bazavov:2009zn}.  Therefore, we use a
lattice-inspired EoS \cite{Laine:2006cp} 
that is consistent with both the current 
consensus and perturbative QCD; also, since it resembles \cite{Bazavov:2009zn},
we expect that using a different lattice EoS will have a 
minor effect on our results.  In fact, as a preview of work in progress, some of the calculations were also done with an equation of state taken from Ref.~\cite{Bazavov:2009zn} (\autoref{EoSplots} at the end of the chapter) and the results compared (see \autoref{plot2}).

%\begin{figure}[t]
%\includegraphics[width = .5\linewidth]{EoSe.eps}
%\includegraphics[width = .5\linewidth]{EoSp.eps}\\
%\begin{center}
%\includegraphics[width = .5\linewidth]{EoSc2.eps}
%\end{center}
%\caption[Equation of state from Laine and Schr\"oder, and from Bazavov \textsl{et al.}]{\label{EoSplots} Equation of state from Laine and Schr\"oder, and from Bazavov.  Above are the energy density and pressure, scaled by $T^4$ as a function of temperature.  Below is the speed of sound squared.}
%\end{figure}

Once a given fluid cell has cooled down to the decoupling temperature
$T_f$, its energy and momentum are converted into particle
degrees of freedom using the Cooper-Frye freeze-out prescription \cite{Cooper:1974mv}.
A value of $T_f=0.14$ GeV was determined by matching to RHIC data %for $\tau_0=1$ fm/c, 
and will also be used for LHC energies,
assuming that it is mostly determined by local conditions, and
less so by initial energy density, system size or collision energy.
The distribution of the particle degrees of freedom may be further
evolved using a hadronic cascade code (as in Ref.~\cite{Bass:2000ib}), or
in a more simple approach the unstable particle resonances
are allowed to decay, without further evolving the stable particle
distributions. In both cases, the total multiplicity and particle
correlations (such as the elliptic flow coefficient) are then calculated
from the stable particle distribution (cf.~\cite{Luzum:2008cw}). 
Surprisingly, it was found in Ref.~\cite{Kolb:1999it,Luzum:2008cw} (see \autoref{RHIChydro}) that 
the momentum integrated elliptic flow coefficient for charged hadrons---to good approximation---is equal to half the momentum anisotropy,
\begin{equation}
v_2\simeq \frac{1}{2} e_p=\frac{1}{2}\frac{\int dx dy\, T^{xx}-T^{yy}}{\int dx dy\, T^{xx}+T^{yy}}\, .
\label{v2def}
\end{equation}
Since
the momentum anisotropy is a property of the fluid, it is independent on
the details of the freeze-out procedure and only mildly dependent on the choices of
$\tau_0,T_f$.
Unlike at RHIC where pairs of $\tau_0$ and $T_f$ could be fine-tuned to fit the particle spectra
at central collisions, no such extra information is available for the LHC.
Hence \autoref{v2def} may provide the most reliable way of determining
the elliptic flow of charged hadrons, and will be used in the following.
Similarly, one can use the total entropy per unit spacetime rapidity $\frac{dS}{d\xi}$
in the fluid as a proxy for the total (charged hadron) multiplicity per unit rapidity 
$\frac{dN}{dY}$
($\frac{dN_{\rm ch}}{dY}$)
with a proportionality factor \cite{Gyulassy:1983ub,Kolb:2001qz}
\begin{equation}
\frac{d S}{d\xi}\sim\frac{d S}{dY}\simeq 4.87 \frac{dN}{dY}\simeq 7.85 \frac{dN_{\rm ch}}{d Y}.
\label{svsn}
\end{equation}
Note that for a gas of massive hadrons in thermal equilibrium 
at $T_f=0.14$ GeV the ratio of entropy to particle density is 
$\sim 6.41$, but the decay of unstable resonances 
produces additional entropy, resulting in \autoref{svsn}.
Since results from RHIC suggest there is only approximately $10$\% viscous entropy production
during the hydrodynamic phase \cite{Romatschke:2007jx,Song:2008si}, the entropy 
$\frac{d S}{dY}$ at $\tau=\tau_0$ can be used
to estimate the final particle multiplicity. In the case of the LHC,
the world average for the predicted charged hadron multiplicity 
for central Pb+Pb collisions at $\sqrt{s}=5.5$ TeV \cite{Armesto:2008fj},
$\frac{dN_{\rm ch}}{d Y}\simeq 1800$, can be used to estimate the 
total entropy at $\tau=\tau_0$, and hence the overall normalization $T_i$
of the initial energy density (see \autoref{tab:parLHC}). 

Using Equations \ref{v2def} and \ref{svsn} for the multiplicity and elliptic flow
allows to make predictions for the LHC without having to model
the hadronic freeze-out, which should make the results more robust.
However, as a consequence one does not get information about
the momentum dependence of the elliptic flow coefficient, prohibiting
detailed comparison with predictions by other groups 
\cite{Chaudhuri:2008je,Abreu:2007kv}.
\begin{figure}[t]
\begin{center}
\includegraphics[width=.75\linewidth]{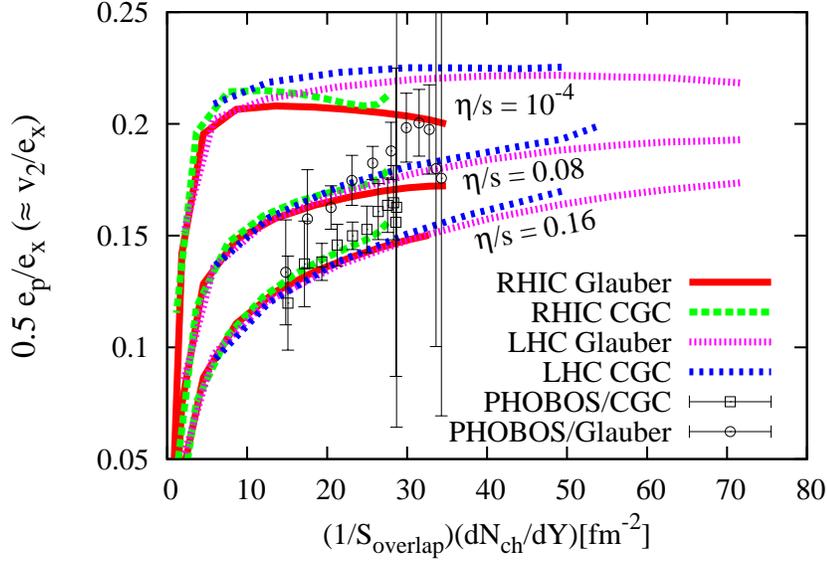}
\end{center}
\caption[Final momentum anisotropy divided by initial spatial eccentricity as a function of entropy divided by overlap area]{\label{plot1}
%Charged hadron integrated elliptic flow $v_2$ divided by initial state eccentricity $e_x$.
Anisotropy (\ref{v2def}) divided by (\ref{exdef}), as a function of 
initial entropy (\ref{svsn}) divided by (\ref{Sodef}).
Shown are results from hydrodynamic simulations for $\sqrt{s}=200$ GeV Au+Au 
(RHIC) and $\sqrt{s}=5.5$ TeV Pb+Pb collisions (LHC). 
For comparison, experimental data for $v_2$ from RHIC \cite{Alver:2006wh}, 
divided by
%\cite{Back:2004mh,Alver:2006wh}, divided by 
% $e_x$ calculated from a model \cite{Drescher:2007cd}, %(including fluctuations)
$e_x$ from two models \cite{Drescher:2007cd},
is shown as a function of measured
$\frac{dN_{\rm ch}} {dY}$ \cite{Adler:2003cb} divided by (\ref{Sodef}). See text for details.}
\end{figure}
\section{Results}
With the initial energy density distribution fixed at $\tau_0$,
the hydrodynamic model then gives predictions for the ratio
of $v_2/e_x$ at the LHC. In \autoref{plot1}, the results are shown
for three different values of shear viscosity, for two different
initial conditions and two different beams/collision 
energies (Au+Au at $\sqrt{s}=200$ GeV, Pb+Pb at $\sqrt{s}=5.5$ TeV). 
%As can be seen from this figure, the 
The resulting values for
$v_2/e_x$ seem to be quasi-universal functions of the total multiplicity scaled
by the overlap area $S_{\rm overlap}$, only depending on the value of $\eta/s$
(and, to a lesser extent, the collision energy). The deviations of the RHIC
simulations from the universal curve can be argued to arise from 
a combination of the finite lifetime of the hydrodynamic phase at $\sqrt{s}=200$ GeV 
and the presence of the QCD phase transition, and is strongest 
for ideal hydrodynamics, in agreement with earlier findings \cite{Song:2008si}.

Also shown in \autoref{plot1} is experimental data for the elliptic flow coefficient
for Au+Au collisions at RHIC, normalized
by $e_x$ from a Monte-Carlo calculation (including fluctuations) 
in Glauber and CGC models 
(see Ref.~\cite{Drescher:2007cd} for details). Since $e_x$ is not 
directly measurable, the differently normalized data gives an estimate
of the overall size of $v_2/e_x$ at RHIC. Directly matching experimental
data on $v_2$ using a hydrodynamic model with an initial $e_x$
specified by the Glauber or CGC model, a reasonable fit was achieved
for a mean value of $\eta/s\simeq0.08$ and $\eta/s\simeq0.16$, respectively 
\cite{Luzum:2008cw,Luzum:2008cwErr}. 
\begin{figure}[t]
\begin{center}
\includegraphics[width=.7\linewidth]{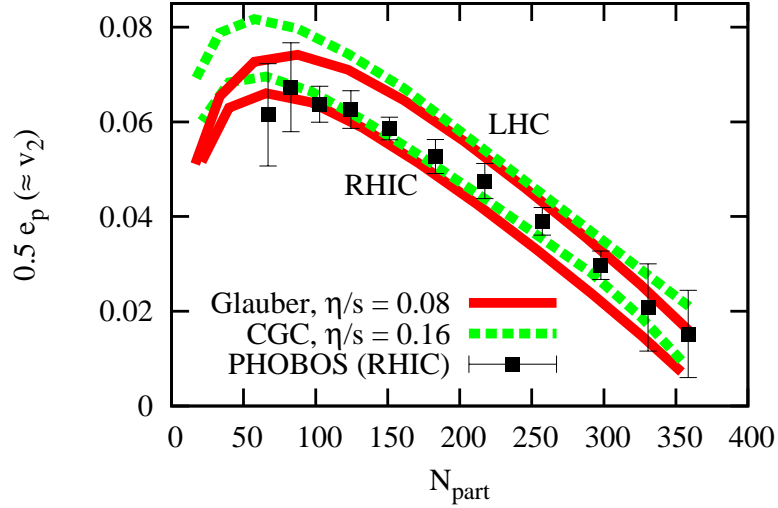}
%\hspace{-1.3cm}
%\includegraphics[width = .56\linewidth]{EoSv2-2.eps}
\end{center}
\caption[Anisotropy prediction for LHC collisions as a function of centrality]{\label{plot2}
Anisotropy (\ref{v2def}) prediction for $\sqrt{s}=5.5$ TeV Pb+Pb collisions (LHC),
as a function of centrality.
Prediction is based on values of $\eta/s$ for the Glauber/CGC model that
matched $\sqrt{s}=200$ GeV Au+Au collision data from PHOBOS
at RHIC (\cite{Alver:2006wh},  shown for comparison).}  %On the right are the LHC predictions using our usual equation of state from Ref.~\cite{Laine:2006cp}(curves) compared to the p4+HRG EoS from Ref.~\cite{Bazavov:2009zn}(circles)}
\end{figure}

\begin{figure}[t]
%\includegraphics{Plot2.eps}
%\begin{center}
\includegraphics[width=.56\linewidth]{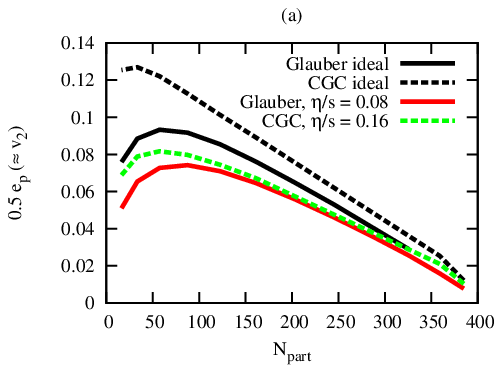}\hspace{-1.3cm}
\includegraphics[width = .56\linewidth]{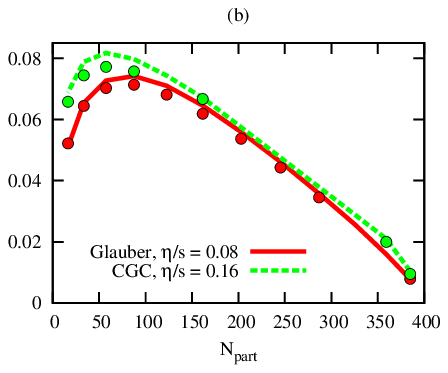}
%\end{center}
\caption[Predicted anisotropy (from \autoref{plot2}), in comparison to ideal hydrodynamics, and with simulations using an alternate lattice QCD equation of state.]{\label{plot3}
Predicted anisotropy from \autoref{plot2}, in comparison to the value when $\eta/s = 0.0001$, corresponding to ideal hydrodynamics (a) (upper black curves), and using an alternate lattice QCD equation of state (b) (circles)---see \autoref{EoSplots}.}
\end{figure}

\begin{figure}
\includegraphics[width = .55\linewidth]{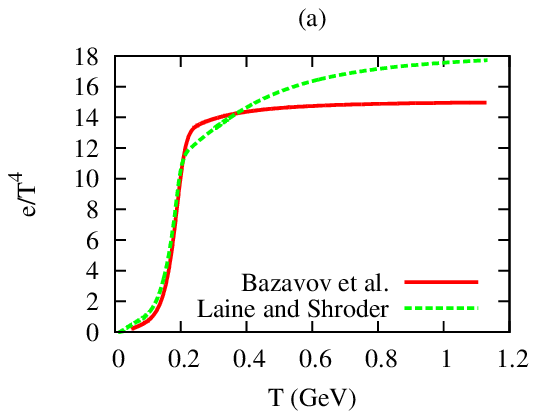}
\hspace{-1.3cm}
\includegraphics[width = .55\linewidth]{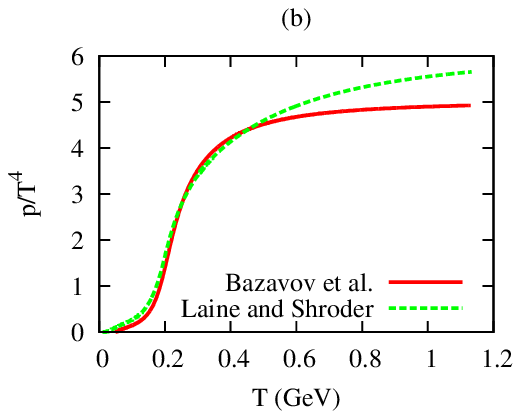}\\
\begin{center}
\includegraphics[width = .55\linewidth]{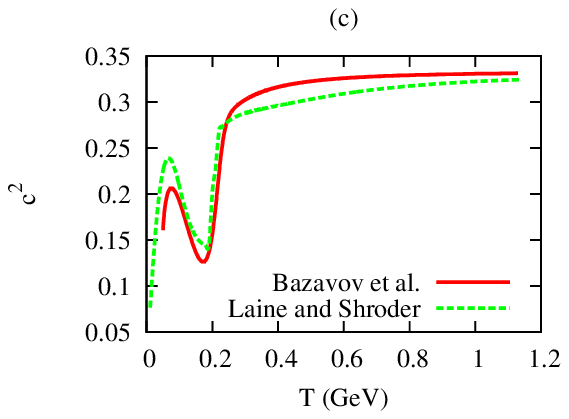}
\end{center}
\caption[Equation of state from Laine and Schr\"oder, and from Bazavov \textsl{et al.}]{\label{EoSplots} Equation of state from Laine and Schr\"oder \cite{Laine:2006cp}---used for all the main results of chapters \ref{RHIChydro} and \ref{LHC}---and from Bazavov \textsl{et al.} (\cite{Bazavov:2009zn}, p4+HRG)---used to generate the circles in \autoref{plot3}(b).  Above are the energy density (a) and pressure (b)---scaled by $T^4$---as a function of temperature.  Below (c) is the speed of sound squared.}
\end{figure}

Under the assumption that the average $\eta/s$ is similar for
collisions at RHIC and the LHC (along with the assumptions 
%about the multiplicity 
discussed above), one can make a prediction for the integrated elliptic flow
coefficient for charged hadrons as a function of impact parameter 
(or more customarily the number of participants $N_{\rm part}$, cf.~\cite{Luzum:2008cw}).
The result is shown in  \autoref{plot2}. As can be seen, we expect
integrated $v_2$ at the LHC to be about ten percent larger than at RHIC, which is less
than the increase predicted by ideal hydrodynamics \cite{Niemi:2008ta},
and in agreement with the extrapolations by Drescher \textsl{et al.}~\cite{Abreu:2007kv}.  For comparison, \autoref{plot3}(a) shows these LHC prediction curves along with those with $\eta/s$ set to 0.0001, corresponding to ideal hydrodynamics and illustrating the larger value of $v_2$ predicted by ideal hydrodynamics.  Also, as can be seen in \autoref{plot3}(b), remaining uncertainty in the equation of state seems to have little effect on this prediction.

\begin{figure}[h]
\begin{center}
\includegraphics[width=.75\linewidth]{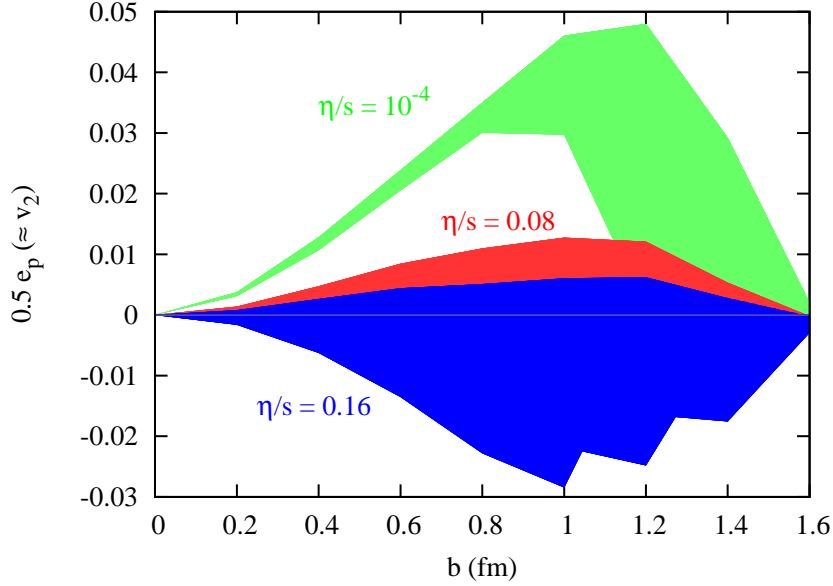}
\end{center}
\caption[$v_2$ for LHC proton-proton collisions]{\label{pp}
Bands encompassing the calculated anisotropy curves for $\sqrt{s}=14$ TeV p+p collisions.   The relaxation time $\frac{\tau_\pi}{\eta} (\epsilon+p)$ was varied between $2.6$ and $6$ and
$\tau_0$ from 0.5 fm to 2.0 fm for each value of $\eta/s$.  (Note that much of the $\eta/s = 0.08$ band is obscured by the $\eta/s = 0.16$ band, as both have significant dependence on the relaxation time and $\tau_0$.)
}
\end{figure}

Finally, using the charge density parametrization of the proton $\rho(b)$ 
in Ref.~\cite{Miller:2007uy} as an equivalent of 
the nuclear thickness function in the Glauber model (cf.~\cite{Luzum:2008cw})
one obtains an estimate for the shape of the transverse energy density
following a relativistic p+p collision. Using the 
predicted multiplicity at mid-rapidity $\frac{dN}{dY}\sim 6$ 
\cite{Kharzeev:2004if,Busza:2007ke} for $\sqrt{s}=14$ TeV p+p collisions at the LHC, one 
can again use \autoref{svsn} to infer the overall normalization
of the energy density (or $T_i$) at $\tau=\tau_0$ (see \autoref{tab:parLHC}).
As a ``Gedankenexperiment'' one can then ask how much elliptic flow
would be generated in LHC p+p collisions if the subsequent evolution
was well approximated by boost-invariant viscous hydrodynamics.
One finds that for ideal hydrodynamics $\frac{e_p}{2}\sim v_2\sim 0.035$ 
for integrated $|v_2|$ in minimum bias collisions (cf.~(28) in \cite{Luzum:2008cw}), 
while for $\eta/s\ge 0.08$, $v_2$ typically changes by 
almost 100 percent when varying the relaxation time
$\frac{\tau_\pi}{\eta} (\epsilon+p)$ between $2.6$ and $6$ and varying
$\tau_0$ by a factor of two (see \autoref{pp}). 
This indicates that for $\eta/s\ge 0.08$, the hydrodynamic
gradient expansion does not converge and as a consequence 
it is unlikely that 
elliptic flow develops in p+p collisions at top LHC energies.
If experiments find a non-vanishing value for integrated $|v_2|>0.02$ in 
minimum bias p+p collisions, 
this would be an indication for an extremely small viscosity $\eta/s<0.08$
in deconfined nuclear matter.

To conclude, viscous hydrodynamics can be used to make predictions
for the ratio of $v_2/e_x$ as a function of multiplicity and $\eta/s$. Assuming
a multiplicity of $\frac{dN_{\rm ch}}{d Y}\simeq 1800$ 
for the matter created in Pb+Pb collisions at LHC, as well as $\eta/s$
% shear viscosity coefficient 
similar to RHIC, we predict the integrated
elliptic flow for charged hadrons to be $10$\% larger at the LHC than at RHIC.
%We do not expect a significant $v_2\neq 0$ measurement in p+p collisions, unless
%the shear viscosity is extremely small ($\eta/s<0.08$).
We expect $v_2$ measurements in p+p collisions to be consistent with zero,
unless the shear viscosity is extremely small ($\eta/s<0.08$).
 \cite{Laine:2006cp} 

%
% ========== Chapter 5 - DWEF
% ==========   defs
\newcommand{\eq}[1]{Eq.~(\ref{#1})}
\newcommand{\be}{\begin{equation}}
\newcommand{\ee}{\end{equation}}
\newcommand{\bea}{\begin{eqnarray}}
\newcommand{\eea}{\end{eqnarray}}
\newcommand{\bfr}{\bf r}
\newcommand{\GV}{G_V}
\newcommand{\psidagop}{\hat{\psi}^\dagger}
\newcommand{\psiop}{\hat{\psi}}
\newcommand{\bra}{\langle}
\newcommand{\ket}{\rangle}
\newcommand{\fm}{\, \text{fm}}
\newcommand{\fmi}{\, \text{fm}^{-1}}
\newcommand{\mev}{\, \text{MeV}}
\def\bfq {{\bf q}}
\def\bfr {{\bf r}}
	\def\bfa{{\bf a}}
\def\bfK{{\bf K}}
\def\bfL{{\bf L}}
\def\bfk{{\bf k}}
\def\bfp{{\bf p}}  
\def\bfx{{\bf x}}\def\bfy{{\bf y}}\def\calU{{\cal U}}
\def\bfz{{\bf z}}
% ==========   defs
%
\chapter{Final State Interactions and the Distorted Wave Emission Function}
\label{DWEF}
Although much of the work in this chapter was performed before the viscous hydrodynamic simulations described previously, it is most naturally described here, after some understanding of hydrodynamic simulations has been obtained.  %Nevertheless, it is useful to keep in mind the chronological order of the work.
This is not meant to be a comprehensive discussion of final state interactions.  It is only a description of one line of inquiry with which the author of this dissertation participated---namely the distorted wave emission function (DWEF) model.
\section{The RHIC HBT Puzzle}
Despite the success of early ideal hydrodynamical simulations of heavy ion collisions at RHIC, they had much difficulty fitting two-particle correlations while simultaneously matching single-particle data.  For example, when the simulation parameters were set such that the experimental multiplicity and mean transverse momentum were at least roughly reproduced, the predictions for Hanbury Brown and Twiss radii (recall \autoref{HBT}) $R_O$ and $R_L$ are too large, while $R_S$ is too small.  In all reasonable cases, it seemed, the emission of particles occurred over a relatively long time period, causing the ratio $R_O/R_S$ to be large.  The experimental result, however, showed $R_O/R_S \approx 1$.   

This difficulty of describing HBT data with otherwise successful methods was dubbed the ``RHIC HBT puzzle'' \cite{Heinz:2002un}.  It should be noted that---although adding viscosity improves the agreement of $R_O/R_S$ in particular---even the most recent viscous hydrodynamic simulations do not completely resolve this puzzle \cite{Romatschke:2007jx}.  Shedding light on this was the original motivation behind the development of the distorted wave emission function (DWEF) model, in which final state interactions are introduced in the form of a relativistic optical model formalism \cite{Cramer:2004ih, Miller:2005ji}. %described as follows. 
Recent discussions of this puzzle can be found in Refs.~\cite{Pratt:2008qv,Pratt:2008bc}.
%
% A more recent review of the status of this puzzle can be found in Ref.~\cite{Pratt:2008qv,Pratt:2008bc}.   The idea was to
% 
%% Test \cite{2008PhRvC..78c4915L}
% 
%\cite{Romatschke:2007jx}
%is part of what has been dubbed ``the RHIC HBT puzzle'' \cite{Heinz:2002un}.  Recall section \ref{HBT} test \autoref{HBT} page test \pageref{HBT} test test \autopageref{HBT}
%
\section{The DWEF Model}
The medium produced in an ultrarelativistic heavy ion collision is very hot and dense.  It may be that the particles emitted from this opaque source experience significant refractive effects, which in turn affect measured quantities such as HBT radii in a way that cannot be captured by hydrodynamic models with only a simple Cooper-Frye freeze out mechanism.  With this in mind, let us introduce a model for these final state interactions and develop the formalism for calculating various quantities within this model.  Then by varying free parameters in the model one can show that it is possible to fit both single- and two-particle data for pions at RHIC, and then analyze the meaning of the values of the parameters necessary to fit the data.
%
%\subsection{Formalism}

The DWEF formalism (along with many of the results) is laid out in detail in Ref.~\cite{Miller:2005ji}.  Here we first present a general derivation of the formalism, in the manner of  Ref.~\cite{Luzum:2008tc}, and then we will describe the specific choices for the analytic form of the optical potential and emission function.
%
%
%
%
% ============== Plane wave formalism
\subsection{Plane wave formalism}
For comparison, it is useful to start by deriving the formalism for calculating HBT radii in the absence of final state interactions, given some source function that represents particles that are emitted (incoherently) from the collision medium and which then travel without interaction into the detectors.  In principle one could obtain this source function from some Cooper-Frye freeze out surface, but here it will be just a given function (and later an analytic parametrization with tunable parameters).  We follow one of the previous derivations \cite{Pratt:1984su} and then we can alter it appropriately to add an optical potential that the emitted particles interact with.  For simplicity we will specifically consider pions, the dominate hadron produced in a heavy ion collision.  The extension to other particles is straight forward.

We want to calculate the correlation function $C(\bfp,\bfq)$%, which is defined as
\bea C(\bfp,\bfq)\equiv {P(\bfp,\bfq)\over P(\bfp)P(\bfq)},\eea
where $P(\bfp_1,\cdots\bfp_n)$ is the probability of observing pions of momentum \{$\bfp_i$\} all in the same event. The identical nature of all pions of the same charge cause $C(\bfp,\bfp)=2$. %As mentioned in \autoref{HBT}, the width of the correlation function is related to the space-time extent of the source.
%
%space\;space\,space\.space\ space

A state created by a random pion source $|\eta\rangle$
 is described by \cite{Gyulassy:1979yi}
\bea |\eta\rangle=\exp\left[\int d^4x\,\eta(x)\gamma(t)\hat{\psi}^\dagger(x)\right]|0\rangle=\exp\left[\int d^3p\, dt\, \eta(\bfp,t)\gamma(t)
c^\dagger(\bfp)e^{-iE_pt}\right]|0\rangle, \label{etadef}
\eea where $\hat{\psi}^\dagger$ is the pion  creation operator in the Heisenberg representation,  $\gamma(t)$ is the random phase factor that 
accounts for the chaotic nature of the source and $c^\dagger(\bfp)$ is the creation operator for a pion of momentum $\bfp$. In particular, 
 an average over collision  events gives
\bea \langle \gamma^*(t)\gamma(t')\rangle&=&\delta(t-t'),\\ \langle \gamma^*(t_1)\gamma^*(t_2)\gamma(t_3)\gamma(t_4)\rangle&=&\delta(t_1-t_3)\delta(t_2-t_4)+\delta(t_1-t_4)\delta(t_2-t_3).\label{random}\eea
We note that as written, the state $|\eta\rangle $ is not normalized to one. However, the normalization constant will divide out of the numerator and denominator of the correlation function.
Therefore we do not make the normalization factor explicit here, but 
note that it enters when calculating the
pion spectrum.

For $\psiop$ and its time derivative to obey the Heisenberg commutation relation one has 
\bea\sqrt{E_pE_{p'}}[c(\bfp),c^\dagger(\bfp')]=\delta^{(3)}(\bfp-\bfp').\eea
Furthermore, we define
\bea \eta(\bfp,t)\equiv\int d^3x e^{-i\bfp\cdot\bfx}\eta(x).\label{ft}\eea
%With this normalization of $\gamma(t),\;|\eta(x)|^2$ is the probability of emitting a pion from the space-time point $x$,
%and $|\eta(\bfp,t)|$ (the three-dimensional spatial Fourier transform of $\eta(x)$) is the probability per unit time of emitting a pion of momentum $\bfp$ and energy $E_p$. 
The state $|\eta\rangle$  is an eigenstate of the destruction operator in the Schroedinger representation, $c(\bfp)$:
\bea c(\bfp)|\eta\rangle=\int dt\, e^{iE_p t}{\eta(\bfp,t)\over E_p}\gamma(t)|\eta\rangle.\label{cex}\eea 
The correlation function is 
\bea C(\bfp,\bfq)={\langle\eta|c^\dagger(\bfp)c^\dagger(\bfq)c(\bfq)c(\bfp)|\eta\rangle\over
\langle\eta|c^\dagger(\bfp)c(\bfp)|\eta\rangle \langle \eta|c^\dagger(\bfq)c(\bfq)|\eta\rangle}
%=1+
%{{|}\langle\eta|c^\dagger(\bfp)c(\bfq)|\eta\rangle{|}^2\over
%\langle\eta|c^\dagger(\bfp)c(\bfp)|\eta\rangle \langle \eta|c^\dagger(\bfq)c(\bfq)|\eta\rangle}
.\label{corr}\eea
The use of \autoref{random} and \autoref{cex} in the numerator of \autoref{corr} yields
\bea \langle\eta|c^\dagger(\bfp)c^\dagger(\bfq)c(\bfq)c(\bfp)|\eta\rangle
=\langle\eta|c^\dagger(\bfp)c(\bfp)|\eta\rangle
\langle\eta|c^\dagger(\bfq)c(\bfq)|\eta\rangle +|\langle\eta|c^\dagger(\bfp)c(\bfq)|\eta\rangle|^2.\label{top}\eea
Furthermore
\bea \langle\eta|c^\dagger(\bfp)c(\bfq)|\eta\rangle =\int dt \exp[-i(E_p-E_q)t]{\eta^*(\bfp,t)\eta(\bfq,t)\over E_pE_q}.
\label{matel}\eea
The quantity  $g(x,\bfp)$ is denoted the emission function and is defined as 
\bea g(x,\bfp)=\int d^3x'\, \eta^*(\bfx+{1\over2}\bfx',t)\eta(\bfx-{1\over2}\bfx',t)e^{i\bfp\cdot\bfx'},
%=\int d^3p
%\eta^*(\bfp+{1\over2}\bfp',t)\eta(\bfp-{1\over2}\bfp',t)e^{-i\bfp'\cdot\bfx}
\eea
so that
\bea \int {d^3p\over(2\pi)^3} g(x,\bfp)e^{-i\bfp\cdot\bfz}=\eta^*(\bfx+{1\over2}\bfz,t)\eta(\bfx-{1\over2}\bfz,t)\\
 \int {d^3p\over(2\pi)^3} g((\bfy+\bfy')/2,t,\bfp)e^{-i\bfp\cdot(\bfy-\bfy')}=\eta^*(\bfy,t)\eta(\bfy',t).
\label{good}\eea
The second expression appears in the right-hand-side of \autoref{matel}  
(if one uses \autoref{ft}) so that we may write 
\bea\langle\eta|c^\dagger(\bfp)c(\bfq)|\eta\rangle =
\int d^4x {\exp[-i(p-q)\cdot x]\over E_pE_q}g(x,{(p+q)\over2}).\label{etag}\eea
Using \autoref{etag} with $\bfp=\bfq$ shows that the function $g(x,\bfp)/E_p^2$ is the probability of emitting a pion of momentum $\bfp$ from a space-time point $x$. Using \autoref{top} and \autoref{etag} in \autoref{corr} gives the desired expression:
\bea  C(\bfp,\bfq)=1+{\int d^4x\;d^4x'g(x,{1\over2}\bfK)g(x',{1\over2}\bfK)\exp[-ik\cdot(x-x')]\over\int d^4x\;d^4x'g(x,\bfp)g(x',\bfq)},\label{corr2}\eea where $\bfK\equiv\bfp+\bfq$ and $k\equiv(E_p-E_q,\bfp-\bfq)$, and the factors of ${1\over E_pE_q}$ have canceled out.

From a formal point of  view, a 
key step in the algebra is the relation between the Heisenberg representation pion creation operator
$\hat{\psi}^\dagger(x)$ and its momentum-space Schroedinger representation counterpart $c^\dagger(\bfp)$ that appears in \autoref{etadef}:
\bea\hat{\psi}^\dagger(x)=\int d^3p \;c^\dagger(\bfp)\;{e^{-i\bfp\cdot\bfx}\over(2\pi)^{3/2}}\;e^{i E_p t}\label{complete}\\
\hat{\psi}(x)=\int d^3p \;c(\bfp)\;{e^{i\bfp\cdot\bfx}\over(2\pi)^{3/2}}\;e^{-i E_p t}
\eea
The operators $c^\dagger(\bfp)$ ($c(\bfp)$) are coefficients 
of a plane wave expansion for $\hat{\psi}^\dagger(x)\;(\psiop(x))$, with the
  plane wave functions  ${e^{i\bfp\cdot\bfx}/(2\pi)^{3/2}}$ being the complete set of basis functions.
However, one could re-write $\hat{\psi}^\dagger(x)\; (\psiop(x))$ as an expansion using   any set  of complete   wave functions. We shall exploit this feature below.
%
%
%
%
% ============== Distorted waves---real potential
\subsection{Distorted waves---real potential}
\label{real}
We represent the random classical source, emitting pions that interact with a real, time-independent external potential ${\cal U}$ by the Lagrangian density:
\bea-{\cal L}=\hat{\psi}^\dagger (-\partial^2+{\cal U}+m^2)\psiop+j(x)\psiop.\label{lang}\eea 
The current operator $j(x)$ is closely related to the emission function $g$ 
\cite{Gyulassy:1979yi}. In this Lagrangian
 the terms ${\cal U}$ and $j(x)$ are independent. Thus the relation between
the emission function   and ${\cal U}$  derived in \cite{Danielewicz:1983we} need not be satisfied.
Also note that the medium---and therefore also the potential---is in principle a time-dependent quantity.  Nevertheless, for simplicity we take ${\cal U}$ to be time independent and it can be interpreted as a time-averaged quantity.

The field operator $\psidagop$  can be expanded in the mode functions 
$\psi^{(-)}_\bfp$ that satisfy:
\bea \label{waveequation} (-\nabla^2+{\cal U})\psi^{(-)}_\bfp(\bfx)=p^2\psi^{(-)}_\bfp(\bfx).\label{mode}\eea
These wave functions obey the usual completeness and orthogonality relations
\bea \int d^3p\, \psi^{(-)*}_\bfp(\bfx)\psi^{(-)}_\bfp(\bfy)=\delta^{(3)}(\bfx-\bfy)\label{comp}
\\
 \int d^3x\, 
\psi^{(-)*}_\bfp(\bfx)    %\eea \end{document}
\psi^{(-)}_{\bfp^\prime}(\bfx)
=\delta^{(3)}(\bfp-\bfp'),\label{orth}\eea
so that one may use the field expansion
\bea \psiop(x)=\int d^3p\;\psi^{(-)}_\bfp(\bfx,t)e^{-iE_pt}d(\bfp),\label{dexp}\eea
with $d^\dagger(\bfp)$ being the creation 
operator for pions of momentum $\bfp$ in the basis of \autoref{mode}.  These mode functions are the 
eponymous distorted waves which replace the plane waves of the previous section.
The expansion  \autoref{dexp} assumes that ${\cal U}$ produces no bound states. 
If they did exist, the integral term would be augmented
by a term involving a sum over discrete states.

The availability of mode expansions when distortion effects are included
means that the simplification of the correlation function can 
proceed as in the previous section. We again use \autoref{etadef} and \autoref{random}.
The use of the field expansion \autoref{dexp} enables a generalization of the function  $\eta(x)$:
\bea \eta(x)=\int d^3p\, \psi^{(-)*}_\bfp(\bfx,t)\tilde{\eta}(\bfp,t),\eea with
\bea\tilde{\eta}(\bfp,t)\equiv \int d^3x\, \psi^{(-)}_\bfp(\bfx,t)\eta(x),\eea
so that
\bea |\eta\rangle=\exp\left[\int d^3p\, dt \tilde{\eta}(\bfp,t)\gamma(t)
d^\dagger(\bfp)\right]|0\rangle. \label{etatilde}\eea Note that the ability to obtain a
 relation between the $\tilde{\eta}(\bfp,t)$ and $\eta(x)$ rests on the relations \autoref{comp} and \autoref{orth}.

The state $|\eta\rangle$ is an eigenstate of $d(\bfp)$.  Thus the result
\bea C(\bfp,\bfq)=1+
{{|}\langle\eta|d^\dagger(\bfp)d(\bfq)|\eta\rangle{|}^2\over
\langle\eta|d^\dagger(\bfp)d(\bfp)|\eta\rangle \langle \eta|d^\dagger(\bfq)d(\bfq)|\eta\rangle},\label{corrnew}\eea
very similar to \autoref{corr},
is  obtained.
We need the matrix elements appearing in the numerator and find
\bea\langle\eta|d^\dagger(\bfp)d(\bfq)|\eta\rangle =  
\int d^4x\, d^3x'\,  
{\exp[-it(E_p-E_q)]\over E_pE_q}\psi_p^{(-)}(\bfx)\psi_q^{(-)*}(\bfx')\eta(x)\eta(\bfx',t).\label{etag1}\eea
and the use of \autoref{good} allows us to obtain
\bea\langle\eta|d^\dagger(\bfp)d(\bfq)|\eta\rangle =&{1\over E_pE_q}\int dt\,d^3x\,d^3x' {d^3p'\over (2\pi)^3}e^{it(E_q-E_p)}
e^{-i\bfp'\cdot\bfx'}\nonumber\\
&\times\ \psi_\bfp^{(-)}(\bfx+\bfx'/2)\psi_\bfq^{(-)*}(\bfx-\bfx'/2)g(x,\bfp').%\nonumber\\
\label{newer}\eea
This result, which can be applied for $\bfp\ne\bfq $ and for $\bfp=\bfq$, specifies the evaluation of the correlation function
of \autoref{corrnew} with the result %\end{document}
\be C(\bfp,\bfq)=1+ \frac {\left\vert S(K,k)\right\vert^2} {S(p)S(q)}  \label{newcc}
\ee %\end{document}
where \bea  S(K,k)\equiv %
\int d^4x\,d^3x'{d^3p'\over(2\pi)^3} %\eea\end{document}
e^{it(E_q-E_{p'})} 
e^{-i\bfp'\cdot\bfx'} 
\psi_\bfp^{(-)}(\bfx+\bfx'/2)%\eea\end{document}
\psi_\bfq^{(-)*}(\bfx-\bfx'/2)g(x,\bfp'),\label{corr3}\eea
and
\bea \label{emission}S(p)\equiv
\int d^4x\,d^3x'{d^3p'\over(2\pi)^3}e^{-i\bfp'\cdot\bfx'}
\psi_\bfp^{(-)}(\bfx+\bfx'/2)\psi_\bfp^{(-)*}(\bfx-\bfx'/2)g(x,\bfp').\eea
This is the expression that is used in  the DWEF formalism
 \cite{Cramer:2004ih,Miller:2005ji}.
In principle one could use either \autoref{corr2} or \autoref{newcc} 
to analyze data, but the extracted 
space time properties of the source $\eta(x)$ would be different.

A comment should be made on the possible momentum and energy dependence 
of the optical potential.
The completeness and orthogonality relations are obtained with any Hermitian 
${\cal U}$ which can therefore be momentum dependent, but not energy dependent.
As explained in section~5 (Equation 43) of Ref.~\cite{Miller:2005ji}, the real part of the 
potential  can and should be 
thought of as a momentum-dependent, but energy-independent potential.  If there were
true energy dependence a factor depending on the derivative of the potential with
respect to energy \cite{Pratt:2007pf}  
would enter into the orthogonality and completeness relations.
%
%
%
%
% ============== Coupled channels
\subsection{Coupled channels}
If the optical potential ${\cal U}$ from the previous section is complex, the derivation above fails.
%The optical potential used in previous work  \cite{Cramer:2004ih,Miller:2005ji}
%is complex. 
Using the necessary completeness and orthogonality relations to relate 
$\eta(x)$ to $\tilde{\eta}(\bfp,t)$  requires the use of a real potential.  If we would like to include the effects of an imaginary part of the potential, we should investigate possible corrections to the above formalism.

The optical potential or pion self-energy
is an effective interaction between the
pion and the medium. 
The medium is not an eigenstate of the Hamiltonian, but rather of $H_0$, which is
  the full
Hamiltonian minus the Hermitian operator representing the 
pionic final state interactions. 
 Eliminating the infinite number of possible states of $H_0$
and representing  these by a single state leads to a self-energy 
that is necessarily complex. Our procedure here is
to specifically consider the infinite number of states of the medium, 
obtain a Lagrangian density that involves  Hermitian interactions, and 
derive the optical potential formalism and any  corrections to it. 

Let $P_n$ denote a projection operator for the medium to be in a given eigenstate 
of $H_0$,  $n$. 
These obey 
\bea \sum_n P_n=1,\nonumber\\ P_nP_m=\delta_{n,m}P_n.\eea
For the case of
 $\pi$-nuclear scattering, $n$ would represent
the nuclear eigenstates. Here $n$ represents states of the medium in the absence of its interactions with pions. 
The correlation function is now given by 
\bea C(\bfp,\bfq)={\sum_n P_n(\bfp,\bfq)\over \sum_n P_n(\bfp)\sum_m P_m(\bfq)}\eea
where $P_n(\bfp)$ is the probability for emission of a pion of momentum $\bfp$ 
from the medium in a  
state $n$. Similarly $P_n(\bfp,\bfq)$ is the probability 
for emission of a pair of pions of 
momentum $\bfp,\bfq$ from the medium in a state $n$.  The sums over $n$ account for the inclusive nature of the process of interest.

It is convenient to 
 define the product of the field operator with the projection operator $P_n$:
\bea\psiop_n(x)\equiv \psiop(x) \;P_n,\eea with
\bea \psiop(x)=\sum_n \psiop_n(x),\label{newpsi}\eea using 
the complete nature of the set $n$. The Lagrangian density is given by
\bea-{\cal L}=\sum_{n}\partial\hat{\psi}_n^\dagger\cdot\partial\hat{\psi}_n
+\sum_{n,m}{\psi}_n^\dagger \left((m_\pi^2 +M^2_m)\delta_{nm} +{\cal U}_{nm}\right)\psiop_m+\sum_n j_n(x)\psiop_n(x)\label{lang1},\eea
where  \bea {\cal U}_{nm}={\cal U}^*_{mn}\equiv (\hat{\calU})_{nm}\eea 
and  $\hat{\calU}$ is the 
Hermitian interaction operator and $M_m^2$, the $m$ matrix element of the diagonal operator $M^2$, represents the effects of the different energies of the states labeled by $m$.
The field operator $\psiop_n$  can be expanded in the mode functions 
$\psi^{(-)}_{\bfp,n}$:
\bea  \sum_{m\ne n}{\cal U}_{nm}(\bfx)\psi^{(-)}_{\bfp,m}(\bfx,t)=(p^2+\nabla^2-M_n^2-\calU_{nn}(\bfx))
\psi^{(-)}_{\bfp,n}(\bfx,t).\label{mod1}\eea 
Here the potential ${\cal U}$ is taken as a local operator in the position space of the outgoing pion.

To see the correspondence between the formulation of \autoref{lang1} and \autoref{mod1},
let $\psiop_1$ correspond to the field operator (and state) of the previous section and solve formally for
 $\psi^{(-)}_{\bfp,m}$ in terms of $\psi^{(-)}_{\bfp,1}$.
It is convenient to define the operator $\widetilde\calU$ with matrix elements given by
\bea\tilde{\calU}_{n,n'} \equiv (1-\delta_{n,1})(1-\delta_{n',1})\calU_{n,n'}\eea
Then 
\bea \psi^{(-)}_{\bfp,n\ne1}=\sum_{m\ne1}\left({1\over \nabla^2+p^2-M^2-\widetilde{\calU}-i\epsilon}\right)_{nm}\calU_{m1}\psi^{(-)}_{\bfp,1},
\label{onec}\eea where $(\nabla^2+p^2-M^2)_{nm}\propto\delta_{n,m},$ and $M^2$ is an operator giving 
$M_n^2$ when acting on the state $n$.
 Then rewrite \autoref{mod1} % \end{document}
in  terms of $\psi^{(-)}_{\bfp,1}$ as %\end{document}
\bea \calU_{11}\psi^{(-)}_{\bfp,1}+\sum_{m,n\ne1}\calU_{1n}\left({1\over \nabla^2+p^2-M^2-\calU-i\epsilon}\right)_{nm}
\calU_{m1}\psi^{(-)}_{\bfp,1}=(p^2+\nabla^2-M_1^2)\psi^{(-)}_{\bfp,1}\eea

The complex object $$\calU_{11}+\sum_{m,n\ne1}\calU_{1m}\left({1\over \nabla^2+p^2-M^2-\widetilde{\calU}-i\epsilon}\right)_{m,n}
\calU_{n1},$$ a non-local operator in coordinate space, 
 can be identified with the optical potential, given by the operator 
%\end{document}
$V(Z)$ as a function of a complex variable $Z$:
\bea V(Z)=\calU_{11}+\sum_{m,n\ne1}\calU_{1m}\left({1\over \nabla^2+Z-M^2-\calU }\right)_{m,n}
\calU_{n1}.\label{vopt}\eea

We proceed by employing 
\autoref{newpsi} and \autoref{lang1} to compute the correlation function. The solutions of \autoref{mod1} form a complete orthogonal set:
\begin{align} 
\sum_n\int d^3p\, \psi^{(-)*}_{\bfp,n}(\bfx)\psi^{(-)}_{\bfp,n}(\bfy)=\ &\delta^{(3)}(\bfx-\bfy)\label{comp1}
\\
 \sum_n\int d^3x \;
\psi^{(-)*}_{\bfp,n}(\bfx)
\psi^{(-)}_{\bfp',n}(\bfx)
=\ &\delta^{(3)}(\bfp-\bfp').\label{orth1}
\end{align}
The field expansion is now 
\bea\psiop(x)=\int d^3p \,\sum_n \; a(\bfp) P_n\psi^{(-)}_{\bfp,n}(\bfx)e^{-iE_pt},\eea
so that
\bea |\eta\rangle=\exp\left[\sum_n\int d^4x \eta_n(x)\gamma(t)\int d^3p\;a^\dagger(\bfp)P_n\psi^{(-)*}_{\bfp,n}(\bfx)e^{iE_pt}\right]\sum_m
|0,m\rangle\label{etatilde1},\eea 
where the state $|0,m\rangle$ is the pionic vacuum if the medium is in the state $m$, and $\eta_n(x)$ represents the  source for the state $n$. 
These state vectors obey the relations
\bea \langle 0,n|0,m\rangle =\delta_{n,m}=
\langle 0,n|P_n|0,m\rangle .\eea
Define  %\end{document}
\bea \eta_n(\bfp,t)
\equiv\int d^3x\;\eta_n(\bfx,t)\psi^{(-)*}_{\bfp,n}(\bfx),\eea 
%\end{document}
%and completeness and orthogonality give
%\bea \eta(\bfy,t)=\sum_n\int d^3p\;\psi^{(-)}_{\bfp,n}(\bfy)\eta_n(\bfp,t)\eea
so that
\bea |\eta\rangle=
\exp\left[\int d^3p\, dt\, \gamma(t)
\sum_n\eta_n(\bfp,t)a^\dagger(\bfp)P_ne^{iE_pt}\right]\sum_m|0,m\rangle,\eea %\end{document}
\bea a(\bfp)|\eta\rangle=\int dt\;\gamma(t)\sum_n{\eta_n(\bfp,t)\over E_p}P_ne^{iE_pt}|\eta\rangle.\eea
The emission probability is given by
\bea E_pE_q\sum_n\langle\eta|a^\dagger(\bfp)P_n a(\bfq)|\eta\rangle=&\sum_{n}\int dt\,d^3x\,d^3y\,\eta_n^*(\bfx,t)\eta_n(\bfy,t) \nonumber\\
&\times \psi_{\bfp,n}^{(-)*}(\bfx)\psi_{\bfq,n}^{(-)}(\bfy)\;e^{i(E_p-E_q)t}\eea or using \autoref{good}
\bea E_pE_q\langle\eta|a^\dagger(\bfp)a(\bfq)|\eta\rangle=&\sum_{n}\int dt\,d^3x\,d^3y\int d^3p'\;g_n((\bfx+\bfy)/2,t,\bfp')
e^{-i\bfp'\cdot(\bfx-\bfy)} \nonumber\\
&\times \psi_{\bfp,n}^{(-)*}(\bfx)\psi_{\bfq,n}^{(-)}(\bfy)\;e^{i(E_p-E_q)t},\label{newnn}\eea 
where \bea   g_n(x,\bfp)=\int d^3x' \eta_n^*(\bfx+{1\over2}\bfx',t)\eta_n(\bfx-{1\over2}\bfx',t)e^{i\bfp\cdot\bfx'}.\eea
If pionic final state interactions are ignored, the term $\sum_n g_n$ enters and
this may be identified with the emission function, $g$ of previous sections.

The expression \autoref{newnn} is the same as \autoref{newer} except that now we sum over the channels $n$. These sums
 may be expressed in
terms of the optical model wave functions of \autoref{onec}.
The term of \autoref{newnn} with $n=1$ corresponds to the DWEF formalism, and the  terms
with $n>1$ are corrections. As an example of a correction term,
suppose part of the imaginary part of the optical potential arises from a pion-nucleon interaction that makes an intermediate $\Delta$. Then a term corresponding to 
one of $n>1$ involves the emission of a pion from a nucleon that makes an 
intermediate $\Delta$.

It is difficult to assess the importance of the second term in a general way. 
The only obvious limit is that if states with $n>1$ are not excited then  
$Im({ V})$ of \autoref{vopt} must vanish. Conversely, if 
 $Im({V})$=0, the states $n>1$ must be 
above the threshold energy and the propagators that appear in the  correction terms
correspond to virtual propagation over a  small distance with limited effect.

Therefore in the following, results will be presented with the imaginary part of of the optical potential set at a vanishing value, and separately with any value allowed.  While the former is correct within the model, the latter will have unknown corrections.  Nevertheless, it will still be illustrative to look at both in the hope that it will give some idea of the effect of an imaginary part of the optical potential in addition to the more reliable information concerning the real part.
\subsection{Complete DWEF formalism}
The key unknown pieces in the expressions above are the emission function $g$ and the optical potential ${\cal U}$.  The wavefunctions $\psi^{(-)}_\bfp$ can be calculated from ${\cal U}$, and then integrals can be performed to obtain the quantities of interest.  Reiterating the results of \autoref{real}, the correlation function for determining HBT radii is given by (recall \autoref{newcc})
\begin{equation}
C(\bfp,\bfq) = \frac{\left\vert\int d^4x\,d^3x'{d^3p'\over(2\pi)^3}
e^{it(E_q-E_{p'})} 
e^{-i\bfp'\cdot\bfx'} 
\psi_\bfp^{(-)}(\bfx+\bfx'/2)%\eea\end{document}
\psi_\bfq^{(-)*}(\bfx-\bfx'/2)S_0(x,\bfp')\right\vert^2}
{\left(\int d^4x\,d^3x'{d^3p'\over(2\pi)^3}e^{-i\bfp'\cdot\bfx'}
\psi_\bfp^{(-)}(\bfx+\bfx'/2)\psi_\bfp^{(-)*}(\bfx-\bfx'/2)S_0(x,\bfp')\right)(\bf p\to \bf q)},
\end{equation}
and single particle observables can be derived from one of the factors in the denominator:
\begin{equation}
E \frac {dN} {d^3p} = \int d^4x\,d^3x'{d^3p'\over(2\pi)^3}e^{-i\bfp'\cdot\bfx'}
\psi_\bfp^{(-)}(\bfx+\bfx'/2)\psi_\bfp^{(-)*}(\bfx-\bfx'/2)S_0(x,\bfp').
\end{equation}
(In this section, the notation $S_0(x,p) \equiv g(p,x)$ is used to make contact with the notation of Ref.~\cite{Miller:2005ji}).

We proceed by using an analytic parametrization that is inspired by hydrodynamic freeze out.  A more detailed discussion can be found in Ref.~\cite{Miller:2005ji}.
%
%To make contact with more common notation, define $$ S(p,x) \equiv
%\int d^3x'{d^3p'\over(2\pi)^3}e^{-i\bfp'\cdot\bfx'}
%\psi_\bfp^{(-)}(\bfx+\bfx'/2)\psi_\bfp^{(-)*}(\bfx-\bfx'/2)g(x,\bfp')$$ so that $$S(p) = \int d^3x\, S(p,x).$$

The form used is:

\begin{align}
g(p,x) &\equiv S_0(x,p) = \frac{\cosh \eta}{(2 \pi)^3}\; e^{\frac{-\eta^2}{2\Delta\eta^2}} \frac{1}{\sqrt{2\pi\; \Delta\tau^2}}\; e^{\frac{-(\tau - \tau_0)^2}{2 \Delta\tau^2}} \frac{M_{\perp}\: \rho (\textbf{b})}{e^{(p \cdot u - \mu_{\pi})/T}-1},\\
 U(\textbf{b}) &= -(w_0+w_2\ \textbf{p}^2)\ \rho(\textbf{b}).
\end{align}
%The correlation function for determining HBT radii is then given by (recall \autoref{corr1})
%\begin{equation}
%C(\bfp,\bfq) = \frac{\left(\int d^4x\,d^3x'{d^3p'\over(2\pi)^3}
%e^{it(E_q-E_{p'})} 
%e^{-i\bfp'\cdot\bfx'} 
%\psi_\bfp^{(-)}(\bfx+\bfx'/2)%\eea\end{document}
%\psi_\bfq^{(-)*}(\bfx-\bfx'/2)S_0(x,\bfp')\right)}
%{\left(\int d^4x\,d^3x'{d^3p'\over(2\pi)^3}e^{-i\bfp'\cdot\bfx'}
%\psi_\bfp^{(-)}(\bfx+\bfx'/2)\psi_\bfp^{(-)*}(\bfx-\bfx'/2)S_0(x,\bfp')\right)(\bf p\to \bf q)},
%\end{equation}
%and single particle observables can be derived from one of the factors in the denominator:
%\begin{equation}
%E \frac {dN} {d^3p} = \left(\int d^4x\,d^3x'{d^3p'\over(2\pi)^3}e^{-i\bfp'\cdot\bfx'}
%\psi_\bfp^{(-)}(\bfx+\bfx'/2)\psi_\bfp^{(-)*}(\bfx-\bfx'/2)S_0(x,\bfp')\right).
%\end{equation}
%
%The notation $S_0$ is used to make contact with the notation of Ref.~\cite{Miller:2005ji}.
%
$p$ is the asymptotic pion momentum and $M_\perp = \sqrt{{\textbf p}_\perp^2 + m_\pi^2}$.
Just as in previous chapters, it is natural to use Milne coordinates, although here we will use radial coordinates in the transverse plane:
\begin{align}%\begin{split}
\eta =& {\rm arctanh}(z/t)
&\tau =& \sqrt{t^2-z^2} 
&b = \sqrt{x^2 + y^2} \nonumber \\
\phi =& \arctan(y/x)
&\textbf{b} =& (b,\phi). &
%test = test
%\end{split}
\end{align}
Also as above, we will restrict ourselves to mid-rapidity data.

$\rho(\bf b)$ represents the transverse density of the medium and is used for the transverse shape of both the emission function and the optical potential.  It is normalized as $\rho(0) = 1$.  The original DWEF model was restricted to rotationally symmetric systems (corresponding to central collisions) and used
\begin{equation}
\rho (\textbf {b}) = \rho (b) = \frac {\left[ e^{-R_{\rm {WS}}/a_{\rm {WS}}}+1\right]^2} {\left[ e^{(b-R_{\rm {WS}})/a_{\rm {WS}}}+1\right]^2}.
\end{equation}
%$$
%\rho (\textbf {b}) = \rho (b) = \left( \frac {e^{-R_{\rm {WS}}/a_{\rm {WS}}}+1} {e^{(b-R_{\rm {WS}})/a_{\rm {WS}}}+1}\right)^2.
%$$
This distribution has the correct exponential fall-off at large distance, and different choices of the parameters $R_{\rm WS}$ and $a_{\rm WS}$ allow for a variety of shapes.
To calculate the elliptic flow coefficient $v_2$, it will be necessary to generalize this form for non-rotationally-symmetric systems (see \autoref{DWEFv2}).

The velocity field that describes the dynamics of the expanding source in a central collision event is parametrized by a fluid rapidity $\eta_t(\textbf{b})$
\begin{equation}
u^{\mu} (x) = (\cosh \eta \cosh \eta_t,\ \sinh \eta_t \cos \phi,\ \sinh \eta_t \sin \phi,\ \sinh \eta \cosh \eta_t).
\end{equation}
The flow rapidity is taken to have a linear radial profile with strength $\eta_f$
\begin{equation}
 \eta_t(\textbf{b}) = \eta_f \frac {b} {R_{\rm WS}}.
\end{equation}
This also will have to be modified when calculating elliptic flow for a peripheral collision.

The free parameters, then, are $\Delta\eta$, $\Delta\tau$, $\tau_0$, $\mu_\pi$, $T$, $w_0$, $w_2$, $R_{\rm WS}$, $a_{\rm WS}$, and $\eta_f$.  These parameters were varied (with various of them occasionally held fixed) to reproduce the single- and two-particle pion data for $\sqrt s = 200$ GeV Au-Au collisions at RHIC \cite{Miller:2005ji}.
%	
%Several approximations are then made to make the problem more feasible numerically.
%
%
%\eqref{emission} \ref{emission} \autoref{emission}
%
%\begin{eqnarray}
%S(p,x) &=& \frac{\cosh \eta}{(2 \pi)^3}\; e^{\frac{-\eta^2}{2\Delta\eta^2}} \frac{1}{\sqrt{2\pi\; \Delta\tau^2}}\; e^{\frac{-(\tau - \tau_0)^2}{2 \Delta\tau^2}} \frac{M_{\perp}\: \rho (\textbf{b})}{e^{(p \cdot u - \mu_{\pi})/T}-1}
%\vert \psi_p^{(-)}(x)\vert^2,\\
% U(\textbf{b}) &=& -(w_0+w_2\ \textbf{p}^2)\ \rho(\textbf{b}).
%\end{eqnarray}
%\begin{equation}
%S(p,x) = \frac{\cosh \eta}{(2 \pi)^3}\; e^{\frac{-\eta^2}{2\Delta\eta^2}} \frac{1}{\sqrt{2\pi\; \Delta\tau^2}}\; e^{\frac{-(\tau - \tau_0)^2}{2 \Delta\tau^2}} \frac{M_{\perp}\: \rho (\textbf{b})}{e^{(p \cdot u - \mu_{\pi})/T}-1}
%\vert \psi_p^{(-)}(x)\vert^2,
%\end{equation}
%\begin{equation}
% U(\textbf{b}) = -(w_0+w_2\ \textbf{p}^2)\ \rho(\t4extbf{b}).
%\end{equation}
%
%
%
%
%
 \subsection{Results for Central Collisions}
% 
% Here some results of the calculation of central collision observables are summarized.
% Should I summarize the results of Ref.~\cite{Miller:2005ji} here?
%
%
Calculations for central RHIC collisions were originally presented in Refs.~\cite{Cramer:2004ih,Miller:2005ji}.  With the above insight, it is instructive to assess the possible importance of the imaginary part of the optical potential.
%
%We proceed by assessing 
% the possible importance of the correction term for the work of
%\cite{Cramer:2004ih,Miller:2005ji} by seeing what happens if the optical potential is 
%taken to be purely real with no imaginary optical potential. 

\begin{figure}%Fig 1
\includegraphics[width=\linewidth]{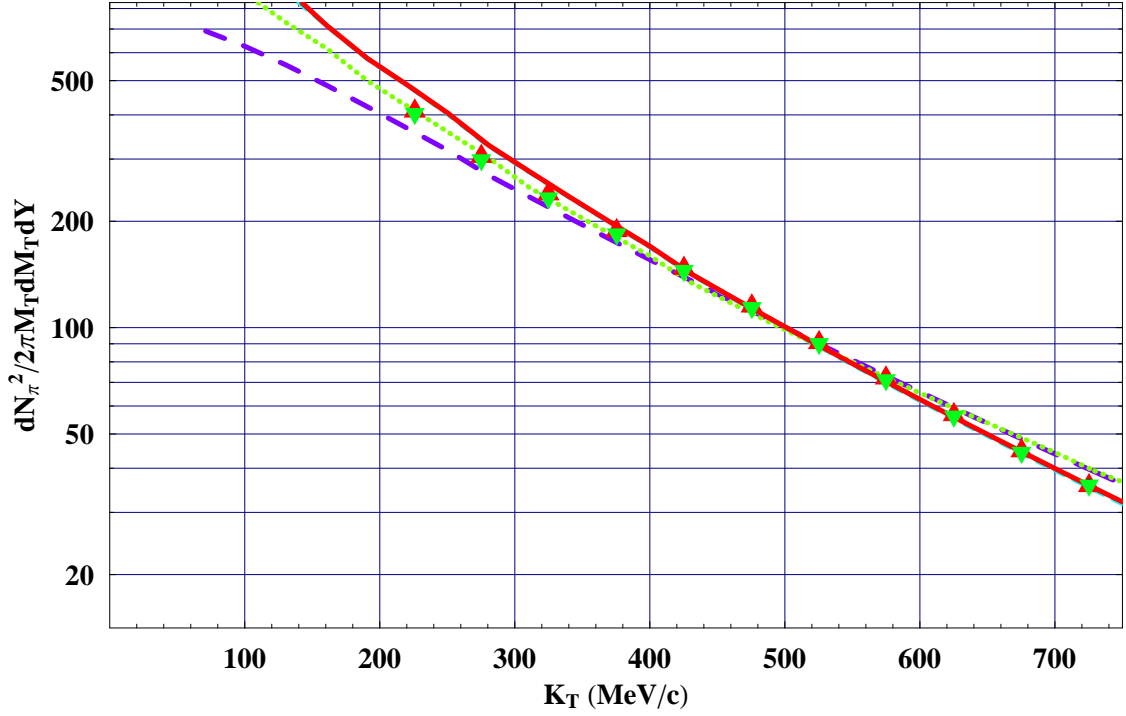}
\caption[DWEF calculated pionic spectrum]{\label{spec} Computed pionic spectrum.
Red upright triangles:  $\pi^-$ spectrum  (STAR)
Green inverted triangles:  $\pi^+$ spectrum points (STAR) \cite{Adams:2003xp}
Red solid line:  DWEF fit with vanishing 
imaginary part of the optical potential,
% $\gamma=0.0001, \;\mu_\pi=M_\pi. $ 
first line of \autoref{tab}. % [X2a] 
Green dotted line: DWEF fit including search on the imaginary part of the optical
potential, %$\mu_\pi=M_\pi$ 
second  line of \autoref{tab}. %[X3]
Blue dashed line(almost entirely covered by the red solid curve): 
DWEF fit with vanishing 
imaginary part of the optical potential,$\Delta\tau$ =  1.5 fm/c, % $\gamma=0.0001$, 
%$\mu_\pi=M_\pi$
 third line of \autoref{tab}.   % [X3c]
Violet long dashed line:  DWEF fit including search on $ \mu_\pi$, setting the optical
potential to essentially 0, %and 
%StR0=StR2=StI2=0.0001,
%$ \gamma=0.0001,$% [X4c]
fourth line of  \autoref{tab}.
% (a middle line on the spreadsheet).
 }
\end{figure}

\begin{figure}%Fig 2
\includegraphics[width=\linewidth]{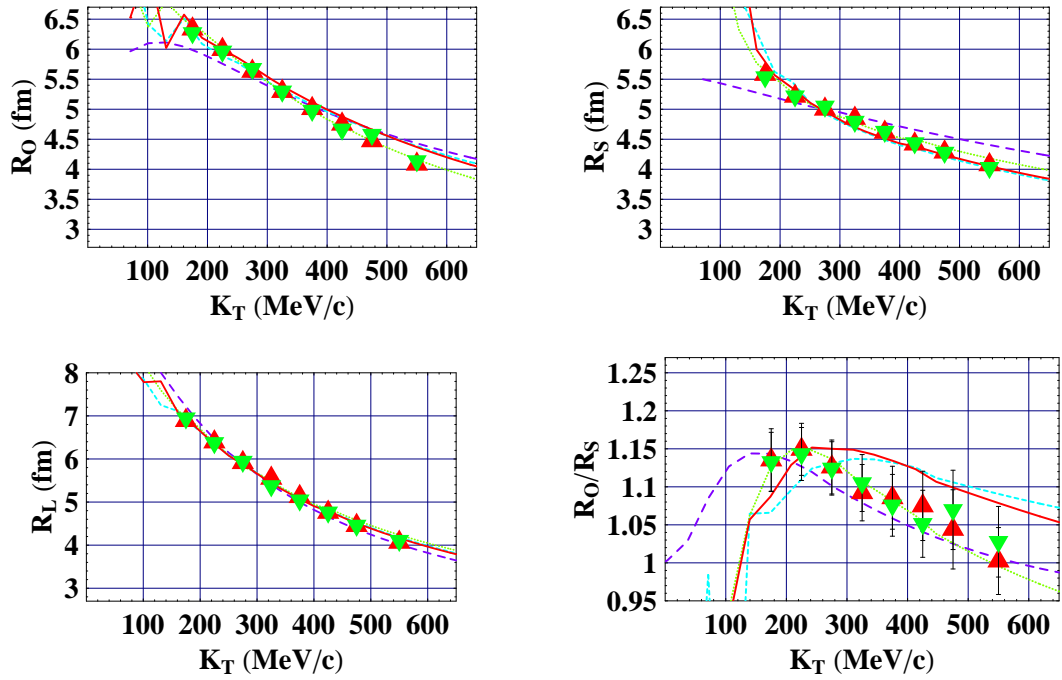}
\caption[DWEF calculated HBT radii]{\label{radii} HBT radii. 
Curves are labeled as in \autoref{spec}. STAR data \cite{Adams:2004yc}}
\end{figure}

 A variety DWEF fits are performed, see \autoref{tab}. 
In Ref.~\cite{Miller:2005ji}
the imaginary part of the 
optical potential as represented by the term $w_2$  is about 
one tenth of the real potential. It is therefore possible that, 
in the limit that $Im(w_2)=0$, there would be
no significant correction term, so we try to understand if removing the imaginary part of
the optical potential can be done without degrading the quality of the fit.
The results are shown in Figures \ref{spec} and \ref{radii}. 
An example of the previous calculations 
\cite{Cramer:2004ih,Miller:2005ji} is shown as the green dotted curve (second line of 
\autoref{tab}). The red solid
curve (first line of \autoref{tab})
shows the result of setting the imaginary potential to a vanishingly small
value. 
This results in only 
a slightly worse description of the data.
The changes in the imaginary part of the optical potential 
$w_2$ are largely compensated by a reduction of the temperature 
from about 160 MeV to about 120 MeV. We also point out that 
 the length of the flux tube as represented by 
$\Delta \eta$ is vastly increased, providing greater justification to our previous
procedure of taking the length of the flux tube 
to be infinitely long in the longitudinal 
direction. However, the emission duration is reduced to
0 fm/c, which is similar to the results of the blast wave model
 \cite{Retiere:2003kf}. This means that all of the pionic emission occurs at a
single proper time. This value
justifies the use of a time-independent optical potential, but does 
seem to be difficult to understand because some spread of emission times
 is expected for a long-lived plasma. The results shown by the blue dashed curves 
(third line of \autoref{tab})
are obtained with fixing the emission duration to 1.5 fm/$c$, which is our previous
value \cite{Cramer:2004ih,Miller:2005ji}. The description of the  spectrum is basically
unchanged but 
 the
radii are less precisely described.
 The violet long-dashed curves (fourth line of \autoref{tab})
show the DWEF fit using a vanishing optical potential.
 This does not give a good description of the momentum dependence of the radii and 
is associate with the largest deviation between our calculations and the data as
represented by the $\chi^2$ values of \autoref{tab}.

It is clear that the precision of our description of the data is improved by
including the imaginary part of the optical potential. However, 
this is a quantitatively
but  not a qualitatively important effect. It is also true that including the real
part of the optical potential is a qualitatively important effect. These
 results suggest that the correction terms embodied by the terms with $n\ne1$ of 
\autoref{newnn} are not very important, but non-negligible. It is also possible
that an optical potential with a different geometry than the volume
form that we have assumed might be able to account for the the neglected terms.
However, an accurate assessment  would require the  development 
a theory that involves dealing  with explicit  models for $g_n,j_n$ and $\calU$.

\begin{table}
  \caption[Parameter sets for DWEF central collision calculations]{\label{tab} Four  parameter sets % with variances. 
obtained with slightly different procedures \cite{Luzum:2008tc}.  The values of $\chi^2$ represents the accuracy of the description of the data.}
  \vspace{0.1cm}
  \begin{tabular}{ccccccccccccc}\hline
   $T$ & $\eta_f$ & $\Delta\tau$ & $R_{WS}$ & $a_{WS}$ & $w_0$& $~w_2$
& $\tau_{0}$ &  $\Delta\eta$ & $\mu_\pi$&$\chi^2$\\
\tiny{(MeV)} & & \tiny{$(\frac {\rm{fm}} {c})$} & \tiny{(fm)} & \tiny{(fm)} & \tiny{$(\rm{fm}^{-2})$} & & \tiny{$(\frac {\rm{fm}} {c})$} & & \tiny{(MeV)} & \\
\hline
121   & 1.05  & 0 & 11.7 & 1.11 & 0.495 & 0.762 +0.0001$i$ & 9.20 &  70.7& 139.57&300\\
162 &   1.22 & 1.55& 11.9 &1.13 &0.488 &1.19+0.13$i$& 9.10&1.68&139.57&117\\
121   &  1.04 & 1.5 & 11.7 & 0.905 & 0.564 & 0.595 +0.0001$i$ &8.85  &70.7 & 139.57&451\\
144 &	0.990 &	2.07 &	12.57 &	0.876&	0.0001&	0.0001+0.0001$i$	&6.85 &$\infty$& 83.5&1068\\
  \hline
      \end{tabular}
      \end{table}
%
% ========== Chapter 6 - v_2 in DWEF model
% ==========   defs
\def\poinc{Poincar\'{e} }
\def\bfq {{\bf q}}
\def\bfqp {{\bf q}_\perp}
\def\bfK{{\bf K}}
\def\bfKp{{\bf K}_\perp}
\def\bfL{{\bf L}}
\def\bfk{{\bf k}}
\def\bfp{{\bf p}}  
\def\bfu{{\bf u}}
\def\bfr{{\bf r}} 
\def\bfy{{\bf y}} 
\def\bfx{{\bf x}}  
\def\be{\begin{equation}}
 \def \ee{\end{equation}}
\def\bea{\begin{eqnarray}}
  \def\eea{\end{eqnarray}}
\def\non{\nonumber\\}
\newcommand{\eqn} {Eq.~(\ref )}
\newcommand{\bb}{\langle}
\newcommand{\kk}{\rangle}
\newcommand{\bk}[4]{\bb #1\,#2 \!\mid\! #3\,#4 \kk}
\newcommand{\kb}[4]{\mid\!#1\,#2 \!\mid}
\newcommand{\kx}[2]{\mid\! #1\,#2 \kk}
\def\notp{{\not\! p}}
\def\notk{{\not\! k}}
\def\up{{\uparrow}}
\def\down{{\downarrow}}
\def\bfb{{\bf b}}
% ==========   defs
%
 %
% \subsection{$\lowercase{v}_2$ in the DWEF Model}
%
\chapter{$\lowercase{v}_2$ in the DWEF Model}
\label{DWEFv2}
Once results for central collisions have been calculated, the next interesting quantity is the elliptic flow coefficient $v_2$ in non-central collisions.  %If one assumes the medium is in the form of a very long tube, such that the optical potential is approximately independent of the beam ($z$) direction, the applicable expression to calculate is
%%
% v_2 = \langle \cos(2 \phi_p) \rangle =  \frac {} {\int d^4x S_0(x,\bfp')}}
%
%
A few pieces need to be generalized for a non-rotationally symmetric system.  For the transverse density $\rho$ we take the modified Woods-Saxon profile from Ref.~\cite{Retiere:2003kf}.
%
% Whereas in the original DWEF formalism only a rotationally symmetric transverse density $\rho(b)$ was needed (corresponding to central collisions), here we are interested in azimuthal anisotropy and so we need to consider a more general form.  Specifically we take the modified Woods-Saxon profile from Ref.~\cite{Retiere:2003kf}
%
\begin{equation}
\label{rho}
 \rho (\textbf{b}) = \frac{(\exp[(-1) \frac{R_{\rm{WS}}}{a_{\rm{WS}}}] + 1)^2}{(\exp[(b \sqrt{\frac{\cos^2 \phi}{R_x^2} + \frac{\sin^2 \phi}{R_y^2}}-1) \frac{R_{\rm{WS}}}{a_{\rm{WS}}}] + 1)^2},
\end{equation}
with $R_{\rm{WS}} = \sqrt{\frac 1 2 (R_x^2 + R_y^2)}$.  Thus lines of constant density in the transverse plane form ellipses with semimajor to semiminor axis ratio $\frac {R_y}{R_x}$.

Next we must generalize the fluid velocity $u$, for which we again defer to Ref.~\cite{Retiere:2003kf}.  %It is parametrized %(in $(\tau,x,y,\eta)$ coordinates) 
%using a transverse fluid rapidity $\eta_t(\textbf{b})$
%too wide
\begin{equation}
u^{\mu} (x) = (\cosh \eta \cosh \eta_t,\ \sinh \eta_t \cos \phi_b,\ \sinh \eta_t \sin \phi_b,\ \sinh \eta \cosh \eta_t).
\end{equation}
The transverse direction is taken to be perpendicular to lines of constant density.  It can be shown that the angle of such a  fluid velocity, $\phi_b$, obeys \cite{Retiere:2003kf}
\begin{equation}
\phi_b (\phi) = \tan^{-1} (\frac{R_x^2}{R_y^2} \tan \phi).
\end{equation}

The transverse fluid rapidity $\eta_t(\textbf{b})$ is first taken to have the same elliptic symmetry as the density, increasing linearly with the ``radial'' coordinate $\tilde{b} \equiv \sqrt{\frac{(b\cos(\phi))^2}{R_x^2}+\frac{(b\sin\phi))^2}{R_y^2}}$.  Then added to this is a term proportional to $\cos(2 \phi)$ representing the amount of elliptic flow built up before freezeout
\begin{equation}
\label{rapidity}
 \eta_t(\textbf{b}) = \eta_f\; b \sqrt{\frac{\cos^2 \phi}{R_x^2} + \frac{\sin^2 \phi}{R_y^2}} (1 + a_2\; \cos(2 \phi)).
\end{equation}

The momentum in these coordinates takes the form
\begin{equation}
p^\mu = (M_\perp \cosh Y , p_\perp \cos \phi_p , p_\perp \sin \phi_p , M_\perp \sinh Y).
\end{equation}
Again we choose to focus on data at midrapidity, $Y = 0$, and so 
\begin{equation}
 p \cdot u = M_{\perp} \cosh \eta\; \cosh \eta_t - p_{\perp} \sinh \eta_t\; \cos(\phi_b - \phi_p).
\end{equation}

Thus there are two extra parameters that characterize the departure from cylindrical symmetry.  In all, the parameters involved in this model are: $\Delta\eta$, $\Delta\tau$, $\tau_0$, $\mu_\pi$, $T$, $w_0$, $w_2$, $R_x$, $R_y$, $a_{\rm{WS}}$, $\eta_f$, and $a_2$.   Rather than rerunning the fit for peripheral collisions (which would be prohibitively difficult numerically) we will here be interested in the effect of an optical potential like those found to give the best fit to central collision data, and therefore we will only adjust adjust $\frac {R_y} {R_x}$ and $a_2$ to give reasonable results for non-central collisions.
%
%
%We are interested in the effect of an optical potential like the one found to give the best fit in Ref.~\cite{Miller:2005ji} and so we will keep all of these parameters fixed to those best-fit values, and only adjust $\frac {R_y} {R_x}$ and $a_2$ to give reasonable results for non-central collisions.
%
%It should be noted that the formalism developed is not strictly correct when the optical potential is complex.  (See the discussion in Ref.~\cite{Luzum:2008tc}.)  We therefore also investigate the best fit values of Ref.~\cite{Luzum:2008tc} for a vanishing imaginary part of the optical potential.
%
\section{Calculating $v_2$}
\label{sec:calculation}
This section outlines how the calculation of $v_2$ is carried out.  A set of coupled differential equations must be solved numerically to obtain the wavefunctions $\psi_p^{(-)}$, and then a five-dimensional integral must be performed (two of which can be done analytically with suitable approximations.)
\subsection{The Wavefunctions $\psi_p^{(-)}(x)$}
$\psi_p^{(-)}$ satisfies \autoref{waveequation}.
% \begin{equation}
%  \left( \nabla^2 - \frac{\partial^2}{\partial t^2}-U(\textbf{b})-m_{\pi}^2\right) \psi_p^{(-)}(x) = 0
% \end{equation}
Since $U(\textbf{b})$ is independent of $t$ and $z$, we can write
\begin{equation}
 \psi_p^{(-)}(x) \equiv e^{- i \omega_p t} e^{i p_z z} \psi_p^{(-)}(\textbf{b}),
\end{equation}
and \autoref{waveequation} becomes
\begin{equation}
  \left( \nabla_{\perp}^2 - U(\textbf{b})+p_{\perp}^2 \right) \psi_p^{(-)}(\textbf{b}) = 0,
\end{equation}
or
\begin{equation}
  \left( \frac{\partial^2}{\partial b^2} + \frac{1}{b} \frac{\partial}{\partial b} + \frac{1}{b^2} \frac{\partial^2}{\partial \phi^2} - U(\textbf{b})+p_{\perp}^2 \right) \psi_p^{(-)}(\textbf{b}) = 0.
\end{equation}

Decomposing $\psi_p^{(-)}$ and $U(\textbf{b})$ into angular moments
\begin{align}
 \psi_p^{(-)}(\textbf{b}) &= \sum_{m = -\infty}^{\infty} f_m(p,b){(-i)}^m e^{i m\; (\phi - \phi_p)}, \\
\label{Um}
  U(\textbf{b}) &\equiv \sum_n U_n(b) e^{i n \phi},
\end{align}
results in
%too wide
\begin{equation}
\sum_m \left[  \left( \frac{\partial^2}{\partial b^2} + \frac{1}{b} \frac{\partial}{\partial b} - \frac{m^2}{b^2} + p_{\perp}^2 \right) f_m - \sum_n U_n\ f_{m-n}\; i^n e^{i n \phi_p} \right] e^{i m\phi}e^{-i m \phi_p} = 0.
\end{equation}

So the term in brackets vanishes identically for each $m$, and we must solve a set of coupled differential equations. In practice, every $f_m$ above a certain $m_{max}$ is set to zero, and a finite set of coupled equations is solved numerically.

The boundary conditions are the same as for the cylindrically symmetric case---far outside the medium one should have a canonically normalized plane wave plus an outgoing wave, i.e.
\begin{equation}
\label{boundary}
 f_m (b \gg R_{\rm{WS}}) = J_m (p\ b) + T_m H_m^{(1)} (p\ b)
\end{equation}
with $J_m$ and $H_m^{(1)}$ Bessel functions and Hankel functions of the first kind, respectively.

Details of this calculation can be found in appendix \ref{details}.  The program used to calculate the wavefunctions was tested in part by comparing to a semi-analytic solution described in appendix \ref{analytic}.
\subsection{Integration}
Once the wavefunctions are found, the integrals must be performed:
\begin{equation}
\label{v2}
 v_2 \equiv  \langle \cos(2 \phi_p)\, \rangle =  \frac{\int d\phi_p\, \cos(2\, \phi_p)  S(p)}{\int d\phi_p\,  S(p)}.
\end{equation}
with
\begin{equation}\begin{split}
 S(p) &= {\int d^4x\,d^3x'{d^3p'\over(2\pi)^3}e^{-i\bfp'\cdot\bfx'}
\psi_\bfp^{(-)}(\bfx+\bfx'/2)\psi_\bfp^{(-)*}(\bfx-\bfx'/2)S_0(x,\bfp')} \\
&= {\int \tau\, d\tau\, d\eta\, b\, db\, d\phi\,d^3x'{d^3p'\over(2\pi)^3}e^{-i\bfp'\cdot\bfx'}
\psi_\bfp^{(-)}(\bfx+\bfx'/2)\psi_\bfp^{(-)*}(\bfx-\bfx'/2)S_0(x,\bfp')}
\end{split}\end{equation}
%
%If one assumes the medium is in the form of a very long tube, such that the optical potential is approximately independent of the beam ($z$) direction, and uses the large source approximation \cite{Miller::2005ji}
%
%
%
%Once the wavefunctions are found, a ten-dimensional integral must be performed:
%\begin{equation}
% v_2 = \frac {\int d\phi_p\, \cos(2\, \phi_p) \int d^4x\,d^3x'{d^3p'\over(2\pi)^3}e^{-i\bfp'\cdot\bfx'}
%\psi_\bfp^{(-)}(\bfx+\bfx'/2)\psi_\bfp^{(-)*}(\bfx-\bfx'/2)S_0(x,\bfp')}
%{\int d\phi_p\, \int d^4x\,d^3x'{d^3p'\over(2\pi)^3}e^{-i\bfp'\cdot\bfx'}
%\psi_\bfp^{(-)}(\bfx+\bfx'/2)\psi_\bfp^{(-)*}(\bfx-\bfx'/2)S_0(x,\bfp')}
%\end{equation}

%\begin{equation}
% v_2 = \frac {\int d\phi_p\, \cos(2\, \phi_p) \int d^4x\,d^2b'{d^3p'\over(2\pi)^3}e^{-i\bfp'\cdot\bfx'}
%\psi_\bfp^{(-)}({\bf b}+{\bf b}'/2)\psi_\bfp^{(-)*}({\bf b}-{\bf b}'/2)S_0(x,\bfp')}
%{\int d\phi_p\, \int d^4x\,d^2b'{d^3p'\over(2\pi)^3}e^{-i\bfp'\cdot\bfx'}
%\psi_\bfp^{(-)}({\bf b}+{\bf b}'/2)\psi_\bfp^{(-)*}({\bf b}-{\bf b}'/2)S_0(x,\bfp')}
%\end{equation}
%
Several approximations can make this more numerically tractable.  
%
%
%\begin{equation}
%\label{v2}
% v_2 =  \frac{\int d\phi_p \; \cos(2\; \phi_p)  \int d^4x\; S(p,x)}{\int d\phi_p  \int d^4x\; S(p,x)}.
%\end{equation}
%
If one assumes the the optical potential is approximately independent of the beam direction $z$ as well as time, the $\tau$ integral can be done analytically
\begin{equation}
\int \tau d\tau e^{\frac{-(\tau - \tau_0)^2}{2 \Delta\tau^2}} = \sqrt{2 \pi} \tau_0 \Delta\tau.
\end{equation}
The $\eta$ integral can also be done analytically with the following approximations (as in Ref.~\cite{Miller:2005ji})
\begin{align}
e^{\frac{-\eta^2}{2\Delta\eta^2}} \approx&\ e^{\frac{1}{\Delta\eta^2}}e^{-\frac{\cosh \eta}{\Delta\eta^2}} \\
\frac{1}{e^{(p \cdot u - \mu_{\pi})/T}-1} \approx& \sum_{j = 1}^{j_{max}} e^{(-p \cdot u + \mu_{\pi})j/T},
\end{align}
where the Bose-Einstein distribution is approximated by a sum over Boltzmann distributions truncated at some $j_{max}$, and so
%too wide
\begin{equation}
 \int d\eta\; \cosh \eta\; e^{-\cosh \eta (\frac{1}{\Delta\eta^2}+ \frac{M_{\perp}j}{T} \cosh \eta_t)} = 2 K_1 \left(\frac{1}{\Delta\eta^2}+ \frac{j}{T}M_{\perp}\; \cosh \eta_t \right).
\end{equation}
Finally, we use the large source approximation \cite{Miller:2005ji}
\bea 
\psi_{\bfp_i}^{(-)}(\bfb+\bfb'/2) 
\psi_{\bfp_j}^{(-)*}(\bfb-\bfb'/2)\;g(\bfb'^2)\approx
\psi_{\bfp_i}^{(-)}(\bfb)
\psi_{\bfp_j}^{(-)*}(\bfb)\;g(\bfb'^2)\exp{(i\bfK_\perp\cdot\bfb')},\label{LSA}
\eea 
with
\begin{equation}
g(\bfb'^2)=2 \int d^2K_\perp\;M_\perp\exp\left[{- M_\perp \cosh\eta 
               \over T}\right]\;\exp\left[-i\bfK_\perp\cdot\bfb'\right].
\end{equation}
After implementing all these approximations,  for the numerator we have
%too wide
\begin{eqnarray}
\label{integrals}
\lefteqn{\int d\phi_p\;\cos(2 \phi_p) \int d^4x\; S(p,x)} \nonumber \\
 &=&  \frac{2\; \tau_0 M_\perp }{(2\pi)^3}\; e^{\frac{1}{\Delta\eta^2}}\sum_{m,n,j} i^{n-m}e^{\frac{\mu_\pi j}{T}} \nonumber \\
&& \times  \int d^2b\; \rho(\textbf{b}) f_m(p,b)\;f_n^*(p,b)\; e^{i(m-n)\phi} K_1 \left(\frac{1}{\Delta\eta^2}+ \frac{j}{T}M_{\perp}\; \cosh \eta_t \right)\nonumber \\
&& \times  \int d\phi_p\;\cos(2 \phi_p) e^{-i(m-n)\phi_p} e^{\frac j T p_\perp \sinh(\eta_t)\cos(\phi_b-\phi_p)},
\end{eqnarray}
and similarly for the denominator.  The final three integrals are done numerically.

More details of this part of the calculation can also be found in appendix \ref{details}.
\section{Results}
\label{sec:results}
\begin{table*}
%  \centering
  \caption[DWEF model best fit parameter values used in $v_2$ calculation]{Best fit parameter sets.  The top line (Fit 1) is a general fit \cite{Miller:2005ji} while the bottom line (Fit 2) is from a fit where $Im(w_2)$ is held at 0.0001 \cite{Luzum:2008tc}. \label{table}}
  \vspace{0.1cm}
  \begin{tabular}{|c|ccccccccc|}\hline
  & $T$ & $\eta_f$ & $\Delta\tau$ &$R_{\rm{WS}}$ & $a_{\rm{WS}}$& $w_0$& $w_2$ &$\tau_{0}$ &$\Delta\eta$\\% & $\mu_\pi$\\
  & (MeV) & & $(\frac {\rm{fm}} {c})$ &(fm) & (fm)& $(\rm{fm}^{-2})$& & $(\frac {\rm{fm}} {c})$ & \\%& (MeV)\\
\hline
 Fit 1: & 156.58 & 1.310 & 2.0731 & 11.867 & 1.277 & 0.0693 & 0.856+$i$0.116 & 9.04 & 1.047\\% & 139.57\\
 Fit 2: & 121 & 1.05 & 0 & 11.7 & 1.11 & 0.495 & 0.762+$i$0.0001 & 9.20 & 70.7\\% & 139.57\\
     \hline
 \end{tabular}
\end{table*}
We would like to determine the effect of adding final state interactions to hydrodynamic fits.  To gain insight into this, we consider an emission function with parameter values taken from Refs.~\cite{Miller:2005ji,Luzum:2008tc}, which give the best description of the single particle data in general, and also with the imaginary part of the optical potential held at zero (see Table~\ref{table}.  Also note that in both fits the chemical potential was fixed at the pion mass).  

We must make alterations to this central collision model to approximate a more peripheral collision.  The results for a central collision do not unambiguously imply what a peripheral collision will look like without appealing to a particular model for the dynamics of the system.  We therefore choose reasonable parameters to approximately represent a collision with impact parameter $\sim 7$ fm, and then see how the resulting $v_2$ depends on the strength of the optical potential.  In principle one could vary all the parameters and do a separate fit of all the relevant experimental data (multiplicity, HBT radii, $v_2$, etc.) for each of various collision centralities.  However, the computing time to do so would be prohibitive, and here we are most interested in investigating only the effect of the interactions, so we proceed as follows.

First, as in Ref.~\cite{Miller:2005ji}, we scale down $R_{\rm{WS}}$, $a_{\rm{WS}}$, and $\tau_0$ by the number of participants to the 1/3 power, with $N_{part}$ taken from the Glauber model (with the same parameters used in Ref.~\cite{Luzum:2008cw}) for an impact parameter of 0 and 7 fm ($N_{part} = 377.5$ and $171.544$).  Specifically $R_{\rm{WS}} \rightarrow 0.7688 R_{\rm{WS}}$.   Then we adjust the ratio $\frac{R_y}{R_x}$ such that the spatial eccentricity
\begin{equation}
\label{ecc}
 \epsilon \equiv \frac{\langle y^2\rangle - \langle x^2\rangle}{\langle y^2\rangle + \langle x^2\rangle} = \frac{R_y^2 - R_x^2}{R_y^2 + R_x^2}
\end{equation}
has a value of 0.035.  This is a reasonable value corresponding to the spatial eccentricity at freezeout of hydrodynamic fits of peripheral collisions with impact parameter $\sim7$ fm.  Note that the brackets in \autoref{ecc} indicate a spatial average with weight given by \autoref{rho}, while the spatial eccentricity in hydrodynamic simulations are typically given with respect to, e.g., energy density.  We nevertheless keep the eccentricity from \autoref{ecc} fixed at this value with an understanding that it is only a rough but still realistic guide to the shape.

\begin{figure}
 \includegraphics{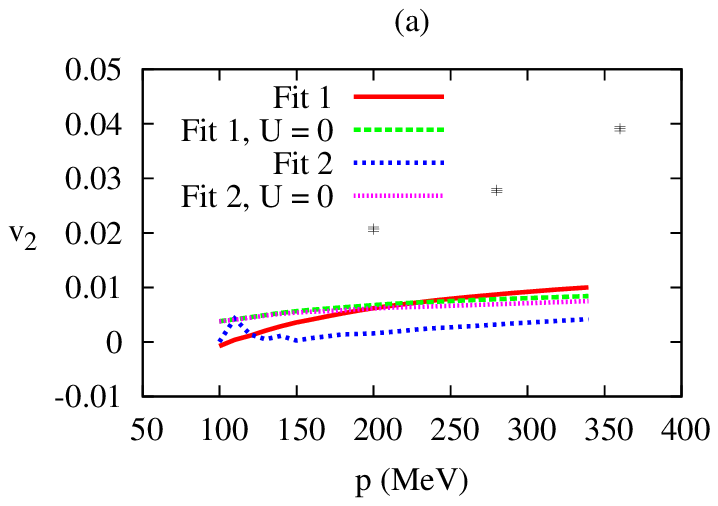}
 \includegraphics{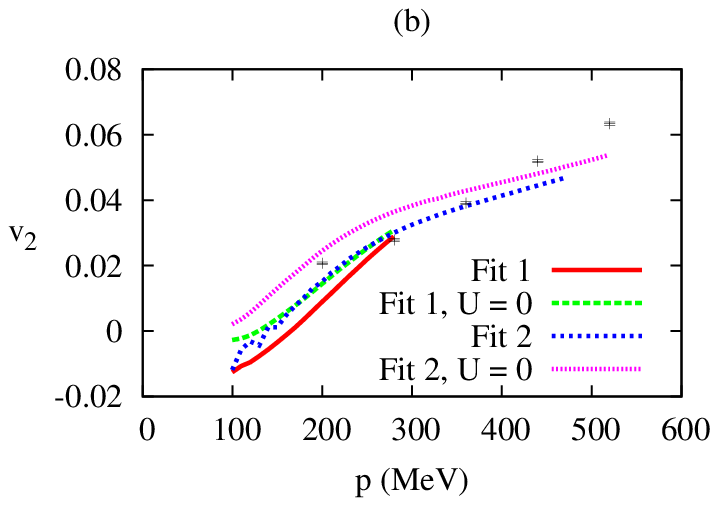}
 \caption[DWEF model calculated $v_2$ as a function of momentum]{\label{a0}Calculated $v_2$ as a function of momentum with $a_2 = 0$ (a) and $a_2 = 0.11, 0.10$ (Fit 1, 2) (b).  Points with error bars are experimental data for pions at 20--30\% centrality from the STAR Collaboration \cite{Adams:2004bi}.}
\end{figure}

Lastly we must specify how much elliptic fluid flow is built up in earlier stages of the collision, represented by the value of $a_2$ (recall \autoref{rapidity}).  First we set $a_2 = 0$ and see what $v_2$ is generated by interactions with the optical potential in the absence of significant elliptic fluid flow (\autoref{a0}(a)).  The calculated elliptic flow coefficient $v_2$ is plotted as a function of momentum, along with the relevant experimental data.  (Note that $\textbf{p}$ in our calculation is the momentum of an asymptotically free pion detected far outside the medium, not the momentum of a particle as it is emitted inside the medium, and can therefore be compared directly to experiment.)  Although we are only able to calculate up to a limited momentum, it is clear that final state interactions alone do not generate an appreciable value for $v_2$ for either the general best-fit parameters (Fit 1) or those with a vanishing imaginary part of the optical potential (Fit 2).

% The calculated $v_2$ is plotted against the relative ``strength'' of the optical potential for pions detected with asymptotic momentum $p = 250$ MeV.  That is, the optical potential (specifically the parameters $w_0$ and $w_2$ given in the appropriate line of Table~\ref{table}) is multiplied by a coefficient that ranges from zero to one.  Thus the effect of the interactions can be visualized as the optical potential is ranged from zero to its full best-fit value.

% \begin{figure}
%  \includegraphics{v2a0.eps}
%  \includegraphics{v2.eps}
%  \caption{\label{data}Experimental $v_2$ for pions at 20--30\% centrality from STAR collaboration \cite{Adams:2004bi} }
% \end{figure}

% The relevant experimental data is plotted in \autoref{data}.  At $p = 250$ MeV, and 20--30\% centrality (corresponding to $\sim7.6$ fm impact parameter), $v_2 \approx .025$. One can see that in the case where little elliptic flow has built up ($a_2 = 0$), final state interactions with the optical potential do not generate an appreciable value for $v_2$ for either the general fit parameters (Fit 1) or those with a vanishing imaginary optical potential (Fit 2)..

Next we increase $a_2$ such that the experimental value for $v_2$ is roughly obtained  (\autoref{a0}(b)).  A value of $a_2 = 0.11$ was required for the parameters from Fit 1, while $a_2 = 0.10$ was sufficient to bring the emission function from Fit 2 into the physical regime.  One can see that the optical potential has a small but non-negligible effect---it decreases $v_2$ on the order of 10--25\% of its zero-interaction value with a slightly smaller effect as momentum increases.
\section{Conclusion of DWEF $v_2$ Calculation}
\label{sec:conclusion}
Final state interactions in the DWEF model were found to have a small, though not entirely insignificant effect on the elliptic flow coefficient $v_2$.  
% It is reasonable to conclude that final state interactions can affect the calculated value of $v_2$ by as much as ~20\%, and so must be properly taken into account to have confidence in the quantitative predictions of hydrodynamic simulations at that level of precision.
This is in addition to the indirect effect of adding final state interactions.  For example, adding an optical potential changes other observables such as the multiplicity, which would alter parameters in a hydrodynamic fit such as freezeout temperature, which would then in turn have an effect on the calculated value of $v_2$.

The precise size of these effects in general can only be determined with a better understanding of the model fits (e.g. Fit 1 versus Fit 2) in addition to a more detailed analysis---a full parameter search using all the relevant experimental data, or perhaps even by adding final state interactions directly into hydrodynamic simulations
(i.e. a hydrodynamic afterburner in the vein of, e.g., Refs.~\cite{Bass:2000ib, Hirano:2005xf, Nonaka:2006yn, Teaney:2001av, Petersen:2008dd}).  It is reasonable, however, to conclude that final state interactions can affect the calculated value of $v_2$ by as much as $\sim20$\% (in agreement with other investigations of final state interactions, e.g., Ref.~\cite{Teaney:2001av}), and so must be properly taken into account to have confidence in the quantitative predictions of hydrodynamic simulations at that level of precision.
\begin{appendices}
% ========== Appendix A -- DWEF
%\appendix
%\begin{appendices}

\chapter{Details of DWEF $\lowercase{v}_2$ Calculation}
\label{v2appendix}
\section{Numerical Implementation}
\label{details}
A program was written in C++, making use of the \href{http://www.gnu.org/software/gsl/}{GNU Scientific Library (GSL)} version 1.9, to do the calculation of $v_2$, as detailed here.% for anyone who may be interested.

The integral over the azimuthal angle of the pion momentum, $\phi_p$ is done as a sum using a simple trapezoid rule.  This is because for each different value of $\phi_p$, a new set of differential equations must be solved.  This also allows for the numerator and denominator of \autoref{v2} to be solved simultaneously, with just a factor of $\cos(2 \phi_p)$ multiplied to the numerator when adding terms to the sum.

For each term in the sum, then, first the wavefunctions $\psi_p^{(-)}$ are obtained.  They obey a set of coupled differential equations of the form
\begin{equation}
\left( \frac{\partial^2}{\partial b^2} + \frac{1}{b} \frac{\partial}{\partial b} - \frac{m^2}{b^2} + p_{\perp}^2 \right) f_m - \sum_n U_n\ f_{m-n}\; i^n e^{i n \phi_p} = 0
\end{equation}
for all integers $m$.  This set is truncated, since large $m$ moments ($f_m$ for $m > p_\perp R_{ws}$) contribute little to the wavefunction.  Therefore, all $f_m$ for $m$ greater than some $m_{max}$ are set to zero, leaving a finite ($2 m_{max} + 1$) number of coupled ordinary differential equations.  These are solved by calling a GSL solver.  Using an embedded Runge-Kutta-Fehlberg method seemed to give the best performance.  For these solutions, \autoref{rho} is integrated numerically to find the moments $U_n$.  This is done with the GSL adaptive integration routine for oscillatory functions.

To match to the proper boundary conditions, one must find ($2 m_{max} + 1$) linearly independent solutions to this set of equations and take the correct linear combination of these solutions that matches the desired boundary conditions.  The straightforward choice for these linearly independent solutions is to sequentially solve for the case where only one of the partial waves is non-zero near the origin.  For example, for the n'th solution let:
\begin{eqnarray}
 f_m (b = b_{min} << \frac 1 p) &=& \delta_{m,n} \nonumber \\
 f_m'(b_{min}) &=& \frac m b \delta_{m,n}
\end{eqnarray}
and then solve the set of differential equations up to some arbitrarily large $b_{max}$ far outside the potential. We can then match each partial wave in this $n^{th}$ solution to the form:
\begin{equation}
 f_{m,n} (b_{max}) =  A_{m,n} J_m (p\ b) + B_{m,n} H_m^{(1)} (p\ b).
\end{equation}

The final wavefunction is then given by the linear combination of these solutions that matches the form of \autoref{boundary} at $b_{max}$:
\begin{equation}
 f_m (b) = \sum_n C_n f_{m,n}(b).
\end{equation}

This part of the program was tested with the trivial case of zero optical potential, in addition to comparing to a separately written program that calculates only the cylindrically symmetric case, as well as to the results of the semi-analytical test case described in appendix \ref{analytic}. %\autoref{analytic}
 
Once these wavefunctions are obtained and stored in memory, the integral over $b$ and $\phi$ in \autoref{integrals} can be performed in addition to the sum over Boltzmann factors.  The integrations are done with two GSL adaptive integration routines, one embedded in the other.  The sum is done inside the argument of the integrals.
%
%\chapter{Semi-Analytic Test Case}
\section{Semi-Analytic Test Case}
\label{analytic}
To test the numerics, the case of a pion moving through an elliptically-shaped step-function potential was solved (semi-)analytically making use of elliptic coordinates.  This can be compared to the case of $a_{ws}\rightarrow 0$ (see  \autoref{DWEFv2}).

We want to solve \autoref{waveequation} with $U{(\textbf{b})}$ an elliptically shaped step function---a finite potential inside an ellipse in the transverse plane, with zero potential outside.

It is useful to change to elliptic (cylindrical) coordinates, denoted $u$ and $v$.  Think of $u$ as a 'radial' coordinate that runs from 0 to $\infty$ and $v$ as an 'angular' coordinate that runs from 0 to $2 \pi$
\begin{eqnarray}
x = a\ \cosh(u)\ \cos(v) \nonumber \\ 
y = a\ \sinh(u)\ \sin(v).
\end{eqnarray}
Note the major and minor axes of the resulting confocal ellipses are reversed from the shape of the density used in the main calculation (which is larger in the $y$ direction).  This is to maintain consistency with the conventional definition of elliptic coordinates.   At the end one can simply take $\phi_p \to (\phi_p + \pi)$ to match the usual convention in RHIC papers.

Consider the case
\begin{equation}
U(\textbf b) = U(u) = U_0\ \Theta(u_0 - u).
\end{equation}
The sharp boundary at $u = u_0$ is an ellipse with major and minor axes
\begin{eqnarray}
R_x = a\ \cosh(u_0) \nonumber \\
R_y = a\ \sinh(u_0).
\end{eqnarray}

In this coordinate system the Laplacian is
\begin{equation}
\nabla^2_\perp = \frac{1}{a^2 \left( \sinh^2(u)+\sin^2(v) \right)} \left( \frac{\partial^2}{\partial u^2} + \frac{\partial^2}{\partial v^2} \right)
\end{equation}
and so \autoref{waveequation} becomes
%too wide
\begin{equation}
\left[ \frac{1}{a^2 \left( \sinh^2(u)+\sin^2(v) \right)} \left( \frac{\partial^2}{\partial u^2} + \frac{\partial^2}{\partial v^2} \right) - U(u) + p^2 \right] \psi_p ({\bf b}) = 0
\end{equation}
or equivalently
%too wide
\begin{equation}
\left[ \frac{\partial^2}{\partial u^2} + 2q(u)\ \cosh(2u)  + \frac{\partial^2}{\partial v^2} - 2 q(u)\ \cos(2v) \right] \psi_p ({\bf b}) = 0
\label{el}
\end{equation}
with 
\begin{equation}
q(u) = \frac{a^2}{4} \left( p^2 - U(u) \right).
\end{equation}
On the inside of the potential and on the outside separately, $q(u)$ does not depend on $u$ and these cases can be solved with separation of variables and the solutions patched together at $u = u_0$.
Let
\begin{eqnarray}
q_{in}& = &\frac{a^2}{4} \left( p^2 - U_0 \right) \nonumber \\
q_{out}& = &\frac{a^2}{4} p^2.
\end{eqnarray}

Start by expanding $\psi_p ({\bf b})$ in terms of so-called elliptic sines and cosines of the 'angular' variable $v$.  They are solutions of `Mathieu's equation' \cite{gradshteyn_table_2000}:
%\cite{Gradshteyn:2000}:
\begin{equation}
\left( - \frac{\partial^2}{\partial v^2} + 2 q\ \cos(2v) \right) C(\alpha,q,v) = \alpha \ C(\alpha,q,v) \label{Mathieu}.
\end{equation}
The general solutions are called `Mathieu functions,' usually denoted $C(\alpha,q,v)$ for solutions even in the coordinate $v$ and $S(\alpha,q,v)$ for odd.  Demanding periodicity of the variable $v$ allows only certain discreet eigenvalues $\alpha$ (denoted here $\alpha_n$ for the even functions and  $\beta_n$ for the odd functions).  This (complete) set of periodic solutions is commonly called elliptic sines and elliptic cosines:

\begin{eqnarray}
C(\alpha_n,q,v) \equiv ce_n (v,q) \nonumber \\
S(\beta_n,q,v) \equiv se_n (v,q).
\end{eqnarray}

The general solution of \autoref{el} can be written in terms of these elliptic sines and cosines:
\begin{equation}
\psi_p ({\bf b}) \equiv \sum_{n=0}^\infty \left[ f_{c_n}(u) ce_n (v,q) + f_{s_n}(u) se_n (v,q) \right].
\end{equation}

Plugging this in to \autoref{el} gives
\begin{eqnarray}
\left[ \frac{\partial^2}{\partial u^2} + 2q\ \cosh(2u)  - \alpha_n \right] f_{c_n} (u) = 0 \\
\left[ \frac{\partial^2}{\partial u^2} + 2q\ \cosh(2u)  - \beta_n \right] f_{s_n} (u) = 0.
\end{eqnarray}

This is called the modified Mathieu equation, which can be obtained from \autoref{Mathieu} by replacing $v \to (i\ u)$. Note that the eigenvalues are different for the functions corresponding to $ce_n$ and $se_n$ ($f_{c_n}$ and $f_{s_n}$ above, respectively). The general solution is then the same as for the original Mathieu equation, analytically continued with $v \to (i\ u)$, though typically they are organized by boundary conditions analogous to Bessel and Neumann functions (denoted $Je_n(u,q)$, $Ne_n(u,q)$, etc.) \cite{visual}:

\begin{eqnarray}
f_{c_n} (u) = C_{c_n} Je_n(u,q) + S_{c_n} Ne_n(u,q) \\
f_{s_n} (u) = C_{s_n} Jo_n(u,q) + S_{s_n} No_n(u,q).
\end{eqnarray}

Note that there are many different sets of so-called Mathieu functions, each being a complete orthogonal basis.  Replacing $q_{in}$ with $q_{out}$ results in a different basis, and there are separate sets of modified Mathieu functions corresponding to the eigenvalues of the elliptic sines and elliptic cosines ($\alpha_n$ and $\beta_n$).

By requiring continuity at the $u=0$ line segment one finds that the general solution inside the potential is:
%too wide
\begin{equation}
\psi_p^{in}(u,v) = \sum_n \left[ Ce^{in}_n Je_n(u,q_{in}) ce_n(v,q_{in}) + Co^{in}_n Jo_n(u,q_{in})se_n(v,q_{in}) \right]
\end{equation}
with undetermined coefficients $Ce^{in}_n, Co^{in}_n$.

Outside, we write the solution as the sum of a plane wave and an outgoing wave \cite{McLachlan:1947}
%too wide
\begin{eqnarray}
\psi_p^{out}(u,v) = \nonumber \\
\sum_n [ \left( \frac{1}{p_n} Je_n(u,q_{out}) + Ce^{out}_n He_n^{(1)}(u,q_{out})\right) ce_n (v,q_{out}) ce_n(\phi_p,q_{out}) \nonumber \\
+\left( \frac{1}{s_n} Jo_n(u,q_{out}) + Co^{out}_n Ho_n^{(1)}(u,q_{out})\right) se_n (v,q_{out}) se_n(\phi_p,q_{out})],
\end{eqnarray}
% \begin{align}
% \psi_p^{out}(u,v) =& \nonumber \\
% &\sum_n [ \left( \frac{1}{p_n} Je_n(u,q_{out}) + Ce^{out}_n He_n^{(1)}(u,q_{out})\right) ce_n (v,q_{out}) ce_n(\phi_p,q_{out}) \nonumber \\
% &+\left( \frac{1}{s_n} Jo_n(u,q_{out}) + Co^{out}_n Ho_n^{(1)}(u,q_{out})\right) se_n (v,q_{out}) se_n(\phi_p,q_{out})]
% \end{align}
where the H's are analogous to Hankel functions
\begin{eqnarray}
He^{(1)}_n (u,q) \equiv Je_n (u,q) + i\ Ne_n (u,q) \\
Ho^{(1)}_n (u,q) \equiv Jo_n (u,q) + i\ No_n (u,q)
\end{eqnarray}
and the plane wave coefficients $p_n$ and $s_n$ are
\begin{eqnarray}
 \frac{1}{p_n}  =  \frac 1 \pi \int_0^{2\pi} dv\ e^{i p\cdot x} ce_n (v,q_{out}) \\
 \frac{1}{s_n}  =  \frac 1 \pi \int_0^{2\pi} dv\ e^{i p\cdot x} se_n (v,q_{out}).
\end{eqnarray}
The coefficients $Ce^{out}_n$ and $Co^{out}_n$, along with the analogous 'inside' coefficients are determined by matching boundary conditions.

To match at the $u = u_0$ boundary, project the 'inside' angular functions\\ (e.g. $ce_n(v,q_{in})$) in terms of the 'outside' ones (e.g. $ce_n(v,q_{out})$).
\begin{eqnarray}
ce_j(v,q_{in}) = \sum_{n=0}^\infty B^c_{jn} ce_n(v,q_{out}) \\
se_j(v,q_{in}) = \sum_{n=0}^\infty B^s_{jn} se_n(v,q_{out}),
\end{eqnarray}
with
\begin{eqnarray}
B^c_{jn} = \frac{1}{\pi} \int_0^{2\pi} dv ce_j(v,q_{in})\ ce_n(v,q_{out}) \\
B^s_{jn} = \frac{1}{\pi} \int_0^{2\pi} dv se_j(v,q_{in})\ se_n(v,q_{out}).
\end{eqnarray}

Then the 'inside' wave functions are
%too wide
\begin{equation}
\psi_p^{in} = \sum_{j,n} \left[ Ce^{in}_j\ Je_j(u,q_{in})\ B^c_{jn} ce_n(v,q_{out}) + Co^{in}_j\ Jo_j(u,q_{in})\ B^s_{jn} se_n(v,q_{out})\right].
\end{equation}
The coefficients ($Ce^{in}_n$, $Co^{in}_n$, $Ce^{out}_n$, $Co^{out}_n$) can then be determined by demanding that $\psi$ and its gradient be continuous at $u = u_0$, which gives the following relations:

%\begin{widetext}

\begin{equation}
\sum_j Ce^{in}_j Je_j(u_0,q_{in}) B^c_{jn} = \frac{1}{p_n} Je_n(u_0,q_{out}) ce_n(\phi_p,q_{out}) + Ce^{out}_n He^{(1)} (u_0,q_{out}) ce_n(\phi_p,q_{out})
\end{equation}
\begin{equation}
\sum_j Co^{in}_j Jo_j(u_0,q_{in}) B^s_{jn} = \frac{1}{s_n} Jo_n(u_0,q_{out}) se_n(\phi_p,q_{out}) + Co^{out}_n Ho^{(1)} (u_0,q_{out}) se_n(\phi_p,q_{out})
\end{equation}
\begin{equation}
\sum_j Ce^{in}_j Je'_j(u_0,q_{in}) B^c_{jn} = \frac{1}{p_n} Je'_n(u_0,q_{out}) ce_n(\phi_p,q_{out}) + Ce^{out}_n He'^{(1)} (u_0,q_{out}) ce_n(\phi_p,q_{out})
\end{equation}
\begin{equation}
\sum_j Co^{in}_j Jo'_j(u_0,q_{in}) B^s_{jn} = \frac{1}{s_n} Jo'_n(u_0,q_{out}) se_n(\phi_p,q_{out}) + Co^{out}_n Ho'^{(1)} (u_0,q_{out}) se_n(\phi_p,q_{out}).
\end{equation}

%\end{widetext}

The plane wave coefficients ($p_n,s_n$) as well as the coefficients from the projection ($B^c_{jn}$, $B^s_{jn}$) must be solved numerically.  In addition, to compare to the $f_m$ in the main calculation, the resulting wavefunctions are integrated to project out the usual angular moments.  Hence the description as a ``semi-analytical'' test case.  In fact, this implementation (done in Mathematica) saves no time over the original numerical version, but it does provide an independent check.
%\end{appendices}
% ========== Appendix B -- Notation and Conventions
\chapter{Notation, Conventions, and Definitions}
\label{notation}
All notational definitions are defined when first introduced, but frequently used notation is collected here for easy reference (or at least notation that is used in well-separated parts of the manuscript).
\begin{itemize}
\item
All quantities are reported using a system of units such that $c = \hbar = k_B = 1$ (``natural units").  I.e., all velocities are measured as fractions of the speed of light $c$, etc.  
\item
The space-time metric in flat space is taken as  $g_{\mu\nu}$ = diag(1, -1, -1, -1), such that timelike 4-vectors have positive norm and spacelike vectors negative.
\item
Projectors:
\begin{align}
\Delta^{\mu \nu} & \equiv g^{\mu \nu}-u^\mu u^\nu\,,\\
P^{\mu \nu}_{\alpha \beta} & \equiv \Delta^{\mu}_\alpha \Delta^\nu_\beta+\Delta^{\mu}_\beta \Delta^\nu_\alpha
-\frac{2}{3} \Delta^{\mu \nu} \Delta_{\alpha \beta}\,,
\end{align}
such that $u_\mu\Delta^{\mu \nu}=u_\mu P^{\mu \nu}_{\alpha \beta} = g_{\mu \nu} P^{\mu \nu}_{\alpha \beta} = 0$.  Projecting with $\Delta^{\mu \nu}$ makes a quantity transverse to a fluid velocity $u^\mu$, while $P^{\mu \nu}_{\alpha \beta}$ makes it transverse, traceless, and symmetric under interchange of indices.
\item
Derivatives:
\begin{align}
D & \equiv u^\mu \partial_\mu\,,\\
\nabla_\mu & \equiv  \Delta^\alpha_\mu \partial_\alpha\,,
%\partial_\mu & \equiv  u_\mu D+\nabla_\mu\,,
\end{align}
so that $\partial_\mu  =  u_\mu D+\nabla_\mu$.  In the fluid rest frame, these are the time derivative and spatial gradient, respectively.  I.e., in the non-relativistic limit
\begin{align}
D & \approx \partial_t+\vec{v}\cdot\vec{\partial}+{\cal O}\left(|\vec{v}|^2\right)\,,\\
\vec{\nabla} & \approx -\vec{\partial}+{\cal O}\left(|\vec{v}|\right)\,.
\end{align}
\item
Brackets:
\begin{align}
A^{(\alpha}B^{\beta)} & \equiv \frac{1}{2}\left(A^\alpha B^\beta+A^\beta B^\alpha\right)\,,\\
A^{[\alpha}B^{\beta]} & \equiv \frac{1}{2}\left(A^\alpha B^\beta-A^\beta B^\alpha\right)\,, \\
A^{\langle\alpha}B^{\beta\rangle} & \equiv P^{\alpha \beta}_{\mu \nu} A^\mu B^\nu\,,
\end{align}
which are used to define $\sigma^{\mu\nu} \equiv \nabla^{\langle\mu} u^{\nu\rangle} $ and the fluid vorticity $\omega^{\mu \nu} \equiv -\nabla^{[\mu} u^{\nu]}$.
\end{itemize}
\end{appendices}
\printendnotes
%
% ==========   Bibliography
%
\nocite{*}   % include everything in the uwthesis.bib file
\bibliographystyle{bst/utphys}
\bibliography{RHIC}
%
% ==========   Appendices
%
%\appendix
%\raggedbottom\sloppy
%
%\begin{appendices}
%% ========== Appendix A -- DWEF
%\include{v2appendices}
%% ========== Appendix B -- Notation and Conventions
%\include{notation}
%%
%\end{appendices}
%
% ==========   Vita
%
\vita{
Matt was born in Benson, Minnesota in 1980.  He graduated from Benson High School in 1999 and then attended Saint John's University in Collegeville, Minnesota, where in 2003 he obtained a Bachelor of Arts degree with a major in Physics and minors in Mathematics and Chemistry.  He earned a Master of Science in 2004 and a Doctor of Philosophy in 2009, both in physics from the University of Washington.  He then moved on to a postdoctoral research position at the Institut de Physique Th\'{e}orique in Saclay, France.
}
\end{document}